\documentclass[fleqn,usenatbib,useAMS]{mnras}

\usepackage{graphicx}	
\usepackage{amsmath}	
\usepackage{amssymb}	
\usepackage{multicol}        
\usepackage{bm}		
\usepackage{pdflscape}	
\usepackage{hyperref}

\usepackage{subcaption}

\usepackage{enumitem}

\usepackage{booktabs}
\usepackage{multirow}

\usepackage[T1]{fontenc}
\usepackage{ae,aecompl}

\usepackage{newtxtext,newtxmath}

	\title[Kalman PTA]{Kalman tracking and parameter estimation of continuous gravitational waves with a pulsar timing array}

\author[Kimpson]{Tom Kimpson$^{1,2,3}$\thanks{Contact e-mail: \href{tom.kimpson@unimelb.edu.au}{tom.kimpson@unimelb.edu.au}}, Andrew Melatos$^{1,2}$, Joseph O'Leary$^{1,2}$, Julian B. Carlin$^{1,2}$, Robin J. Evans$^{4}$, \newauthor William Moran$^{4}$, Tong Cheunchitra$^{1,2}$, Wenhao Dong$^{1,2}$, Liam Dunn$^{1,2}$, Julian Greentree$^{3}$, Nicholas J. O'Neill$^{1,2}$, \newauthor Sofia Suvorova$^{4}$, Kok Hong Thong$^{1,2}$, Andrés F. Vargas$^{1,2}$%
\\
$^{1}$School of Physics, University of Melbourne, Parkville, VIC 3010, Australia \\
$^{2}$OzGrav, University of Melbourne, Parkville, VIC 3010, Australia \\
$^{3}$Department of Space and Climate Physics, University College London, Holmbury St. Mary, RH5 6NT, UK \\
$^{4}$Department of Electrical and Electronic Engineering, University of Melbourne, Parkville, Victoria 3010, Australia }

\date{Last updated \today}

\pubyear{2023}

\begin{document}
\label{firstpage}
\pagerange{\pageref{firstpage}--\pageref{lastpage}}
\maketitle

\begin{abstract}	
Continuous nanohertz gravitational waves from individual supermassive black hole binaries may be detectable with pulsar timing arrays. A novel search strategy is developed, wherein intrinsic achromatic spin wandering is tracked simultaneously with the modulation induced by a single gravitational wave source in the pulse times of arrival. A two-step inference procedure is applied within a state-space framework, such that the modulation is tracked with a Kalman filter, which then provides a likelihood for nested sampling. The procedure estimates the static parameters in the problem, such as the sky position of the source, without fitting for ensemble-averaged statistics such as the power spectral density of the timing noise, and therefore complements traditional parameter estimation methods. It also returns the Bayes factor relating a model with a single gravitational wave source to one without, complementing traditional detection methods. It is shown via astrophysically representative software injections in Gaussian measurement noise that the procedure distinguishes a gravitational wave from pure noise down to a characteristic wave strain of $h_0 \approx 2 \times 10^{-15}$. Full posterior distributions of model parameters are recovered and tested for accuracy. There is a bias of $\approx 0.3$ rad in the marginalised one-dimensional posterior for the orbital inclination $\iota$, introduced by dropping the so-called `pulsar terms'. Smaller biases $\lesssim 10 \%$ are also observed in other static parameters. 
\end{abstract}




\begingroup
\let\clearpage\relax
\endgroup
\newpage
\section{Introduction}\label{sec:intro}
The inspiral of supermassive black hole binaries \citep[SMBHBs;][]{Rajagopal1995,Jaffe_2003, Wyithe2003,Sesana2013,McWilliams_2014,Ravi2015MNRAS.447.2772R,Burke2019, Skyes2022} is predicted to emit nHz gravitational waves (GWs). Other GW sources in this low-frequency regime include cosmic strings \citep[e.g.][]{PTAstring} and cosmological phase transitions \citep[e.g.][]{PTAphase}. The detection of nHz GWs has inspired the development of new observational methods, since it is impractical to engineer terrestrial interferometric detectors with sufficiently long baselines. The foremost method is timing an ensemble of pulsars, i.e. a pulsar timing array \citep[PTA;][]{ Tiburzi2018, 2021hgwa.bookE...4V}. A nHz GW influences the trajectory and frequency of individual radio pulses, leaving a characteristic impression on the pulse times of arrival (TOAs) measured at the  Earth. By measuring TOAs from multiple pulsars simultaneously one can effectively construct a detector with a baseline on the scale of parsecs. Multiple PTA detectors have been built over the last few decades, including the North American Nanohertz Observatory for Gravitational Waves \citep[NANOGrav,][]{NANOgrav2023}, the Parkes Pulsar Timing array \citep[PPTA,][]{Parkes2023}, and the European Pulsar Timing Array \citep[EPTA,][]{EPTA2023}. These individual efforts have joined in international collaboration, under the umbrella of the International Pulsar Timing Array \citep[IPTA,][]{2019MNRAS.490.4666P}, along with a number of newer PTAs such as the Indian Pulsar Timing Array Project \citep[InPTA,][]{ipta}, MeerTime \citep{meertime2,Meertime} and the Chinese PTA \citep[CPTA,][]{Hobbs_2019}. \newline 

The incoherent superposition of multiple SMBHB sources leads to a stochastic GW background at nHz frequencies \citep{Allen1997,Sesana10,Christensen2019,Renzini2022}. Previous efforts have mainly focused on detecting the stochastic background by measuring the cross-correlation between the timing residuals from pairs of pulsars as a function of the angular separation between the pulsars -- the Hellings-Downs curve \citep{Hellings}. After multiple non-detections \citep{Lentati2015,NanoGrav2018,2022MNRAS.510.4873A} consilient evidence for the GW background was presented by NANOGrav \citep{2023ApJ...951L...8A}, EPTA/InPTA \citep{2023arXiv230616214A}, PPTA \citep{2023ApJ...951L...6R} and the CPTA \citep{2023RAA....23g5024X}. \newline

Individual SMBHBs that are sufficiently massive and nearby may be resolvable with PTAs, allowing the early stages of their evolution and coalescence to be investigated \citep{Sesana2010,Yardley2010,Zhu10,Babak2012,2013CQGra..30v4004E,Zhupulsarterms}. 
Indeed, the stochastic GW background itself may be dominated by a few individual binary sources \citep{Ravi2012singlesource}. Individual SMBHBs are continuous wave sources; they generate persistent, quasi-monochromatic modulations of a known form in pulsar timing residuals. Consequently, they are detected more efficiently by either a frequentist matched filter, e.g.\ the ${\cal F}$-statistic \citep{Lee2011MNRAS.414.3251L, Ellis2012ApJ,Zhu2014PPTA}, or else Bayesian inference \citep{Ellis2016,Arzoumanian2020A}, rather than by cross-correlating pulsar pairs. However, PTA observational campaigns to detect individual sources have been unsuccessful so far \citep{Jenet2004,Zhu2014PPTA,Babak2016,Arzoumanian2023}. Inconclusive evidence at low significance was presented recently by the EPTA for an individual source at 4--5 nHz \citep{2023arXiv230616226A}. \newline

Intrinsic pulsar timing noise -- i.e.  random, unmodelled, red-spectrum TOA fluctuations due to irregularities in the rotation of the star -- has been identified as a key factor limiting the sensitivity of PTAs to GW signals \citep{Shannon2010,Lasky2015,Caballero2016,Goncharov2021}. This timing noise has multiple theorized causes including free precession \citep{free_precession_kerr,stairs_freeprecession}, microglitches \citep{Alessandro1995,Melatos2008,Espinoza2021}, asteroid encounters \citep{Shannon_2013,Brook_2014}, glitch recovery \citep{Johnston10,Hobbs2010glitch}, fluctuations in internal and external stochastic torques \citep{Cordes1981, 2006MNRAS.370L..76U,Myers2021MNRAS.502.3113M,Meyers2021,Antonelli2023}, variations in the coupling between the stellar crust and core \citep{Jones1990MNRAS.246..364J,Meyers2021,Melatos2023}, magnetospheric state switching \citep{magneto1,Lyne2010L,Stairs2019MNRAS.485.3230S} and superfluid turbulence \citep{Greenstein1970,Peralta2006,Melatos2014}. In order to mitigate the impact of timing noise, PTAs are typically composed of millisecond pulsars (MSPs), which are relatively stable rotators. However, timing noise in MSPs may be a latent phenomenon that will increasingly assert itself as longer stretches of more sensitive data are analysed in the quest to detect nHz GWs \citep{Shannon2010}. In modern Bayesian PTA searches, the power spectral density of the red intrinsic timing noise is modeled (usually as a broken or unbroken power law) and estimated, in an effort to distinguish it from the noise induced by a stochastic GW background (whose spectrum is also red). In addition to the red timing noise there are secondary, white noise sources to consider such as phase jitter noise and radiometer noise \citep{Cordes2010,Lam2019,Parthasarathy2021}. \newline 

In this work we present an alternative and complementary approach to PTA data analysis for individual, quasi-monochromatic, SMBHB sources which self-consistently tracks the intrinsic timing noise in PTA pulsars and disentangles it from GW-induced TOA modulations. The new approach differs from existing approaches in one key respect: it infers the GW parameters conditional on the unique, time-ordered realization of the noisy TOAs observed, instead of fitting for the ensemble-averaged statistics of the TOA noise process, e.g., the amplitude and exponent of its power spectral density. Stated another way, existing approaches seek to detect a GW signal by marginalizing over the ensemble of possible noise realizations summarized by the power spectral density, whereas the new approach delivers the most likely set of GW parameters consistent with the actual, observed noise realization. The new and existing approaches are therefore complementary. In particular, we formulate PTA analysis as a state-space problem and demonstrate how to optimally estimate the state-space evolution using a Kalman filter, a tried-and-tested tool \citep{Kalman1,Meyers2021,Melatos2023}. We combine the Kalman tracking of the pulsars' intrinsic rotational states with a Bayesian nested sampler \citep{Skilling, Ashton2022} to estimate the static GW parameters and calculate the marginal likelihood (i.e. the model evidence) for model selection. The adaptive bias of the Kalman filter tracks timing noise more effectively than alternative techniques such as least-squares estimators. \newline

The paper is organised as follows. In Section \ref{sec:model} we present the state-space model for the rotational states of an array of pulsars falling freely in the curved spacetime of a single-source GW. In Section \ref{sec:detect} we develop a Kalman filter to track the state evolution and deploy the Kalman filter in conjunction with nested sampling to estimate the GW and other system parameters, along with the model evidence. In Section \ref{sec:testing} we describe how we create synthetic validation data to test the method. In Section \ref{sec:rep_example} we test the method on synthetic data for a single representative GW source. In Section \ref{sec:parameter_space} we extend the tests to cover an astrophysically relevant domain of SMBHB source parameters. In Section \ref{sec:bias_and_identifiability} we quantify the bias in the parameter estimates. In Section \ref{sec:computation_costs} we review the computational cost of the method. Conclusions are drawn in Section \ref{sec:discussion}. The data are formulated as pulse frequency time series with Gaussian measurement noise as a proof of principle and to maintain consistency with previous work \citep{Myers2021MNRAS.502.3113M,Meyers2021}. It will be necessary to modify the method to accept pulse TOAs instead of a pulse frequency time series when analysing real, astronomical data, a subtle generalization which is deferred to future work. Throughout the paper we adopt natural units, with $c = G = \hbar = 1$, and metric signature $(-,+,+,+)$. \newline

\section{State-Space Formulation}\label{sec:model}
We formulate the PTA analysis as a state-space problem, in which the intrinsic rotational state of each pulsar evolves according to a stochastic differential equation and is related to the observed pulse sequence via a measurement equation. In this work we take the intrinsic state variable to be the $n$-th pulsar's spin frequency $f_{\rm p}^{(n)}(t)$, as measured in the local, freely-falling rest frame of the pulsar's centre of mass. A phenomenological model for the evolution of $f_{\rm p}^{(n)}(t)$ is presented in Section \ref{sec:psr_frequency}.  We take the measurement variable to be the radio pulse frequency measured by an observer at Earth, $f_{\rm m}^{(n)}(t)$.  The measurement equation relating $f_{\rm m}^{(n)}(t)$ to $f_{\rm p}^{(n)}(t)$ is presented in Section \ref{sec:psr_measured}. The superscript $1\leq n\leq N$ indexes the $n$-th pulsar in the array. The subtle problem of generalizing the measurement variable to pulse TOAs is postponed to future work, as noted in Section \ref{sec:intro}.
\subsection{Spin evolution} \label{sec:psr_frequency}
A predictive, first-principles theory of timing noise does not exist at present; there are several plausible physical mechanisms, referenced in Section \ref{sec:intro}. We therefore rely on an idealized phenomenological model to capture the main qualitative features of a typical PTA pulsar's observed spin evolution, i.e.\ random, mean-reverting, small-amplitude excursions around a smooth, secular trend. In the model, $f_{\rm p}^{(n)}(t)$ evolves according to the sum of a deterministic torque and a stochastic torque. The deterministic torque is attributed to magnetic dipole braking, with braking index $n_{\rm em}=3$ for the sake of definiteness \citep{1969ApJ...157..869G}. Most PTAs involve millisecond pulsars, for which the quadratic correction due to $n_{\rm em}$ in $f_{\rm p}^{(n)}(t)$ is negligible over the observation time $T_{\rm obs} \sim 10 \, {\rm yr}$, and the deterministic evolution $f_{\rm em}^{(n)}(t)$ can be approximated accurately by 
\begin{equation}
 f_{\rm em}^{(n)}(t) = f_{\rm em}^{(n)}(t_1) + \dot{f}_{\rm em}^{(n)}(t_1)t \ , \label{eq:spinevol}
\end{equation} where an overdot denotes a derivative with respect to $t$, and $t_1$ labels the time of the first TOA. The stochastic torque is assumed to be a zero-mean, white noise process. Specifically, the frequency evolves according to an Ornstein-Uhlenbeck process, described by a Langevin equation with a time-dependent drift term \citep{Vargas},
\begin{equation}
	\frac{df_{\rm p}^{(n)}}{dt} = -\gamma^{(n)}	 [f_{\rm p}^{(n)} - f_{\rm em}^{(n)} (t)] + \dot{f}_{\rm em}^{(n)}(t) +\xi^{(n)}(t) \ . 
	\label{eq:frequency_evolution}
\end{equation}
In Equation \eqref{eq:frequency_evolution}, $f_{\rm em}^{(n)}$ is the solution of the electromagnetic spin-down equation given by Equation \eqref{eq:spinevol}, $\dot{f}_{\rm em}^{(n)}$ is the spin derivative, $\gamma^{(n)}$ is a damping constant whose reciprocal specifies the mean-reversion timescale, and $\xi^{(n)}(t)$ is a white noise stochastic process which satisfies
\begin{align}
	\langle \xi^{(n)}(t) \rangle &= 0 \ , \\
	\langle \xi^{(n)}(t) \xi^{(n')}(t') \rangle &= [\sigma^{(n)}]^2 \delta_{n,n'} \delta (t - t') \ .	\label{eq:xieqn}
\end{align}
In Equation \eqref{eq:xieqn} $[\sigma^{(n)}]^2$ is the variance of $\xi^{(n)}$ and parametrizes the amplitude of the noise. Combined with the mean reversion it gives characteristic root mean square fluctuations $\approx \sigma^{(n)} / [\gamma^{(n)}]^{1/2}$ in $f_{\rm p}^{(n)}(t)$ \citep{gardiner2009stochastic}. It is important to note that white noise fluctuations in $\xi(t)$ translate into red noise fluctuations in the rotational phase $\phi(t) = \int_{t_1}^t dt' \, f_{\rm p}(t')$ after being filtered by the terms involving $d/dt$ and $\gamma^{(n)}$ in Equation \eqref{eq:frequency_evolution}, consistent with the observed power spectral density of typical millisecond pulsars in the nHz band relevant to PTA experiments. \newline

Equations \eqref{eq:spinevol}--\eqref{eq:xieqn} represent a phenomenological model, which aims to reproduce qualitatively the typical timing behaviour observed in PTAs, viz.\ a mean-reverting random walk about a secular spin-down trend \citep{NANOgrav2023,EPTA2023,Zic2023arXiv230616230Z}. Equations \eqref{eq:spinevol}--\eqref{eq:xieqn} are not derived from first principles by applying a microphysical theory. As a first pass, they also exclude certain phenomenological elements, which are likely to be present in reality, e.g.\ the classic, two-component, crust-superfluid structure inferred from post-glitch recoveries \citep{Baym1969,vanEysden,Alpar2017MNRAS.471.4827G,Myers2021MNRAS.502.3113M,Meyers2021}. An approach akin to Equations \eqref{eq:spinevol}--\eqref{eq:xieqn} has been followed successfully in other timing analyses in the context of anomalous braking indices \citep{Vargas} and hidden Markov model glitch searches \citep{Melatos2020ApJ...896...78M,Lower2021MNRAS.508.3251L,Dunn2022,Dunn2023MNRAS.522.5469D}. However, Equations \eqref{eq:spinevol}--\eqref{eq:xieqn}  involve significant idealizations, which must be recognized at the outset \citep{Meyers2021,Myers2021MNRAS.502.3113M,Vargas}. First, the white noise driver $\xi(t)$ in Equation \eqref{eq:frequency_evolution} is not differentiable, which makes the formal interpretation of $d^2 f_{\rm p} / dt^2$ ambiguous, even though $d^2 f_{\rm p} / dt^2$ is not used in the PTA analysis proposed in this paper. Second, the white spectrum assumed for $\xi(t)$ may or may not be suitable for millisecond pulsars in PTAs. It is challenging observationally to infer the spectrum of $\xi(t)$ from the observed spectrum of the phase residuals, because the inference is conditional on the (unknown) dynamical model governing $df_{\rm p}/dt$. For small-amplitude fluctuations sampled relatively often, as in millisecond pulsars in PTAs, it is likely that $\xi(t)$ is white to a good approximation over the inter-TOA intervals and generates red phase residuals as observed, but caution is warranted nevertheless. Third, the Brownian increment $dB(t)=\xi(t)dt$ does not include non-Gaussian excursions such as L\'{e}vy flights \citep{Sornette2004}, which have not been ruled out by pulsar timing experiments to date. The above three idealizations are supplemented by other, physical approximations noted above, e.g.\ neglecting $n_{\rm em}$ in Equation \eqref{eq:spinevol} and differential rotation between the crust and superfluid in Equation \eqref{eq:frequency_evolution}.

\subsection{Modulation of pulsar frequency by a GW} \label{sec:psr_measured}
In the presence of a GW, the pulse frequency measured by an observer in the local rest frame of the neutron star's center of mass is different from that measured by an observer on Earth. Specifically, the pulse frequency at the Earth is modulated harmonically at the GW frequency. We derive the nonlinear measurement equation relating $f_{\rm m}(t)$ to $f_{\rm p}(t)$ in this section. The measurement equation is a key input into the Kalman filter in Section \ref{sec:kalman_filter}
\subsubsection{Plane GW perturbation}\label{sec:plane_gw}
We consider a gravitational plane wave from a single, distant SMBHB. The GW perturbs the background Minkowski metric $\eta_{\mu \nu}$ as 
\begin{equation}
	g_{\mu \nu} = \eta_{\mu \nu} + h_{\mu \nu} \ , \label{eq:plane_wave}
\end{equation}
where the metric perturbation $h_{\mu \nu}$ has zero temporal components $h_{0 \nu} = h_{\mu 0} = 0$. For elliptically polarised GWs emitted by a SMBHB, the spatial metric components are \citep{Maggiore}
\begin{equation}
	h_{ij} (t, \boldsymbol{x}) = H_{ij}^{(+)} e^{i \left[\Omega(\boldsymbol{x}\cdot \boldsymbol{n} - t) + \Phi_0\right] } + H_{ij}^{(\times)} e^{i \left[\Omega(\boldsymbol{x}\cdot \boldsymbol{n} - t) + \Phi_0 + \pi/2\right] } \, , \label{eq:binary}
\end{equation} 
written in terms of nearly Lorentz spatial coordinates $\boldsymbol{x}$ and global coordinate time $t$ \citep{schutz2022}. The GW propagates in the $\boldsymbol{n}$-direction (where $\boldsymbol{n}$ is a unit vector), has a constant (see justification below) angular frequency $\Omega$, phase offset $\Phi_0$ and two orthogonal polarisations with amplitude tensors $H_{i j}^{(+,\times)}$. Throughout this paper we work with pulsar TOAs defined relative to the Solar System barycentre (SSB). We are free to choose coordinates such that $\Phi_0$ is the GW phase at $t=0$ at the SSB. The amplitude tensors are given by
\begin{align}
	H_{ij}^{(+)} &= h_+ e_{ij}^{+} \, , \\
	H_{ij}^{(\times)} &= h_{\times} e_{ij}^{\times} \, ,
\end{align}
where $h_{+}$ and $h_{\times}$ are the respective polarisation amplitudes. The plus and cross polarisation tensors $e_{ij}^{+}$ and $e_{ij}^{\times}$ are uniquely defined by the principal axes of the wave, viz.\ the unit 3-vectors $\boldsymbol{k}$ and $\boldsymbol{l}$, according to
\begin{align}
	e_{i j}^{+}(\boldsymbol{n}) =k_i k_j-l_i l_j \ , \\
	e_{i j}^{\times}(\boldsymbol{n}) =k_i l_j+l_i k_j \ .
\end{align}
The principal axes are in turn specified by the location of the GW source on the sky (colatitude $\theta$, azimuth $\phi$) and the polarisation angle $\psi$ according to
\begin{align}
	\boldsymbol{k}  = &(\sin \phi \cos \psi-\sin \psi \cos \phi \cos \theta) \boldsymbol{\hat{x}} \nonumber \\
	& -(\cos \phi \cos \psi+\sin \psi \sin \phi \cos \theta) \boldsymbol{\hat{y}} \nonumber \\
	& +(\sin \psi \sin \theta) \boldsymbol{\hat{z}} \ , \\
	\boldsymbol{l} = &(-\sin \phi \sin \psi-\cos \psi \cos \phi \cos \theta) \boldsymbol{\hat{x}} \nonumber \\
	& +(\cos \phi \sin \psi-\cos \psi \sin \phi \cos \theta) \boldsymbol{\hat{y}}\nonumber  \\
	& +(\cos \psi \sin \theta) \boldsymbol{\hat{z}} \ ,
\end{align}
where e.g. $\boldsymbol{\hat{x}}$ is a unit vector in the direction of the $x$-axis. The direction of GW propagation is related to the principal axes by
\begin{equation}
	\boldsymbol{n} = \boldsymbol{k} \times \boldsymbol{l} \ . 
\end{equation}

In this paper the source is approximated as monochromatic. In reality the gravitational wave frequency $f_{\rm gw}=\Omega / 2 \pi $ increases during the inspiral by an amount \citep[e.g.][]{Sesana2010}
\begin{equation}
	\Delta f_{\rm gw} = 0.05 \, \mathrm{nHz}\left(\frac{M_{\rm c}}{10^{8.5} M_{\odot}}\right)^{5 / 3}\left[\frac{f_{\rm gw}(t=t_1)}{50 \mathrm{~nHz}}\right]^{11 / 3}\left(\frac{T_{\mathrm{obs}}}{10 \mathrm{yr}}\right) \ ,
	\label{eq:f_evolution}
\end{equation}
where $M_{\rm c}$ is the chirp mass of the SMBHB, $f_{\rm gw}(t=t_1)$ is the GW frequency at the time of the first observation, and $T_{\rm obs}$ is the length of time over which $\Delta f_{\rm gw}$ is measured, which for PTAs is $\sim 10$ years. A source can be considered monochromatic, if $\Delta f_{\rm gw}$ is less than the PTA frequency resolution $1/T_{\rm obs}$. Equation \eqref{eq:f_evolution} implies $(M_{\rm c}/10^{8.5} M_{\odot})^{5/3} [ f_{\rm gw}(t=t_1) / 50 \mathrm{~nHz}]^{11/3} (T_{\rm obs}/10 \mathrm{yr})^2 \gtrsim 20$ for $\Delta f_{\rm gw} \gtrsim 1/T_{\rm obs}$. The majority of SMBHBs detectable with PTAs are expected to satisfy $\Delta f_{\rm gw} < 1/T_{\rm obs}$; for a PTA composed of pulsars with a mean distance of 1.5 kpc, 78\% of simulated SMBHBs satisfy this condition for the current IPTA, whilst for the second phase of the Square Kilometer Array this fraction drops to 52\%; see Figure 7 in  \cite{Rosado10.1093/mnras/stv1098}. We are therefore justified in treating the GW source as monochromatic as a first pass in this introductory paper \citep{Sesana10,Sesana2010,Ellis2012ApJ}. \newline

\subsubsection{Measurement equation}
In general radio pulses from a pulsar are transmitted as amplitude modulations of a radio-frequency carrier wave. They are described by the geometric object $\boldsymbol{p}$, which we identify as the momentum 4-vector of the radio pulse. The presence of a GW induces a shift in the temporal component of the associated momentum one-form, while the photon travels from the emitter to the observer, i.e. $\Delta p_t = p_t|_{\rm observer} - p_t|_{\rm emitter} $. One obtains \citep[e.g.][]{Maggiore}
\begin{equation}
 \Delta p_t =  \frac{\pi f_{\rm p} h_{ij} (t; \boldsymbol{x}= 0)q^i q^j }{1 + \boldsymbol{n}\cdot \boldsymbol{q} }  \left[1 -e^{i \Omega (1 + \boldsymbol{n}\cdot \boldsymbol{q})  d}\right] \ ,
	\label{eq:momentum_shift}
\end{equation}
where $f_{\rm p}$ is the pulse frequency measured in the momentarily comoving reference frame of an observer at rest in the nearly Lorentz coordinates $(t,{\boldsymbol{x}})$. In Equation \eqref{eq:momentum_shift} $\boldsymbol{q}$ is the unit vector connecting the observer and the pulsar and $d$ is the distance to the pulsar. We take the pulsar location to be constant i.e.  neither $\boldsymbol{q}$ nor $d$ are functions of time. In practice the pulsar locations vary with respect to the Earth but are constant with respect to the SSB. The barycentering correction is typically applied when generating TOAs, e.g. with {\sc tempo2} \citep{tempo2} and related timing software, and is inherited by the frequency time series. Some pulsars, including some PTA pulsars, do have non-negligible proper motions of order $10^2$ km s$^{-1}$ after the barycentering corrections have been applied \citep[e.g.][]{10.1093/mnras/sty3390}, but we do not consider this effect in this paper.\newline 

Generally the measured frequency of a photon recorded by an observer who is travelling with 4-velocity $\boldsymbol{u}$ is given by the coordinate-independent expression $p_{\alpha} u^{\alpha}$. After barycentering, one has $u^{\alpha} =(1,0,0,0)$ for both the emitter and the observer to leading order in the respective momentarily comoving reference frames. Hence Equation \eqref{eq:momentum_shift} can be written for the $n$-th pulsar as
\begin{equation}
	f_{\rm m}^{(n)}(t) = f_{\rm p}^{(n)}\left [t-d^{(n)} \right ] g^{(n)}(t) +  \varepsilon^{(n)}(t)\ ,
	\label{eq:measurement}
\end{equation}
where $d^{(n)}$ labels the distance to the $n$-th pulsar, and $\varepsilon^{(n)}$ is a Gaussian measurement noise which satisfies 
\begin{align}
	\langle \varepsilon^{(n)}(t) \rangle &= 0 \ , \\
	\langle \varepsilon^{(n)}(t) \varepsilon^{(n')}(t') \rangle &= \sigma_{\rm m}^2 \delta_{n,n'} \delta(t - t') \ ,	\label{eq:vareps}
\end{align}
where $\sigma_{\rm m}$ is the covariance of the measurement noise at the telescope and is shared between all pulsars by assumption. The measurement function $g^{(n)}(t)$ is related to a redshift $z^{(n)}(t)$ through
\begin{equation}
	g^{(n)}(t) = 1 -z^{(n)}(t) \, , \label{eq:g_func_new}
\end{equation}
with
\begin{align}
	z(t) =&  \frac{[q^{(n)}]^i [q^{(n)}]^j }{2 [1 + \boldsymbol{n}\cdot \boldsymbol{q}^{(n)}] } \nonumber\\  
	&\times \left[ 	h_{ij}(t,\boldsymbol{x}=0) - h_{ij} (t,\boldsymbol{x}=0) e^{i \Omega (1 + \boldsymbol{n}\cdot \boldsymbol{q}^{n}) d^{(n)}  } \right] \, , \label{eq:z1}
\end{align}
where $[q^{(n)}]^i$ labels the $i$-th coordinate component of the $n$-th pulsar's position vector $\boldsymbol{q}^{(n)}$. It is also instructive to express Equation \eqref{eq:z1} in a trigonometric form \citep[cf. e.g.][]{Sesana2010,Perrodin2018,2023ApJ...951L..50A} as
\begin{align}
	z(t) =&  \frac{[q^{(n)}]^i [q^{(n)}]^j }{2 [1 + \boldsymbol{n}\cdot \boldsymbol{q}^{(n)}] } & \nonumber \\ 
	&\times \Big \{ \left[H_{ij}^{(+)} \cos \Phi(t) + H_{ij}^{(\times)} \sin \Phi(t)\right] - \nonumber \\
	&\left[H_{ij}^{(+)} \cos \Phi^{(n)}(t) + H_{ij}^{(\times)} \sin \Phi^{(n)}(t)\right]\Big \} \, , \label{eq:z_trigonometric}
\end{align}
where we define the phases 
\begin{align}
 \Phi(t) &= -\Omega t + \Phi_0 \, ,\\
 \Phi^{(n)}(t) &= \Phi(t) + \Omega \left[ 1 + \boldsymbol{n} \cdot \boldsymbol{q}^{(n)} \right] d^{(n)} \, . \label{eq:different_phases}
\end{align}
Equations \eqref{eq:measurement}--\eqref{eq:different_phases}  define a non-linear measurement equation that relates the intrinsic pulsar spin frequency to the pulse frequency measured by an observer on Earth. 

\section{Signal tracking, parameter estimation and model selection} \label{sec:detect}
The set of static parameters $\boldsymbol{\theta}$ of the model outlined in Section \ref{sec:model} can be separated into parameters controlling the intrinsic frequency evolution of the pulsars in the array and the GW source, i.e. 
\begin{equation}
	\boldsymbol{\theta} =  \boldsymbol{\theta}_{\rm psr} \cup \boldsymbol{\theta}_{\rm gw} \ , \label{eq:params1}
\end{equation}
with
\begin{equation}
	\boldsymbol{\theta}_{\rm psr} = \left \{ \gamma^{(n)},\sigma^{(n)}, f_{\rm em}^{(n)}(t_1),\dot{f}_{\rm em}^{(n)}(t_1),d^{(n)}\right\}_{1\leq n \leq N} \ , \label{eq:psrparams}
\end{equation}
and
\begin{equation}
	\boldsymbol{\theta}_{\rm gw} = \left \{h_0, \iota, \delta, \alpha, \psi, \Omega, \Phi_0 \right \} \ ,  \label{eq:params3}
\end{equation}
where $\delta$, $\alpha$, and $\iota$ are the declination, right ascension and inclination of the GW source respectively \footnote{$\iota$ is the angle between the unit normal to the SMBHB orbital plane, $\boldsymbol{L}$, and the observer's line of sight, i.e. $\cos \iota = \boldsymbol{n} \cdot \boldsymbol{L}$.}. In Equation \eqref{eq:params3} we reparameterize the two GW polarisation amplitudes, $h_{+}$ and $h_{\times}$, in terms of $\iota$ and the characteristic wave strain $h_0$ through 
\begin{align}
	h_+ &= h_0(1 + \cos^2 \iota) 	\label{eq:hphx} \ ,\\
	h_{\times} &= -2h_0\cos \iota 	\label{eq:hphx2} \ .
\end{align}
We use the parameterisation in terms of $h_0$ and $\iota$ throughout the remainder of this work. A PTA containing $N$ pulsars comprises $7 + 5N$ parameters to estimate. Typically the pulsar parameters are constrained better \textit{a priori} by electromagnetic observations than the GW parameters. For example estimates of pulsar distances are accurate to $\sim$ 10$\%$ \citep{Cordes2002astro.ph..7156C, Verbiest2012ApJ...755...39V, Desvignes2016,Yao2017}, but we have no prior information about $\delta$ and $\alpha$. \newline 

In this section we present a new method to infer $\boldsymbol{\theta}$ and calculate the marginal likelihood (i.e. the model evidence). In Section \ref{sec:kalman_filter} we outline how noisy measurements of the pulsar frequency, $f_{\rm m}^{(n)}(t)$, can be used to estimate the hidden state sequence, $f_{\rm p}^{(n)}(t)$, using a Kalman filter. In Section \ref{sec:nested_sampling} we demonstrate how to deploy the Kalman filter in conjunction with a nested sampling technique to perform Bayesian inference. Model selection and the specification of the null model are described in Section \ref{sec:model_selection}. A complete summary of the workflow is presented in Section \ref{sec:methodsummary}. The method complements traditional PTA analyses because (i) it harnesses the favourable asymptotic properties of the adaptive gain of the Kalman filter to track timing noise more nimbly than alternatives like least-squares estimators; (ii) it infers ${\boldsymbol{\theta}}$ conditional on the specific, time-ordered, random realization of the noise corresponding to the observed sequence of TOAs rather than on ensemble-averaged quantities like the power spectral density. Points (i) and (ii) are discussed further in Section \ref{sec:35}

\subsection{Kalman filter and likelihood}\label{sec:kalman_filter}

\begin{figure*}
	\includegraphics[width=\textwidth, height =0.5\textwidth]{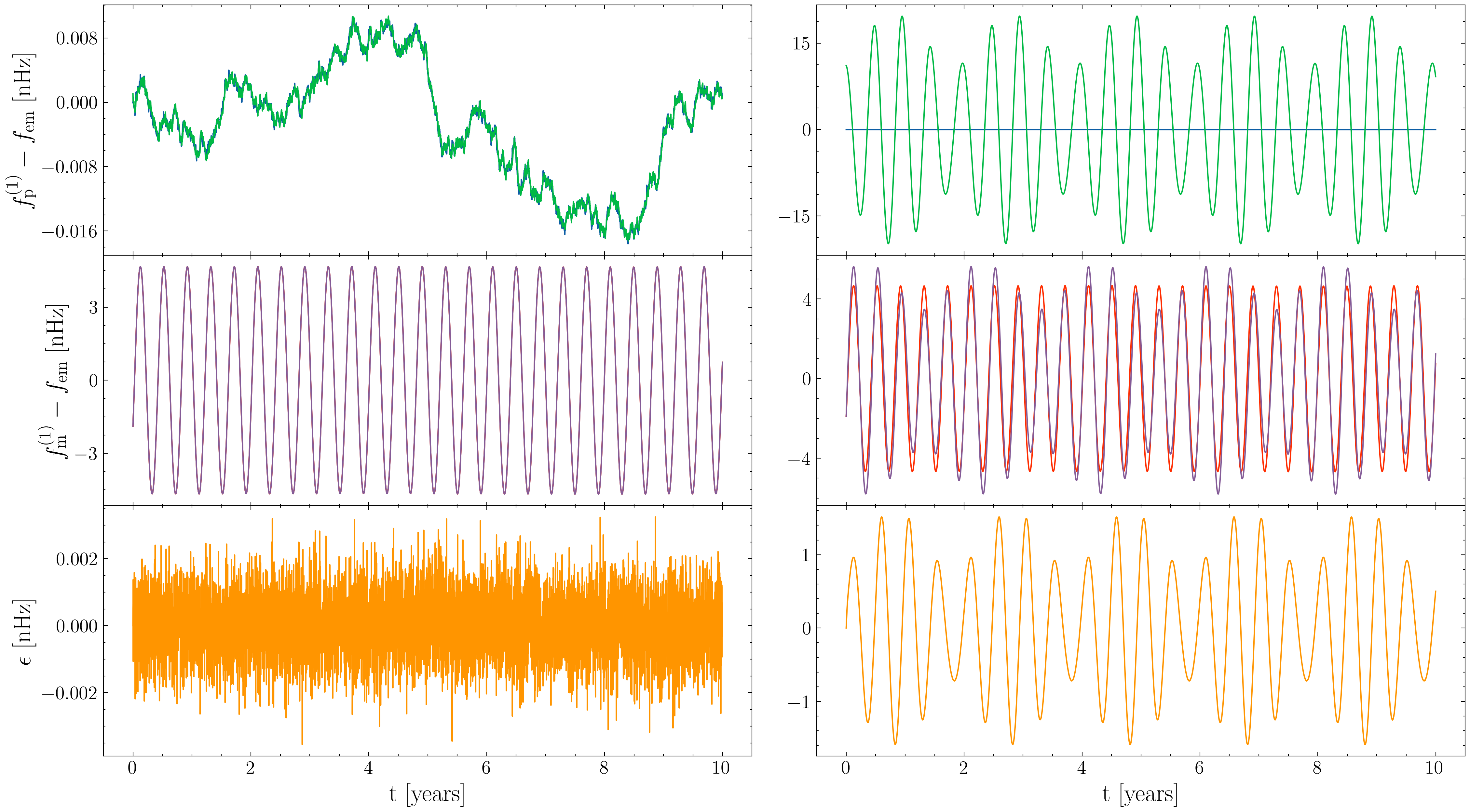}
	\caption{Sample output of the Kalman filter illustrating the accuracy of the reconstructed state sequence $f_{\rm p}^{(1)}(t)$ when the static parameters are correct (${\boldsymbol{\hat\theta}} = {\boldsymbol{\theta}}$, left column) and incorrect (${\boldsymbol{\hat\theta}} \neq {\boldsymbol{\theta}}$, right column). The top panels show the true pulsar state $f_{\rm p}^{(1)}(t) - f_{\rm em}^{(1)}(t)$ (blue curve) and the state estimated by the Kalman filter $\hat{f}_{\rm p}^{(1)}(t) - f_{\rm em}^{(1)}(t)$  (green curve). We subtract $f_{\rm em}^{(1)}(t)$ to better illustrate the stochastic wandering of the pulsar frequency. In the left column, the blue/green solutions overlap almost perfectly; in the right column, they do not. The middle panels show the true measured frequency $f_{\rm m}^{(1)}(t) - f_{\rm em}^{(1)}(t)$ (red curve) and the frequency estimated  by the Kalman filter $\hat{f}_{\rm m}^{(1)}(t) - f_{\rm em}^{(1)}(t)$ (magenta curve). Again we subtract $f_{\rm em}^{(1)}(t)$. In the left column the red/magenta solutions overlap almost perfectly; in the right column, they do not. The bottom panels show the residual or innovation $\epsilon(t) =f_{\rm m}^{(1)}(t) - \hat{f}_{\rm m}^{(1)}(t)$. In the right column $\Omega$ is displaced from its true value by 20\%. Results are shown for a single pulsar. } 
	\label{fig:kalman_example}
\end{figure*}

The Kalman filter \citep{Kalman1} is a Gauss-Markov model used to algorithmically recover a temporal sequence of stochastically evolving  system state variables, $\boldsymbol{X}(t)$, which are not observed directly, given a temporal sequence of noisy measurements, $\boldsymbol{Y}(t)$. It finds common use in engineering applications and has been applied successfully in neutron star astrophysics \citep[e.g.][]{Myers2021MNRAS.502.3113M,Meyers2021,Melatos2023}. In this work we use the linear Kalman filter, which assumes a linear relation between $d{\boldsymbol{X}}/dt$ and ${\boldsymbol{X}}(t)$ (dynamics) and between ${\boldsymbol{Y}}(t)$ and ${\boldsymbol{X}}(t)$ (measurement), with ${\boldsymbol{X}}(t) = \{ f_{\rm p}^{(n)}(t) \}$ and ${\boldsymbol{Y}}(t) = \{ f_{\rm m}^{(n)}(t) \}$. Extension to non-linear problems is straightforward using either an extended Kalman filter \citep{zarchan2000fundamentals}, unscented Kalman filter \citep{882463van} or particle filter \citep{Simon10}. Equations \eqref{eq:measurement} and \eqref{eq:z_trigonometric} are non-linear in the static parameters (e.g.\ $d^{(n)}$), even though they are linear in ${\boldsymbol{X}}(t)$ and ${\boldsymbol{Y}}(t)$. Hence inferring the static parameters is a non-linear exercise, to be tackled separately after the linear Kalman filter operates on the data for a fixed set of static parameters. Inference of the static parameters in Equations \eqref{eq:measurement} and \eqref{eq:z_trigonometric} by nested sampling is discussed in Section \ref{sec:nested_sampling}. \newline 

Implementation instructions for the linear Kalman filter for the PTA state-space model in Section \ref{sec:model}, including the full set of Kalman recursion relations, are presented in Appendix \ref{sec:kalman}. At each discrete timestep indexed by $ 1 \leq i  \leq M$, the Kalman filter returns an estimate of the state variables, $\hat{\boldsymbol{X}}_i = \hat{\boldsymbol{X}}(t_i)$, and the covariance of those estimates, ${\boldsymbol{P}}_i = \langle {\boldsymbol{\hat X}}_i {\boldsymbol{\hat X}}_i^{\rm T} \rangle$, where the superscript T denotes the matrix transpose. The filter tracks the error in its predictions of $\boldsymbol{X}_i$ by converting ${\boldsymbol{\hat X}}_i$ into predicted measurements ${\boldsymbol{\hat Y}}_i$ via Equations \eqref{eq:measurement} and \eqref{eq:z_trigonometric} and comparing with the actual, noisy measurements ${\boldsymbol{\hat Y}}_i$. This defines a residual $\boldsymbol{\epsilon}_i = \boldsymbol{Y}_i  - \hat{\boldsymbol{Y}}_i$, which is sometimes termed the innovation. The Kalman filter also calculates the uncertainty in $\boldsymbol{\epsilon}_i$ via the innovation covariance $\boldsymbol{S}_i = \langle \boldsymbol{\epsilon}_i \boldsymbol{\epsilon}_i^{T} \rangle$. The innovation and the innovation covariance are then used to correct the state estimate ${\boldsymbol{\hat X}}_i$ according to Equation \eqref{eq:kalmangainupdate}. For a fixed set of static parameters, the Kalman filter returns an estimate of the state sequence ${\boldsymbol {\hat X}}_1, \dots , {\boldsymbol{\hat X}}_M$ which minimizes the mean square error. We explain how to use this intermediate output to infer the optimal values of the static parameters ${\boldsymbol{\theta}}$ in Section \ref{sec:nested_sampling} \newline

The Gaussian log-likelihood of obtaining ${\boldsymbol{Y}}_i$ given ${\boldsymbol{\hat X}}_i$ can  then be calculated at each timestep from the Kalman filter output according to
\begin{eqnarray}
	\log \mathcal{L}_i =  -\frac{1}{2} \left (N \log 2 \pi + \log  \left | \boldsymbol{S}_i \right | + \boldsymbol{\epsilon}_i^{\intercal} \boldsymbol{S}_i^{-1}  \boldsymbol{\epsilon}_i \right ) \ .
\end{eqnarray}
The total log-likelihood for the entire sequence is
\begin{eqnarray}
	\log \mathcal{L} =  \sum_{i=1}^{M} \log \mathcal{L}_i \ . \label{eq:likelihood}
\end{eqnarray}
Given ${\boldsymbol{Y}}_1, \dots, {\boldsymbol{Y}}_M$, $\mathcal{L}$ is a function of the estimates ${\boldsymbol{\hat \theta}}$ of the static parameters passed to the Kalman filter, i.e. $\mathcal{L}$ = $\mathcal{L}(\boldsymbol{Y} | \boldsymbol{\hat \theta})$. Similarly the estimates of the state and measurement variables, $\hat{\boldsymbol{X}}$ and $\hat{\boldsymbol{Y}}$, are functions of $\boldsymbol{\hat \theta}$. If $\boldsymbol{\hat{\theta}}$ is close to the true underlying parameters $\boldsymbol{\theta}$, then the errors in $\hat{\boldsymbol{X}}$ and $\hat{\boldsymbol{Y}}$ are minimized and $\mathcal{L}$ is maximised. This is illustrated with synthetic data in Figure \ref{fig:kalman_example}. In the left column, a time series of $f_{\rm m}^{(1)}(t)$ including Gaussian noise (middle panel, red curve) is generated from Equations \eqref{eq:spinevol}-- \eqref{eq:xieqn}, \eqref{eq:measurement}, and \eqref{eq:z_trigonometric} for a single pulsar and fed into the Kalman filter along with the true static parameters ${\boldsymbol{\hat \theta}} = {\boldsymbol{\theta}}$. The Kalman filter recovers the evolution of $f_{\rm p}^{(1)}(t)$ with high fidelity; the estimate of $\hat{f}_{\rm p}^{(1)}(t)$ (left top panel, blue curve) overlaps almost perfectly with the true $f_{\rm p}^{(1)}(t)$ (left top panel, green curve). The predicted state $\hat{f}_{\rm p}^{(1)}(t)$ is converted  into a predicted measurement $\hat{f}_{\rm m}^{(1)}(t)$ (middle panel, magenta curve), which again overlaps almost perfectly with the true measurement. The residuals $\epsilon(t)$ between the true and predicted measurements are small ($\lesssim 0.1\%$) and normally distributed (left bottom panel). By contrast, in the right column, the exercise is repeated while passing incorrect static parameters (${\boldsymbol{\hat\theta}} \neq {\boldsymbol{\theta}}$) to the Kalman filter, where $\Omega$ is displaced from its true value by $20 \%$. In this case the Kalman filter fails to track $f_{\rm p}^{(1)}(t)$ accurately, as the discrepancy between the blue and green curves in the top panel of the right-hand column indicates. It similarly fails to predict $f_{\rm m}^{(1)}(t)$ accurately, as shown by the discrepancy between the red and magenta curves in the middle panel, and the residuals are no longer distributed normally (right bottom panel). In Section \ref{sec:nested_sampling}, we explain how to combine the Kalman filter with a nested sampler to iteratively guide ${\boldsymbol{\hat \theta}}$ towards the true value of ${\boldsymbol{\theta}}$.

\subsection{Nested sampling}\label{sec:nested_sampling}
We can use the likelihood returned by the Kalman filter, Equation \eqref{eq:likelihood}, in conjunction with likelihood-based inference methods to estimate the posterior distribution of $\boldsymbol{\theta}$ by Bayes' Rule,
\begin{equation}
	p(\boldsymbol{\theta} | \boldsymbol{Y}) = \frac{\mathcal{L}(\boldsymbol{Y} | \boldsymbol{\theta}) \pi(\boldsymbol{\theta})}{\mathcal{Z}} \ ,
\end{equation}
where $\pi(\boldsymbol{\theta})$ is the prior distribution on $\boldsymbol{\theta}$ and $\mathcal{Z}$ is the marginalised likelihood, or evidence
\begin{equation}
	\mathcal{Z} = \int d \boldsymbol{\theta} \mathcal{L}(\boldsymbol{Y} | \boldsymbol{\theta})  \pi(\boldsymbol{\theta})  \ . \label{eq:model_evidence}
\end{equation}
 We estimate the posterior distribution and the model evidence through nested sampling \citep{Skilling} in this paper. Nested sampling evaluates marginalised likelihood integrals, of the form given by Equation \eqref{eq:model_evidence}. It also approximates the posterior by returning samples from $p(\boldsymbol{\theta} | \boldsymbol{Y})$. It does so by drawing a set of $n_{\rm live}$ live points from $\pi(\boldsymbol{\theta})$ and iteratively replacing the live point with the lowest likelihood with a new live point drawn from $\pi(\boldsymbol{\theta})$, where the new live point is required to have a higher likelihood than the discarded point. The primary advantage of nested sampling is its ability to compute $\mathcal{Z}$, on which model selection relies. Nested sampling is also computationally efficient and can handle multi-modal problems \citep{Ashton2022}. For these reasons, it has enjoyed widespread adoption in the physical sciences, particularly within the cosmological community \citep{Mukherjee2006,Feroz2008,Handley2015}, neutron star astrophysics \citep{Myers2021MNRAS.502.3113M,Meyers2021,Melatos2023}, particle physics \citep{proceedings2019033014} and materials science \citep{2009arXiv0906materials}. For reviews of nested sampling we refer the reader to \cite{Buchner2021} and \cite{Ashton2022}. Multiple nested sampling algorithms and computational libraries exist \citep[e.g.][]{Feroz2008,Feroz2009,Handley2015,dynesty2020,UltraNest2021}. In gravitational wave research it is common to use the \texttt{dynesty} sampler \citep{dynesty2020} via the \texttt{Bilby} \citep{bilby.507.2037A} front-end library. We follow this precedent and use \texttt{Bilby} for all nested sampling Bayesian inference in this work. \newline 
 
The primary tunable parameter in nested sampling is $n_{\rm live}$. More live points assist with large parameter spaces and multi-modal problems, whilst the uncertainties in the evidence and the posterior scale as $\mathcal{O} (n_{\rm live}^{-1/2})$. However the computational runtime scales as $\mathcal{O}(n_{\rm live})$. \cite{Ashton2022} offered a rule-of-thumb trade-off, where the minimum number of live points should be greater than the number of static parameters. Informal empirical tests conducted as part of this paper support the trade-off suggested by \cite{Ashton2022}; we find typically that the true ${\boldsymbol{\theta}}$ is contained within the 90\% credible interval of the one-dimensional marginalised posteriors of ${\boldsymbol{\hat{\theta}}}$ for $n_{\rm live} > 7 + 5N$ with $N \leq 50$. Unless stated otherwise we take $n_{\rm live} = 1000$ for all results presented in this work. Empirically, we find that values of  $n_{\rm live} > 1000$ do not improve appreciably the accuracy of the results presented in Sections \ref{sec:testing} and \ref{sec:rep_example}.

\subsection{Model selection}\label{sec:model_selection}
The evidence integral $\mathcal{Z}$ returned by nested sampling is the probability of the data $\boldsymbol{Y}$ given a model $\mathcal{M}_i$. We compare competing models via a Bayes factor,
\begin{equation}
	\beta = \frac{\mathcal{Z}(\boldsymbol{Y} | \mathcal{M}_1)}{\mathcal{Z}(\boldsymbol{Y} | \mathcal{M}_0)} \ . \label{eq:bayes}
\end{equation}
Throughout this work we take $\mathcal{M}_1$ to be the state-space model that includes the presence of a GW. $\mathcal{M}_0$ is the null model, which assumes no GW exists in the data, and is equivalent to setting $g^{(n)}(t)=1$ in Equation \eqref{eq:g_func_new}. The Bayes factors we quote in this work therefore quantify whether the data support evidence for a GW signal compared to there being no GW signal present.

\subsection{Summary of workflow}\label{sec:methodsummary}
For the reader's convenience we now summarise the workflow for a representative PTA analysis using the Kalman filter and nested sampler for parameter estimation and model selection:
\begin{enumerate}[leftmargin=2em]
	\item Specify a PTA composed of $N$ pulsars 
	\item Obtain $N$ data inputs $f_{\rm m}^{(n)}(t)$, collectively labelled $\boldsymbol{Y}$
	\item Specify a state-space model $\mathcal{M}$, with static parameters $\boldsymbol{\theta}$
	\item Specify prior distribution $\pi(\boldsymbol{\theta})$
	 \item Sample $n_{\rm live}$ points from $\pi(\boldsymbol{\theta})$ 
	 \item For each live point:
\begin{enumerate}[leftmargin=2em]
	\item Pass the sample $\boldsymbol{\theta}_{\rm sample}$ to the Kalman filter
	\item Iterate over the input data using the Kalman filter and obtain a single $\log \mathcal{L}$ value, Equation \eqref{eq:likelihood}
\end{enumerate}
	\item Remove the live point with the lowest likelihood value, $\log \mathcal{L}_{\rm lowest}$
	\item Sample a new live point from $\pi(\boldsymbol{\theta})$, subject to the requirement that the new likelihood obeys $\mathcal{L}_{\rm new}$ > $\mathcal{L}_{\rm lowest}$, where $\log \mathcal{L}_{\rm new}$ is calculated via steps (vi)(a)--(vi)(b).
	\item Update $p\left(\boldsymbol{\theta}|\boldsymbol{Y}\right)$ and $\mathcal{Z}$ with nested sampler
	\item Repeat steps (vii)--(ix) until convergence criteria are satisfied.
\end{enumerate}
In order to compute $\beta$ the above workflow is repeated for a different $\mathcal{M}$. The resulting $\mathcal{Z}$ values can then be compared. We remind the reader that the above workflow differs from a realistic PTA analysis in one important respect, namely that the data are input as frequency time series $f_{\rm m}^{(n)}(t)$ instead of pulse TOAs. The generalization to TOAs is subtle and will be tackled in a forthcoming paper.

\subsection{Relation to traditional PTA analyses} \label{sec:35}
It is natural to ask how the workflow in Sections \ref{sec:kalman_filter}--\ref{sec:methodsummary} differs from traditional PTA analyses. One superficial difference is the detection statistic: traditional analyses use a frequentist matched filter like the maximum-likelihood ${\cal F}$-statistic \citep{Ellis2012ApJ,2023arXiv230616226A} or Bayesian inference \citep{2023ApJ...951L..50A,2023ApJ...951L..28A} whereas the workflow in Sections \ref{sec:kalman_filter}--\ref{sec:methodsummary} maximizes the Kalman likelihood in Equation \eqref{eq:likelihood}. \newline 

A more subtle difference is the manner in which the intrinsic, achromatic timing noise is tracked and modelled. With respect to noise tracking, traditional analyses usually involve some form of least-squares minimization. The adaptive gain, which measures the fractional amount of new information incorporated into the updated state estimate at each time step, tends to zero in the limit $t\rightarrow \infty$ in least-squares minimization but remains non-zero in the same limit for a Kalman filter. That is, a Kalman filter responds more nimbly to new data than least-squares estimators, which is why it is favoured in many electrical and mechanical engineering applications \citep{Gelb:1974,zarchan2000fundamentals,byrne2005signal,sarka2013bayesian}. The Kalman gain is discussed more fully in Appendix \ref{appendix:brownian} and compared with traditional analyses. With respect to noise modeling, traditional analyses usually introduce a sum of Fourier modes in a TOA-based phase model, with random coefficients drawn from an ensemble-averaged power spectral density, whose form is often a (broken) power law with estimable amplitude and exponent(s). In this paper, the Kalman filter tracks the specific, time-ordered, random realization of the noise in the data consistent with the Ornstein-Uhlenbeck process in Equations \eqref{eq:frequency_evolution}--\eqref{eq:xieqn}, without ensemble averaging or imposing the extra time-domain structure implicit in a finite-term Fourier expansion (e.g.\ continuity from one time step to the next). Nevertheless, the noise models are similar and can be related in certain limits; for example, Equations \eqref{eq:frequency_evolution}--\eqref{eq:xieqn} imply an associated power spectral density, which is a broken power law. A fuller discussion of Equations \eqref{eq:frequency_evolution}--\eqref{eq:xieqn}, including a comparison with traditional noise models, is presented in Appendix \ref{appendix:OU}.

\section{Validation with synthetic data} \label{sec:testing}
In this section we outline how synthetic data are generated in order to validate the analysis scheme in Section \ref{sec:detect}. Synthetic data enable validation to occur systematically and under controlled conditions. In Section \ref{sec:synt_pta} we describe how to construct a representative synthetic PTA, and how to set astronomically reasonable values for the static pulsar parameters $\boldsymbol{\theta}_{\rm psr}$. In Section \ref{sec:gendata} we demonstrate how to solve Equations \eqref{eq:spinevol}--\eqref{eq:xieqn}, and \eqref{eq:measurement}--\eqref{eq:z_trigonometric} for the synthetic PTA so as to generate noisy, frequency time series $f_{\rm m}^{(1)}(t), \dots, f_{\rm m}^{(N)}(t) $. 

\subsection{Constructing a synthetic PTA}\label{sec:synt_pta}
\begin{figure}
	\includegraphics[width=\columnwidth]{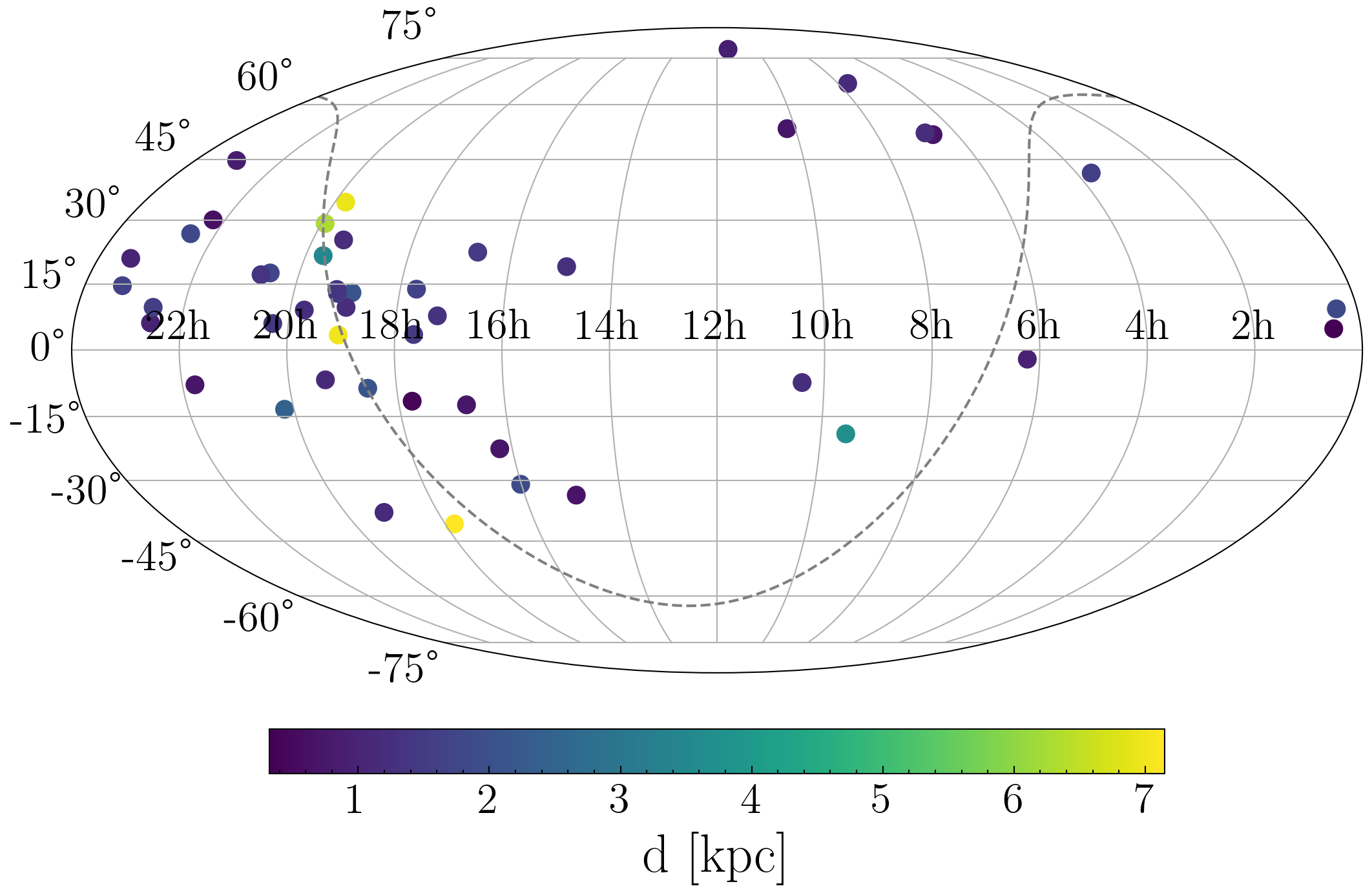}
	\caption{Spatial distribution in Galactic coordinates of 47 pulsars from the 12.5 year NANOGrav data release that make up the synthetic PTA used in this work. The pulsar distances relative to the observer are also indicated, with distance $\leq 2 \, {\rm kpc}$ for $38$ pulsars. The grey dashed curve denotes the Galactic plane.}
	\label{fig:pulsar_distrib}
\end{figure}
We consider, by way of illustration, a synthetic PTA composed of the 47 pulsars in the 12.5 year NANOGrav dataset \citep{2020ApJ...905L..34A}. The NANOGrav pulsars are chosen arbitrarily as being representative of a typical PTA; the analysis below extends unchanged to any other PTA. \newline 

For each pulsar we adopt fiducial values for ${\boldsymbol{q}}^{(n)}$, $d^{(n)}$, $f_{\rm em}^{(n)}(t_1)$, and $\dot{f}^{(n)}_{\rm em}(t_1)$, with the latter two quantities evaluated at the Solar System barycenter. A table of fiducial values is presented in Appendix \ref{appendix_fiducial} for the sake of reproducibility. The sky positions and colour-coded distances of the pulsars are displayed in Figure \ref{fig:pulsar_distrib}. The pulsar parameters are acquired via the Australia Telescope National Facility (ATNF) pulsar catalogue \citep{Manchester2005} using the \texttt{psrqpy} package \citep{psrqpy}. \newline 

The other static pulsar parameters are $\gamma^{(n)}$ and $\sigma^{(n)}$, for which no direct measurements exist. The ratio $\sigma^{(n)} / [\gamma^{(n)}]^{1/2}$ sets the typical root mean square fluctuations in $f_p^{(n)}(t)$, as discussed in Section \ref{sec:psr_frequency}, and the mean reversion timescale typically satisfies $[\gamma^{(n)}]^{-1} \gg T_{\rm obs}$ \citep{Price2012,Myers2021MNRAS.502.3113M,Meyers2021,Vargas}. In this paper, for the sake of simplicity in the absence of independent measurements, we fix $\gamma^{(n)} = 10^{-13}$ s$^{-1}$ for all $n$, consistent with \citep{Vargas}. We follow two complementary approaches in order to  set physically reasonable values for $\sigma^{(n)}$. The first approach relies on the empirical timing noise model from \cite{Shannon2010ApJ...725.1607S} which gives the standard deviation of the pulsar TOAs, $\sigma_{\rm TOA}^{(n)}$, as
\begin{align}
	\ln \left[\frac{\sigma_{\rm TOA}^{(n)}}{\mu s} \right]=& \ln \alpha_1 +  \alpha_2 \ln f_{\rm p}^{(n)} + \alpha_3 \ln \left[\frac{\dot{f}_{\rm p}^{(n)}}{10^{-15} \text{s}^{-2}} \right] \nonumber \\ 
	&+ \alpha_4 \ln \left( \frac{T_{\rm cad}}{ 1 \text{ year}} \right) \ , \label{eq:sigmap}
\end{align}
where $T_{\rm cad}$ is the cadence of the timing observations. For MSPs the best fit parameters are $\ln \alpha_1 = -20 \pm 20$, $\alpha_2 = 1 \pm 2 $, $\alpha_3 = 2 \pm 1$, $\alpha_4 = 2.4 \pm 0.6$ ($\pm 2\sigma$ confidence limits). The uncertainties are broad; for the purpose of generating astrophysically representative synthetic data we adopt the central values in this paper. Throughout this work we assume for simplicity that all pulsars are observed with a weekly cadence, $T_{\rm cad} = 1 \,{\rm week}$. In order to relate Equation \eqref{eq:sigmap} to $\sigma^{(n)}$ in Equation \eqref{eq:spinevol}, we equate $\sigma^{(n)}$ to the root mean square TOA noise accumulated over one week, obtaining
\begin{eqnarray}
	\sigma^{(n)} \approx \sigma_{\rm TOA}^{(n)} f_{\rm p}^{(n)} {T_{\rm cad}}^{-3/2} \ . \label{eq:sigmap_f}
\end{eqnarray}
For the synthetic NANOGrav PTA depicted in Figure \ref{fig:pulsar_distrib}, the median $\sigma^{(n)}$ calculated in this way is $\sigma^{(n)} = 5.51 \times 10^{-24} $ s$^{-3/2}$, with $\min [ \sigma^{(n)} ] = 1.67 \times 10^{-26}$s$^{-3/2}$ for PSR J0645+5158 and $\max [ \sigma^{(n)} ] = 2.56 \times 10^{-19}$ s$^{-3/2}$ for PSR J1939+2134 \newline 

\begin{figure}
	\includegraphics[width=\columnwidth]{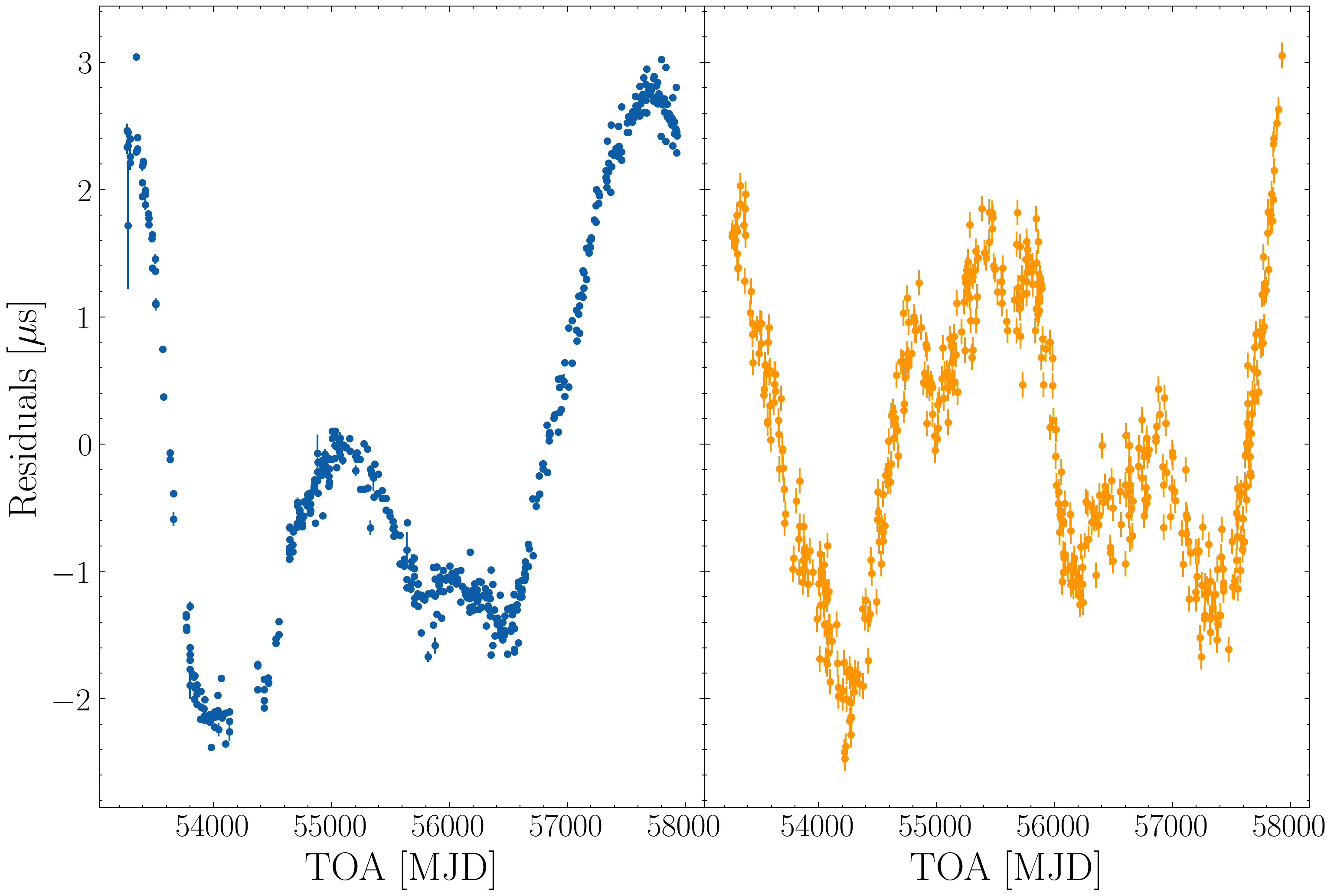}
	\caption{Actual (left panel, blue points) and synthetic (right panel, orange points) phase residuals for NANOGrav millisecond pulsar PSR J1939+2134. The actual residuals are sourced from the NANOGrav 12.5 year wideband timing dataset \citep{pennucci_timothy_t_2020_4312887,nanogravwideband}. The synthetic residuals are generated by numerically solving Equation \eqref{eq:spinevol} with $\gamma^{(n)} = 10^{-13}$ s$^{-1}$ and $\sigma^{(n)} = 2\times 10^{-27}$ s$^{-3/2}$. The error bars indicate the uncertainty in the phase residuals and are generated by propagating the uncertainty in the TOAs through {\sc tempo2}. We set the TOA uncertainty to be a constant, viz. $0.1 \mu$s. The blue and orange residuals are qualitatively similar by inspection. Similar results are obtained for the other 46 pulsars in the synthetic PTA in Section \ref{sec:synt_pta} by tuning the value of $\sigma^{(n)}$.}
	\label{fig:qualitative_compare}
\end{figure}

As a cross-check, we estimate $\sigma^{(n)}$ by solving Equation \eqref{eq:spinevol} numerically using the \texttt{baboo} package \footnote{\url{https://github.com/meyers-academic/baboo}}, generating a synthetic phase solution,
\begin{eqnarray}
	\varphi^{(n)}(t) = \int_0^t dt' f^{(n)}_{\rm p}(t') \ ,
\end{eqnarray}
and adjusting $\sigma^{(n)}$ iteratively to generate phase residuals which qualitatively (i.e.\ visually) resemble empirical phase residuals measured from real pulsars; see Section 2 in \cite{Vargas} for a successful example of this approach. We obtain empirical phase residuals from the NANOGrav 12.5 year wideband timing dataset \citep{pennucci_timothy_t_2020_4312887,nanogravwideband}. A visual cross-check is sufficient for the purposes of this paper, where we seek broadly representative values for $\sigma^{(n)}$. In an analysis involving real PTA data, $\sigma^{(n)}$ would be estimated from the data jointly with the other static parameters in ${\boldsymbol{\theta}}$. In Figure \ref{fig:qualitative_compare} we compare the synthetic and empirical residuals for PSR J1939+2134, one of the 47 NANOGrav pulsars plotted in Figure \ref{fig:pulsar_distrib}. We see that the synthetic and empirical residuals are qualitatively similar. For this pulsar the synthetic residuals are generated using $\sigma^{(n)} = 2 \times 10^{-27}$ s$^{-3/2}$ which is much less than the central value $\sigma^{(n)} = 2.56 \times 10^{-19}$ s$^{-3/2}$ inferred from Equations \eqref{eq:sigmap} and \eqref{eq:sigmap_f}, but is consistent with the wide ranges on $\alpha_1$, $\dots$, $\alpha_4$, e.g.\ $\ln \alpha_1 = -38$ implies $\sigma^{(n)}= 3.90 \times 10^{-27}$ s$^{-3/2}$ for PSR J1939$+$2134.


\subsection{Generating a synthetic sequence of pulse frequencies}\label{sec:gendata}
We generate $N$ synthetic noisy time series of the measured pulse frequency $f_{\rm m}^{(n)}(t)$, one for each pulsar $1\leq n \leq N$, as follows:
\begin{enumerate}[leftmargin=2em]
	\item Integrate Equations \eqref{eq:spinevol}--\eqref{eq:xieqn} numerically for the synthetic PTA described in Section \ref{sec:synt_pta}, to obtain random realizations of $f_{\rm p}^{(n)}(t)$ for $1\leq n \leq N$.
	\item Map from $f_{\rm p}^{(n)}(t)$ to $f_{\rm m}^{(n)}(t)$ via Equations \eqref{eq:measurement} and \eqref{eq:z_trigonometric}.
\end{enumerate}
Equations \eqref{eq:spinevol}--\eqref{eq:xieqn} are solved by a Runge-Kutta It$\hat{\text{o}}$ integrator implemented in the \texttt{sdeint} python package \footnote{\url{https://github.com/mattja/sdeint}}. The static pulsar parameters  $\boldsymbol{\theta}_{\rm psr}$ are completely specified for the synthetic PTA outlined in Section \ref{sec:synt_pta}. For this work we consider all pulsars to be observed for $T_{\rm obs} =10$ years, uniformly sampled with a weekly cadence. \newline

The measurement noise covariance as defined in Equation \eqref{eq:vareps} can be approximately related to the uncertainty in the pulse TOA, $\sigma_{\rm TOA}$, through
\begin{equation}
	\sigma_{\rm m} \approx \sigma_{\rm TOA} f_{\rm p}^{(n)}  \ {T_{\rm cad}}^{-1} \ . \label{eq:sigma_m_eqn}
\end{equation}
Although Equations \eqref{eq:sigmap_f} and \eqref{eq:sigma_m_eqn} superficially resemble one another, they are distinct. Equation \eqref{eq:sigmap_f} deals with the timing noise intrinsic to the pulsar due to rotational irregularities, whereas Equation \eqref{eq:sigma_m_eqn} handles the detector measurement noise. For a millisecond pulsar with $f_{\rm p}^{(n)} \sim 10^2$ Hz observed with weekly cadence and $\sigma_{\rm TOA} \sim 1 \mu$s, Equation \eqref{eq:sigma_m_eqn} implies $\sigma_{\rm m} \sim 10^{-10}$ Hz. The most accurately timed pulsars have $\sigma_{\rm TOA} \sim 10 $ ns and $\sigma_{\rm m} \sim 10^{-12}$ Hz. Throughout this paper we fix $\sigma_{\rm m} = 10^{-11}$ Hz and take it as known \textit{a priori} rather than a parameter to be inferred for the sake of simplicity; this assumption can be relaxed easily when analysing real data. Whilst $\sigma_{\rm m}$ is the same for every pulsar, $f_{\rm m} ^{(n)}(t)$ is constructed from a different random realisation of $\varepsilon^{(n)}(t)$ for each pulsar. \newline 

Finite precision arithmetic leads to numerical errors when solving Equations \eqref{eq:spinevol}--\eqref{eq:xieqn} $ \text{in  the regime } |\sigma^{(n)} dB(t) | \ll f_{\rm p}^{(n)}$ relevant to PTAs, where $dB(t)$ labels an increment of Brownian motion (see Appendix \ref{sec:kalman}). To fix the problem, we subtract the deterministic frequency evolution and track the new variable
\begin{equation}
	f_{\rm p}^{* (n)} = f_{\rm p}^{(n)} - f_{\rm em}^{(n)} \ , \label{eq:hetero1}
\end{equation}  
equivalent to a change of variables. We similarly modify the measurement variable to be
\begin{equation}
	f_{\rm m}^{* (n)} = f_{\rm m}^{(n)} - f_{\rm em}^{\diamond (n)} \ , \label{eq:hetero2}
\end{equation}
where $ f_{\rm em}^{\diamond (n)}$ is a guess of the deterministic evolution based on the pulsar ephemeris returned by {\sc tempo2}  which is measured to high accuracy in practice. For synthetic data we can set $ f_{\rm em}^{\diamond (n)} = f_{\rm em}^{(n)}$ without loss of generality, but this is impossible generally for astronomical observations, because the spin-down ephemeris is only known approximately. Equation \eqref{eq:measurement} is then updated to read 
\begin{equation}
	f_{\rm m}^{* (n)}(t) = f_{\rm p}^{* (n)}(t-d) g^{(n)}(t) -  f_{\rm em}^{(n)}(t-d)\left[ 1-g^{(n)}(t)\right] \ .
	\label{eq:measurement_cov}
\end{equation}
We emphasise that the change of variables in Equations \eqref{eq:hetero1} and \eqref{eq:hetero2} is a convenient device to bring the numerical values into a reasonable dynamic range without having to use excessively long floating point formats (e.g. long double, quadruple). It does not remove any degrees of freedom nor does it involve an approximation. In particular $f_{\rm em}^{(n)}(t_1)$
and $\dot{f}_{\rm em}^{(n)}(t_1)$ remain as static parameters but appear in the measurement equation \eqref{eq:measurement_cov} rather than the dynamical state equations, i.e. Equations \eqref{eq:spinevol}--\eqref{eq:xieqn}.

\section{Representative end-to-end analysis} \label{sec:rep_example}
In this section we apply the analysis scheme in Section \ref{sec:detect} and the validation procedure in Section \ref{sec:testing} to a PTA perturbed by a GW from an individual quasi-monochromatic SMBHB source. The analysis of a stochastic background composed of the superposition of multiple sources is more challenging and is postponed to a forthcoming paper. The goal of this section is to run end-to-end through every step in the analysis for a representative worked example to help the reader implement the scheme and reproduce the results in Sections \ref{sec:parameter_space} and \ref{sec:bias_and_identifiability}. To assist with reproducibility, we present intermediate outputs from each step. In Section \ref{sec:priors} we define the priors on $\boldsymbol{\theta}$. In Section \ref{sec:earth_psr_terms} we define the measurement equation used in the inference model by the Kalman filter. In Section \ref{sec:parameter_estim} we apply the workflow in Section \ref{sec:methodsummary} to estimate $\boldsymbol{\theta}$, for a single realisation of the pulsar process noise and the measurement noise. In Section \ref{sec:multiple_noise} we extend the parameter estimation exercise to multiple noise realizations to quantify the variance in the parameter estimates. In Section \ref{sec:detection} we calculate the detectability of the source as a function of $h_0$.

The static GW source parameters $\boldsymbol{\theta}_{\rm gw}$ used for this mock analysis are selected to be astrophysically reasonable and representative. The injected components of $\boldsymbol{\theta}_{\rm gw}$ and $\boldsymbol{\theta}_{\rm psr}$ are summarised in the second column of Table \ref{tab:parameters_and_priors}.

\subsection{Prior distribution}\label{sec:priors}
The first step is to select a reasonable Bayesian prior, $\pi(\boldsymbol{\theta})$, for the static parameters. For $\pi(\boldsymbol{\theta}_{\rm gw})$ we choose standard non-informative priors \citep[e.g.][]{Bhagwat2021,2023MNRAS.521.5077F} as summarised in Table \ref{tab:parameters_and_priors}. The static parameters $\Phi_0$ and $\psi$ are degenerate; a GW with $(\Phi_0, \psi)$ is identical to a GW with $(\Phi_0+\pi, \psi+\pi/2)$. The degeneracy is well-known in the PTA literature \citep{PhysRevLett.132.061401,2022PhRvD.105l2003B} and results in bimodal posteriors for $\Phi_0$ and $\psi$. For the purposes of this paper it is sufficient to circumvent the issue by restricting the prior on $\psi$ to the domain $0 \leq \psi \leq \pi/2$. A similar approach is taken in targeted searches of continuous GWs by LIGO \citep{2009CQGra..26t4013P}. In the hypothetical event that we do not restrict the prior we obtain bimodal posteriors in $\Phi_0$ and $\psi$ (cf. Section \ref{sec:parameter_estim}) which are out of phase by $(\Phi_0+\pi, \psi+\pi/2)$, as expected, in line with e.g. Figure 2 of \cite{2022PhRvD.105l2003B}. The degeneracy between  $\Phi_0$ and $\psi$ is related to the formal concept of identifiability from the theory of signal processing, which is applied routinely to engineering problems \citep{e5be7c83a0d24500826f6e1b414d1733}. Identifiability refers to whether it is theoretically possible to infer unique and accurate parameter values, given the measured data and the model structure \citep{WALTER1996125,DOBRE20122740,GUILLAUME2019418,casella2021statistical}. \newline 

We now discuss the choice of $\pi(\boldsymbol{\theta}_{\rm psr})$. The parameters that govern the deterministic evolution of $f_{\rm p}^{(n)}(t)$, namely $f_{\rm em}^{(n)} (t_1)$ and $\dot{f}_{\rm em}^{(n)} (t_1)$, are well-determined by radio timing observations. We identify $f_{\rm em}^{(n)} (t_1)$ and $\dot{f}_{\rm em}^{(n)} (t_1)$ with the pulsar barycentric rotation frequency and its time derivative respectively, as quoted in catalogues \citep{Manchester2005}. For the 12.5 year NANOGrav pulsars the median fractional errors on $f_{\rm em}^{(n)}(t_1)$ and $\dot{f}_{\rm em}^{(n)}(t_1)$ are $\pm 2.68 \times 10^{-13} $ per cent and $\pm 2.31 \times 10^{-3}$ per cent respectively. In this paper we adopt uniform priors on $f_{\rm em}^{(n)}(t_1)$ and $\dot{f}_{\rm em}^{(n)}(t_1)$, which extend $\pm 10^3 \eta_f^{(n)}$ and $\pm 10^3 \eta_{\dot{f}}^{(n)}$ respectively about the central, injected values, where $\eta_f^{(n)}$ and $\eta_{\dot{f}}^{(n)}$ denote the errors quoted in the ATNF Pulsar Database \citep{Manchester2005}. Wider-than-necessary priors, such as those above, test the method more stringently than narrow priors. The results below confirm that the method estimates ${\boldsymbol{\theta}}_{\rm gw}$ accurately, whether the priors are wide or narrow. \newline

The pulsar distances $d^{(n)}$ are less constrained than $f_{\rm em}^{(n)} (t_1)$ and $\dot{f}_{\rm em}^{(n)} (t_1)$, with typical uncertainties $\sim 10 \%$ \citep{Yao2017,Arzoumanian2018ApJS..235...37A}. In this paper, however, we omit the term involving $d^{(n)}$ from Equation \eqref{eq:z_trigonometric}, as justified in Section \ref{sec:parameter_estim}, so there is no need to set a prior on $d^{(n)}$. Furthermore we do not set a prior on $\gamma^{(n)}$, because we have $\gamma^{(n)} T_{\rm obs} \sim 10^{-5}$, and $\gamma^{(n)}$ is effectively ``unobservable'' over a decade; that is, for $T_{\rm obs}=10$ years the solution of Equation \eqref{eq:spinevol} is approximately independent of $\gamma^{(n)}$, as long as $[\gamma^{(n)}]^{-1} \gg T_{\rm obs}$ is satisfied. It is therefore sufficient for validation purposes to carry $\gamma^{(n)}$ through the analysis at its injected value. \footnote{We check this empirically in two independent ways for an informal selection of test cases: (i) we set an uninformative prior $\pi\left [\gamma^{(n)} / 1 \, {\rm s^{-1}} \right] \sim$ LogUniform($10^{-15}, 10^{-10}$); and (ii) we deliberately displace $\gamma^{(n)}$ in the inference analysis from its true, injected value in the synthetic data, e.g. $\gamma^{(n)} = 10^{-14} \text{ s}^{-1}$ versus $10^{-13} \text{ s}^{-1}$, respectively. The results from (i) and (ii) are found to be the same as those reported in Section \ref{sec:rep_example}.} \newline 

Most pulsars in the synthetic PTA have $10^{-25} \leq \sigma^{(n)} / (1 \, {\rm s^{-3/2}}) \leq 10^{-23}$ as calculated from Equations \eqref{eq:sigmap} and \eqref{eq:sigmap_f}. For these pulsars we set an uninformative broad prior $\pi[\sigma^{(n)} / (1 \, {\rm s^{-3/2}})] \sim$ LogUniform($10^{-26}, 10^{-22}$). The sole exception is PSR J1939+2134 which has $\sigma^{(n)} \sim 10^{-19} \text{s}^{-3/2}$. For validation purposes, we artificially set $\sigma^{(n)} = 10^{-23} \, {\rm s^{-3/2}}$ for this pulsar, so that $\sigma^{(n)}$ for every pulsar in the synthetic PTA falls within the aforementioned log-uniform prior. This is done purely for testing convenience; it is straightforward to expand the prior when analysing real, astronomical data. \newline

By not setting priors on $\gamma^{(n)}$ and $d^{(n)}$ we reduce the number of parameters to $7 + 3N$. We use the notation $\boldsymbol{\theta}_{\rm psr, reduced}$ to refer to the reduced parameter set, cf. Equation \eqref{eq:psrparams}. Explicitly, we write
\begin{equation}
	\boldsymbol{\theta}_{\rm psr, reduced} = \left \{ f_{\rm em}^{(n)}(t_1),\dot{f}_{\rm em}^{(n)}(t_1),\sigma^{(n)}\right\}_{1\leq n \leq N} \ . \label{eq:psrreduced}
\end{equation}
The injected static parameters and their corresponding priors for the representative analysis in Section \ref{sec:rep_example} are summarised in Table \ref{tab:parameters_and_priors}.
\begin{table*}
	\centering
	\begin{tabular}{lccll}
		\toprule
		&Parameter & Injected value & Units & Prior  \\
		\hline
		\multirow{7}{2mm}{$\boldsymbol{\theta}_{\rm gw}$} & $\Omega$       & $5 \times 10^{-7}$ & Hz & LogUniform($10^{-9}$, $10^{-5}$) \\
	  & $\alpha$          & $1.0$  & rad & Uniform($0, 2 \pi $)\\
	  & $\delta$              & $1.0$  & rad & Cosine($-\pi/2, \pi/2$) \\
	  & $\psi$              & $0.90$ & rad & Uniform($0, \pi/2 $) \\
	  & $\Phi_0$          & $3.30$ & rad & Uniform($0, 2 \pi $) \\
	  & $h_0$            & $5 \times 10^{-15}$ & --- & LogUniform($10^{-15}$, $10^{-9}$) \\
	  & $\iota$             & $1.0$ & rad & Sin($0, \pi$) \\ 
		\hline
		 & $f_{\rm em}^{(n)} (t_1)$       & $f_{\rm ATNF}^{(n)}$ & Hz & Uniform$\left[f_{\rm ATNF}^{(n)} - 10^3 \eta^{(n)}_{f}, f_{\rm ATNF}^{(n)} + 10^3 \eta^{(n)}_{f} \right]$ \\
		& $\dot{f}_{\rm em}^{(n)} (t_1)$       & $\dot{f}_{\rm ATNF}^{(n)}$ & s$^{-2}$ & Uniform$\left[ \dot{f}_{\rm ATNF}^{(n)} - 10^3 \eta^{(n)}_{\dot{f}}, \dot{f}_{\rm ATNF}^{(n)} + 10^3 \eta^{(n)}_{\dot{f}} \right]$ \\
		\vspace{1mm} $\boldsymbol{\theta}_{\rm psr}$ &  $d^{(n)}$       &$d_{\rm ATNF}^{(n)}$  & m & --- \\
		\vspace{1mm} & $\sigma^{(n)}$              & $\sigma_{\rm SC}^{(n)}$ & $s^{-3/2}$ & LogUniform($10^{-26}, 10^{-22}$) \\
		& $\gamma^{(n)}$              & $10^{-13}$ & s$^{-1}$ & --- \\
		\bottomrule
	\end{tabular}
	\caption{Summary of injected static parameters used for generating synthetic data in the representative example of Section \ref{sec:rep_example}, along with the prior used for Bayesian inference on each parameter (rightmost column). The top and bottom halves of the table contain the values for $\boldsymbol{\theta}_{\rm gw}$ and $\boldsymbol{\theta}_{\rm psr}$ respectively. The subscript ``ATNF'' denotes values obtained from the ATNF pulsar catalogue as described in Section \ref{sec:synt_pta}. The subscript ``SC'' indicates that the injected value is calculated using Equations \eqref{eq:sigmap} and \eqref{eq:sigmap_f} in \protect \cite{Shannon2010}. The quantities $\eta^{(n)}_{f}$ and $\eta^{(n)}_{\dot{f}}$ are the errors in $f^{(n)}_{\rm em} (t_1)$ and $\dot{f}^{(n)}_{\rm em} (t_1)$ respectively, as quoted in the ATNF catalogue. No priors are set for $d^{(n)}$ and $\gamma^{(n)}$, as those parameters do not enter the inference model,  as discussed in Section \ref{sec:priors}, cf. Equation \eqref{eq:psrreduced}.
}
	\label{tab:parameters_and_priors}
\end{table*}

\subsection{Earth and pulsar terms}\label{sec:earth_psr_terms}
The general measurement equation used in the inference model, Equation \eqref{eq:z_trigonometric}, separates into four terms. The first two terms, $H_{ij}^{(+)} \cos \Phi(t) + H_{ij}^{(\times)} \sin \Phi(t)$, depend only on the GW source parameters which are shared across all pulsars. The argument of the trigonometric functions corresponds to the GW phase at the Earth.  The next two terms, $H_{ij}^{(+)} \cos \Phi^{(n)}(t) + H_{ij}^{(\times)} \sin \Phi^{(n)}(t)$, depend additionally on $d^{(n)}$ and $\boldsymbol{q}^{(n)}$ and vary between pulsars. The argument of the trigonometric functions corresponds to the GW phase at each individual pulsar. The first and second pairs of terms are commonly referred to as the ``Earth term'' and ``pulsar term'' respectively. Whilst the Earth term is phase coherent between all pulsars, the pulsar terms have uncorrelated phases. They are typically considered as a source of self-noise and dropped from many standard PTA analyses \citep[e.g.][]{Sesana2010,Babak2012,Petiteau2013,Zhu2015,Taylors2016,Goldstein2018,Charisi2023arXiv230403786C} at the expense of a modestly reduced detection probability ($\sim 5 \%$) and the introduction of a bias in the inferred sky position \citep{Zhupulsarterms,Chen2022}. \newline 

In this paper we follow the standard approach and drop the pulsar terms from the inference calculation (but not the source model). Explicitly the measurement equation used in the Kalman filter reduces to
\begin{equation}
		f_{\rm m}^{(n)}(t) = f_{\rm p}^{(n)}(t-d) g^{(n)}_{\rm Earth}(t) \ , 
		\label{eq:measuremen_earth}
	\end{equation}
	with
	\begin{equation}
		g^{(n)}_{\rm Earth}(t) = 1 - z^{(n)}_{\rm Earth}(t) \, ,
		\label{eq:g_func_trig_earth}
	\end{equation}
and 
\begin{align}
	z^{(n)}_{\rm Earth}(t) =  \frac{[q^{(n)}]^i [q^{(n)}]^j }{2 [1 + \boldsymbol{n}\cdot \boldsymbol{q}^{(n)}] }
	&\left[H_{ij}^{(+)} \cos \Phi(t) + H_{ij}^{(\times)} \sin \Phi(t)\right]  \, . \label{eq:z_trigonometric_earth}
\end{align}
	We defer the inclusion of the pulsar terms in the inference calculation to a future paper. As discussed in Section \ref{sec:priors} this choice reduces the dimensionality of the parameter estimation problem because the measurement equation, Equation \eqref{eq:g_func_trig_earth}, is no longer a function of the pulsar distance.  We stress that the pulsar terms are dropped only when doing Bayesian inference, i.e. from the Kalman filter model that feeds into the nested sampling algorithm. The synthetic data include the pulsar terms in full. \newline

	\subsection{Posterior distribution}\label{sec:parameter_estim}
	In this section, we calculate the posterior probability distribution for the static parameters $\boldsymbol{\theta} = \boldsymbol{\theta}_{\rm gw} \cup \boldsymbol{\theta}_{\rm psr}$ and compare it to the known, injected values. The aims are (i) to demonstrate that the analysis scheme works (i.e.\ that it converges to a well-behaved, unimodal posterior), and (ii) to give a preliminary sense of its accuracy. Initially we consider a single noise realisation when generating the synthetic data for the representative example in Table \ref{tab:parameters_and_priors}. We apply the Kalman filter in conjunction with nested sampling in order to infer the joint posterior distribution $p({\boldsymbol{\theta}} | {\boldsymbol{Y}})$. \newline 
	
\begin{figure*}
	\includegraphics[width=\textwidth, height =\textwidth ]{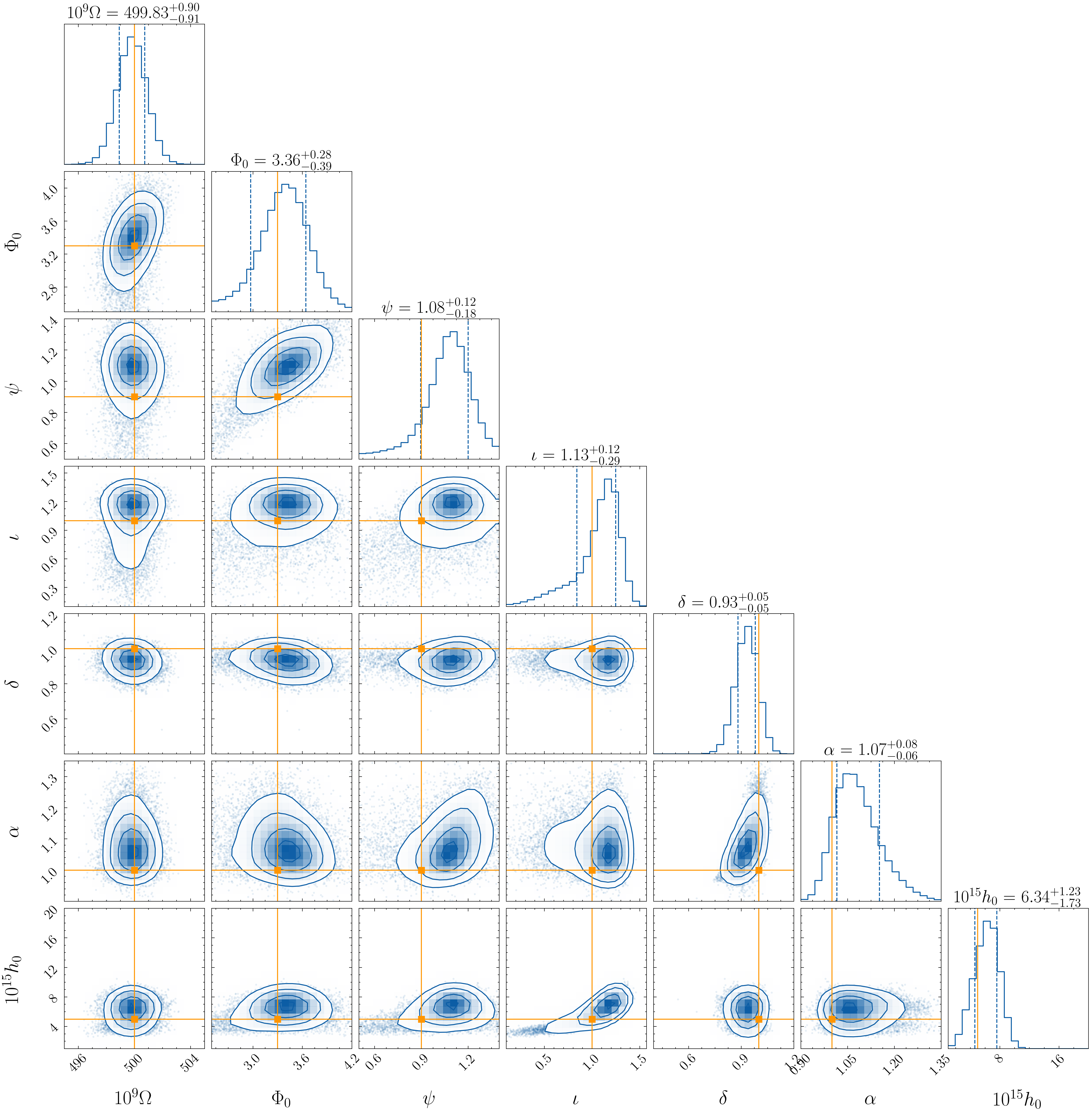}
	\caption{Posterior distribution of the GW source parameters $\boldsymbol{\theta}_{\rm gw}$ for the representative system in Table \ref{tab:parameters_and_priors}, for a single realisation of the system noise. The horizontal and vertical orange lines indicate the true injected values. The contours in the two-dimensional histograms mark the (0.5, 1, 1.5, 2)-$\sigma$ levels after marginalizing over all but two parameters. The one-dimensional histograms correspond to the joint posterior distribution marginalized over all but one parameter. The supertitles of the marginalized one-dimensional histograms specify the posterior median and the 0.16 and 0.84 quantiles. We plot the scaled variables $10^9 \Omega$ (units: ${\rm rad \, s^{-1}}$) and $10^{15} h_0$. The Kalman filter and nested sampler estimate accurately all seven parameters in ${\boldsymbol{\theta}}_{\rm gw}$, although biases are observed in some parameters. The horizontal axes span a subset of the prior domain for all seven parameters.}
	\label{fig:corner_plot_1}
\end{figure*}

Figure \ref{fig:corner_plot_1} displays results for the seven parameters in  $\boldsymbol{\theta}_{\rm gw}$ in the form of a traditional corner plot. The histograms are the one-dimensional posteriors for each parameter, marginalized over the six other parameters. The dashed vertical blue lines mark the 0.16 and 0.84 quantiles; the solid orange line marks the known injected value. The two-dimensional contours mark the (0.5, 1, 1.5, 2)-sigma level surfaces. All histograms and contours are consistent with a unimodal joint posterior, which peaks near the known, injected values. There is scant evidence of railing against the prior bounds. For this representative example, with characteristic strain $h_0 = 5 \times 10^{-15}$, the analysis scheme estimates $\boldsymbol{\theta}_{\rm gw}$ accurately. The injected values are contained within the 90\% credible interval for five of the seven parameters in $\boldsymbol{\theta}_{\rm gw}$ . \newline

The posterior is approximately symmetric about the known injected value for some parameters such as $\Omega$, where the posterior median and the injected value coincide approximately. For other parameters (e.g. $\psi$, $\iota$) the distribution is not symmetric about the injected value and the posterior median and the injected value do not coincide. The median value of the posterior for $\iota$ is shifted by $\approx 0.13$ rad relative to the injected value, although the injected value does still remain inside the 90\% credible interval. Similar effects are seen, albeit with a smaller shift, in other parameters such as $\delta$ and $\alpha$. For a single realisation of the noise it is not clear if this discrepancy is a systematic effect, i.e. a bias, or a random outcome specific to this particular noise realisation. We explore this further in Sections \ref{sec:multiple_noise} and \ref{sec:bias_and_identifiability} using more noise realizations and show that indeed there is a systematic bias from dropping the pulsar terms, similar to that reported by \cite{Zhupulsarterms}. \newline 

Similar results are obtained for the 3$N$ parameters in $\boldsymbol{\theta}_{\rm psr, reduced}$. Again, the parameters are recovered unambiguously, in the sense that the nested sampler converges smoothly to a unimodal joint posterior near the known, injected parameter values, with all $3N$ injected parameters lying within the 90\% credible interval of the $3N$ one-dimensional, marginalized posteriors. We do not display the resulting corner plot because it is too big ($3N = 141$), and because inferring ${\boldsymbol{\theta}}_{\rm gw}$ is the focus of this paper and contemporary PTA analyses.

\subsection{Multiple noise realisations: dispersion of outcomes} \label{sec:multiple_noise}
\begin{figure*}
	\includegraphics[width=\textwidth, height =\textwidth]{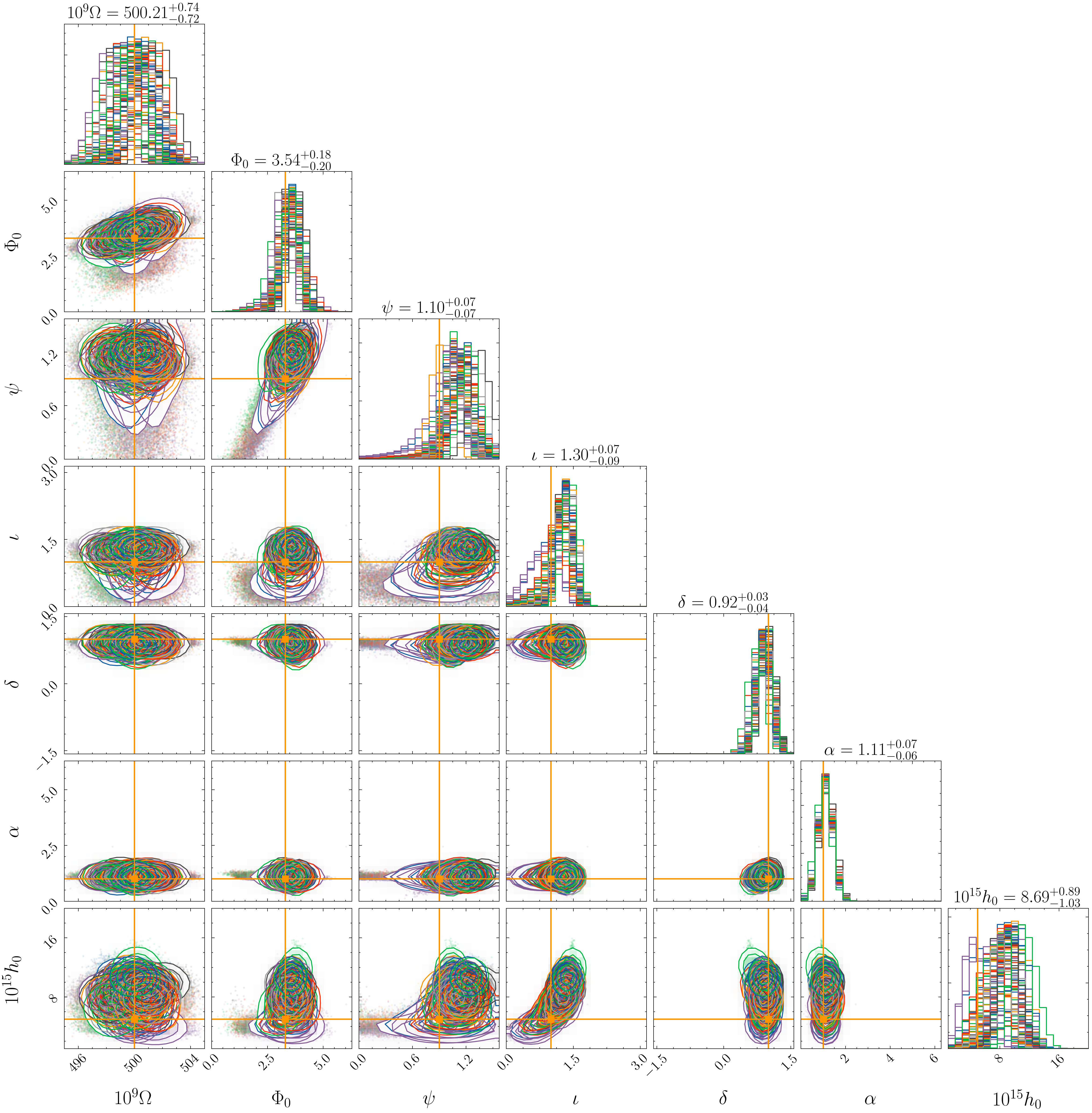}
	\caption{Same as Figure \ref{fig:corner_plot_1} but for 100 realisations of the noise processes, with curves coloured differently. The supertitles of the one-dimensional histograms record the posterior median and the 0.16 and 0.84 quantiles of the median realisation. The known, injected value lies within the 90\% credible interval for 597 out of the 700 combinations of seven parameters and 100 noise realizations. There is an appreciable dispersion among the peaks of the one-dimensional posteriors, with coefficient of variation $\sim10 \%$ across the 700 combinations. A slight skew-bias is apparent in some of the parameters, e.g. $\iota$.} 
	\label{fig:corner_plot_2}
\end{figure*}
\begin{table*}
	\centering
	\begin{tabular}{ccllll}
		\toprule
		Parameter & Injected value & Units & $W_{1, \rm median}$ & $W_{1, \rm median, inj}$ (\%) & $W_{1, \rm median, prior}$ (\%)  \\
		\hline
		$\Omega$     &   $5 \times 10^{-7}$ & Hz & $8.6 \times 10^{-10}$ &$0.2$&$8.6\times10^{-3}$  \\
		$\Phi_0$          & $3.3$ & rad &  $0.20$ &$6.2$ &$3.1$  \\
		$\psi$              & $0.9$ & rad & $0.09$ &$9.9$&$2.8$   \\
		$\iota$             & $1.0$ & rad & $0.11$ & $10.5$ &$3.3$ \\ 
		$\delta$              & $1.0$  & rad & $0.04$ &$4.0$&$1.3$  \\
		$\alpha$          & $1.0$  & rad & $0.06$ &$5.5$& $0.9$  \\
		$h_0$            & $5 \times 10^{-15}$ & ---& $1.3 \times 10^{-15}$  &$25.0$&$1.3\times 10^{-4}$  \\
		\bottomrule
	\end{tabular}
	\caption{Median value of the Wasserstein distance $W_{1, \rm median}$, for each parameter in $\boldsymbol{\theta}_{\rm gw}$, calculated across the $10^3 \choose 2$ pairs of probability posteriors, for the $10^3$ noise realisations in Figure \ref{fig:pairwise_wasserstein}. $W_{1, \rm median, inj}$ is the $W_{1, \rm median}$ value normalised by the injection value. $W_{1, \rm median, prior}$ is the $W_{1, \rm median}$ value normalised by the width of the prior domain (cf. Table \ref{tab:parameters_and_priors}). Both $W_{1, \rm median, inj}$ and $W_{1, \rm median, prior}$ are quoted as percentages. $W_{1, \rm median}$ is generally smaller than the scales set by the injection values and the prior domain.}
	\label{tab:Wasserstein}
\end{table*}
The results in Section \ref{sec:parameter_estim} are obtained for a single realisation of the noise processes $\xi^{(n)}(t)$ and $\varepsilon^{(n)}(t)$. It is important to confirm that the analysis scheme returns accurate answers for arbitrary noise realisations and that the specific realisation of the noisy data used in Section \ref{sec:parameter_estim} is not particularly advantageous by accident. It is also important to quantify, albeit approximately, the natural random dispersion in the one-dimensional posterior medians from one noise realization to the next, as the dispersion is a practical measure of the accuracy of the parameter estimation scheme, when it is applied to real astronomical data, where the true parameter values and specific noise realisation are unknown. \newline 

To this end we start with the representative example in Table \ref{tab:parameters_and_priors} and generate $1000$ realisations of the process noise $\xi^{(n)}(t)$ and measurement noise $\varepsilon^{(n)}(t)$. For each realisation we independently estimate the static parameters, $\boldsymbol{\theta}$. In Figure \ref{fig:corner_plot_2} we plot the estimates of $\boldsymbol{\theta}_{\rm gw}$ for 100 arbitrary realisations. The seven parameters in ${\boldsymbol{\theta}}_{\rm gw}$ are recovered unambiguously. The corner plot is arranged identically to Figure \ref{fig:corner_plot_1}, which stems from one realisation. We plot 100 realisations rather than the full set of 1000 to avoid overcrowding. As in Section \ref{sec:parameter_estim}, $\boldsymbol{\theta}_{\rm psr}$ is also recovered unambiguously across the 1000 noise realisations, but for the sake of brevity we do not show the results here. \newline

Figure \ref{fig:corner_plot_2} confirms two main points: (i) the results from the 100 noise realisations overlap with the single realisation in  Figure \ref{fig:corner_plot_1}; and (ii) the dispersion among the peaks of the one-dimensional posteriors is appreciable, with variations of $\sim10 \%$ (coefficient of variation) across the seven parameters and 100 realizations. Indeed, considering the one-dimensional marginalized posteriors, we find that the injected value is contained within the 90\% credible interval in 597 (i.e. $85 \%$) out of the 700 possible combinations of the seven parameters and 100 realizations. Figure \ref{fig:corner_plot_2}, like Figure \ref{fig:corner_plot_1}, displays tentative signs of bias, where the one-dimensional posteriors are not symmetric about the injected value. For example, the maximum a posteriori probability estimates of $\iota$ appear right-skewed, consistently overestimating the injected value in all 100 realisations. Similar trends are seen in $\psi$ and $h_0$. However, the width of the posteriors is comparable to the putative bias, so it is difficult to draw strong conclusions. Biases are discussed in detail in Section \ref{sec:bias_and_identifiability}. There is no strong evidence for correlations between parameter pairs, e.g.\ banana-shaped contours, except arguably $\iota$-$h_0$. \newline

We now consider the complete set of $10^3$ noise realisations, going beyond the subset of nine realisations in the preceding discussion. We ask the question: how ``similar'' are the $10^3$ marginalized, one-dimensional posteriors computed for each of the seven parameters in ${\boldsymbol{\theta}}_{\rm gw}$? There is no unique way to answer this question. In this paper we appeal to the Wasserstein distance \citep[WD;][]{Wasserstein,Villani2009} from optimal transport theory, which defines an intuitive notion of similarity between probability distributions. The WD is a popular metric in machine learning \citep{2017arXiv170107875A}, climate modelling \citep{2022JCli...35.1215P,2023QJRMS.149..843K}, computational biology \citep{GONZALEZDELGADO2023168053} and geophysics \citep{2023GeoRL..5003880M}; a short review is presented in Appendix \ref{sec:wasserstein}. The WD measures the cost of an optimal strategy for moving probability mass between two distributions from position $x$ to position $y$, with respect to some cost function $c(x,y)$. In this paper we use the first WD moment, $W_1(\mu,\nu)$, which is defined and interpreted in Appendix \ref{sec:wasserstein}. For our purposes, $W_1$ has a convenient physical interpretation given by the Monge-Kantorovich duality \citep{villani2003topics,Villani2009}, viz.
\begin{eqnarray}
	| \boldsymbol{E}(X_{\mu} )-\boldsymbol{E}(Y_{\nu} ) | \leq W_1(\mu, \nu) \ , \label{eq:WDdefn}
\end{eqnarray}
where $X_{\mu}$ and $Y_{\nu}$ are random variates drawn from the distributions $\mu$ and $\nu$ respectively, and $\boldsymbol{E}$ denotes the expected value. That is, $W_1$ bounds the difference in the expectation value of a parameter selected from ${\boldsymbol{\theta}}_{\rm gw}$ with respect to the PDFs $\mu$ and $\nu$. Taking a concrete example, suppose that we infer two one-dimensional posterior distributions $\mu(\iota)$ and $\nu(\iota)$ for $\iota$, for two different realisations of the noise, and calculate $W_1(\mu, \nu) =0.5$ rad. Then we can conclude that $| \boldsymbol{E}(\iota)_{\mu} - \boldsymbol{E}(\iota)_{\nu}\nu | \leq 0.5 \, {\rm rad}$. \newline

Table \ref{tab:Wasserstein} summarizes the WD between the $5\times 10^5$ pairs of one-dimensional posteriors across the $10^3$ realizations, for each of the seven parameters in ${\boldsymbol{\theta}}_{\rm gw}$. The median $W_1$ for each parameter is tabulated in the fourth column and denoted by $W_{1,{\rm median}}$. The $W_{1,{\rm median}}$ normalised by the injected, known, value is denoted by $W_{1,{\rm median, inj}}$ and quoted in the fifth column as a percentage. $W_{1,{\rm median, inj}}$ ranges from 
0.17\% for $\Omega$ to 25\% for $h_0$. These values quantify approximately the natural dispersion in parameter estimates when the analysis scheme is applied to real, astronomical data, where the true parameter values are unknown. The $W_{1,{\rm median}}$ normalised by the width of the prior domain is denoted by $W_{1,{\rm median, prior}}$ and quoted in the sixth column as a percentage. $W_{1,{\rm median, prior}}$ ranges from $1.3 \times 10^{-4}$ \% for $h_0$ to 3.3\% for $\iota$. These values confirm that the nested sampler converges reliably to a single, narrow peak without railing against the prior bounds for $10^3$ noise realizations. The $W_{1,{\rm median}}$ normalised by the injected, known, value is denoted by $W_{1,{\rm median, inj}}$ and quoted in the fifth column as a percentage. $W_{1,{\rm median, inj}}$ ranges from 
0.17\% for $\Omega$ to 25\% for $h_0$.

 Further analysis of the WD is performed in Appendix \ref{sec:wasserstein} where we use it for two separate purposes: (i) to measure the similarity between probability distributions and (ii) as a convenient heuristic for assessing convergence of nested sampling.

\subsection{Detectability versus $h_0$} \label{sec:detection}
\begin{figure}
	\includegraphics[width=\columnwidth]{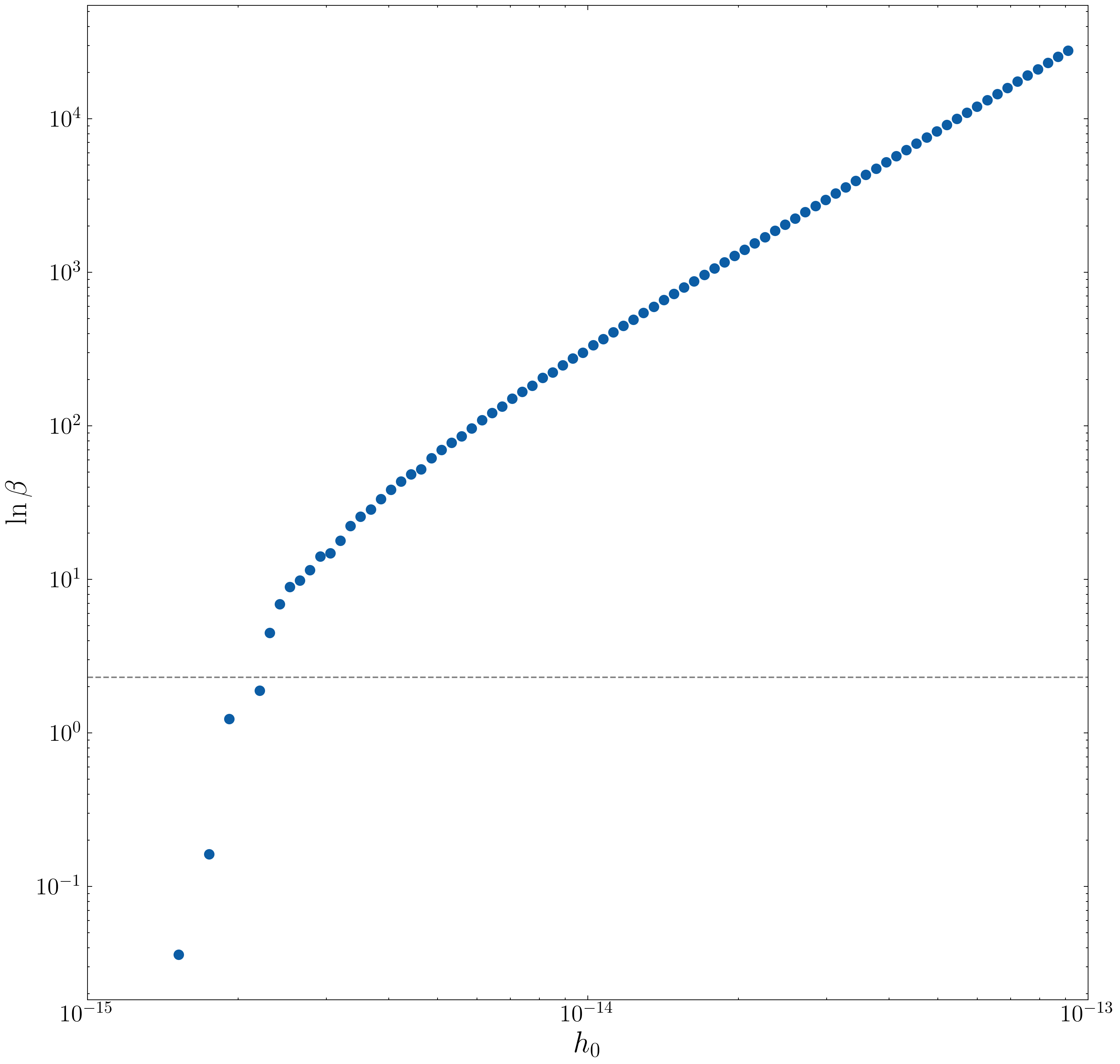}
	\caption{Log Bayes factor (odds ratio) $\ln \beta$ between the competing models $\mathcal{M}_1$ (GW present in data) and $\mathcal{M}_0$ (GW not present in data) at different GW strains, $h_0$, for the representative example in Table \ref{tab:parameters_and_priors}. The horizontal grey dashed line labels an arbitrary detection threshold, $\beta = 10$. The minimum detectable strain, for $\beta < 10$, equals $3 \times 10^{-15}$. Missing points for $\beta \lesssim 10$ occur when noise subverts the hierarchical relationship between $\mathcal{M}_0$ and $\mathcal{M}_1$ (see Section \ref{sec:detection}). Note that the vertical axis features a logarithmic scale for $\ln\beta$ not $\beta$. This is appropriate, because $\beta$ grows exponentially with $h_0$.}
	\label{fig:bayes}
\end{figure}

We frame the problem of detecting a GW in noisy PTA data in terms of the Bayesian model selection procedure described in Section \ref{sec:model_selection}. In Equation \eqref{eq:bayes} $\mathcal{M}_1$ is the Earth-term-only model with a GW present, i.e. the state-space model with a Kalman filter based on Equation \eqref{eq:measuremen_earth}. The Bayes factor, $\beta$, defined in Equation \eqref{eq:bayes} is plotted logarithmically in Figure \ref{fig:bayes} for the representative source in Table \ref{tab:parameters_and_priors}, except that we now vary the source amplitude, $h_0$, from $10^{-15}$ (undetectable) to $10^{-12}$ (easily detectable). To control the test, the noise processes in the synthetic data are identical realisations for each value of $h_0$; the only change from one $h_0$ value to the next is $h_0$ itself. \footnote{Changing the noise realizations as well, from one value of $h_0$ to the next, adds uninformative scatter to the trend in Figure \ref{fig:bayes}.} \newline

We see in Figure \ref{fig:bayes} an approximate quadratic relationship $\ln \beta \propto h_0^2$ for $h_0 \gtrsim 10^{-14}$. The GW source is detectable with decisive evidence ($\beta \geq 10$) for $h_0 \gtrsim 2.2 \times 10^{-15}$. Of course, the minimum detectable strain is particular to the system in Table \ref{tab:parameters_and_priors}. It is influenced in general by $T_{\rm obs}$, ${\boldsymbol{\theta}}_{\rm gw}$, and ${\boldsymbol{\theta}}_{\rm psr}$, as discussed in Section \ref{sec:parameter_space}. Adjusting $\sigma_{\rm m}$ we find that an approximate quadratic relationship $\ln \beta \propto h_0^2 / \sigma_{\rm m}^2$ also exists, where $h_0 / \sigma_{\rm m}$ is the effective signal-to-noise ratio (SNR). \newline

The log odds ratio $\ln \beta$ drops off (i.e. $\beta$ levels off at unity) for $h_0 \lesssim 3 \times10^{-15}$. This happens because the two competing models becoming increasingly indistinguishable once the measurement noise dominates the GW signal. Moreover, the points in Figure \ref{fig:bayes} become sparser for $h_0 \lesssim 3 \times 10^{-15}$. This happens because we obtain $\ln \beta <0$ for the missing, intermediate points. This is a noise artefact of the nested sampler. When the sampler converges sub-optimally, the hierarchical relationship between $\mathcal{M}_0$ and  $\mathcal{M}_1$ fails, and $\mathcal{Z}(\boldsymbol{Y} | \mathcal{M}_1) > \mathcal{Z}(\boldsymbol{Y} | \mathcal{M}_0)$ no longer holds. There is nothing special about the particular missing points; if one recreates the $\beta(h_0)$ curve by rerunning the nested sampler with another random seed, a different set of points are missing. 

%

\section{SMBHB source parameters} \label{sec:parameter_space}
\begin{figure}
	\centering
	\includegraphics[width=\columnwidth]{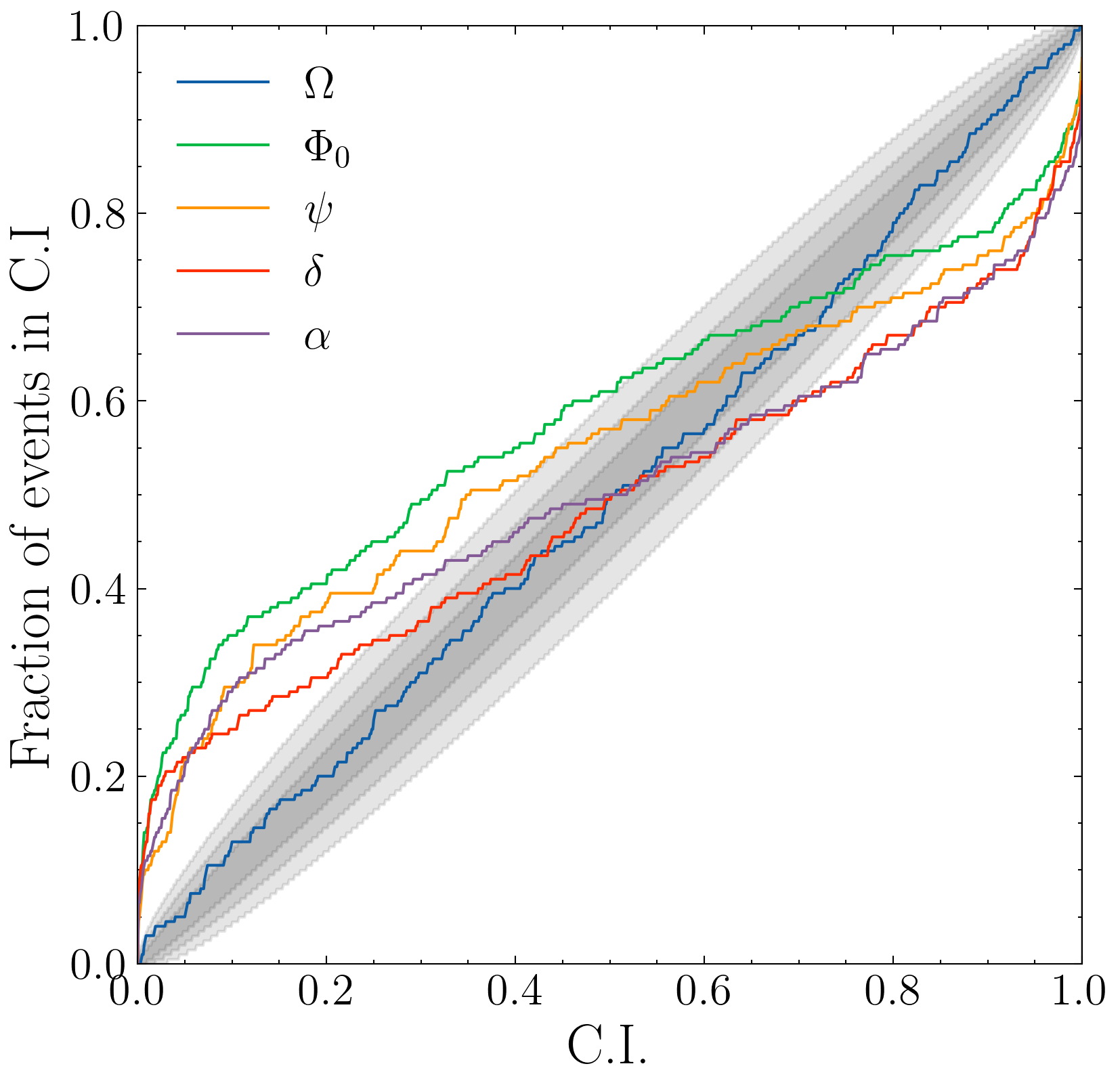}
	\caption{Fraction of injections included within a given credible interval of the estimated posterior, as a function of the credible interval itself (i.e. PP plot). The injections are 200 simulated GW sources generated by drawing randomly five parameters in $\boldsymbol{\theta}_{\rm gw}$ from the prior distributions in Table \ref{tab:parameters_and_priors}. Each coloured curve corresponds to a different parameter (see legend). The parameters $h_0$ and $\iota$ are fixed at $5 \times 10^{-15}$ and 1.0 rad respectively in order to maintain an approximately constant SNR. The grey shaded contours label the $1\sigma$, $2\sigma$ and 3$\sigma$ confidence intervals. For parameters with well estimated posteriors, the PP curve should fall along the diagonal of unit slope. $\Omega$ is generally well-estimated (i.e. it lies close to the unit diagonal) but the four other parameters show evidence of being over-constrained (i.e. the curves lie above the unit diagonal for low credible intervals, and below the unit diagonal for high credible intervals). This is due to a modelling bias whose origin is discussed in Sections \ref{sec:parameter_estim} and \ref{sec:bias_and_identifiability}.}
	\label{fig:parameter_space}
\end{figure}
Section \ref{sec:rep_example} focuses on a single representative system, parameterised in Table \ref{tab:parameters_and_priors}. In this section we test the method for various systems, varying the SMBHB source parameters through astrophysically relevant ranges. \newline 

We analyse 200 injections constructed by fixing $h_0 = 5 \times 10^{-15}$ and $\iota =1.0$ rad and drawing the remaining five elements of $\boldsymbol{\theta}_{\rm gw}$ randomly from the prior distributions defined in Table \ref{tab:parameters_and_priors}. We fix $h_0$ and $\iota$ in order to maintain an approximately constant SNR across the 200 injections. For each injection we compute the posterior distribution of ${\boldsymbol{\theta}}_{\rm gw}$. To summarise the results we use a parameter-parameter (PP) plot \citep{doi:10.1198/106186006X136976}. A PP plot displays the fraction of injections included within a given credible interval of the estimated posterior, plotted as a function of the credible interval itself. In the ideal case of perfect recovery, the PP plot should be a diagonal line of unit slope. \newline 

 Figure \ref{fig:parameter_space} displays the results of the numerical experiment described in the previous paragraph. The shaded grey contours enclose the $1\sigma$, $2\sigma$, and $3\sigma$ significance levels for 200 injections. We see that only $\Omega$ falls wholly within the $3\sigma$ shaded region. The PP curves for the other parameters deviate from the diagonal of unit slope. The deviation is more pronounced for $\alpha$ and $\delta$ and less for $\psi$ and $\Phi_0$. The shape of the graph indicates that the posteriors for these parameters are over-constrained; there are fewer injections contained within higher-value credible intervals than would be expected statistically, and there are more injections contained within lower-value credible intervals. This stems from the bias noted in Section \ref{sec:multiple_noise}. The origin of the bias is discussed in Section \ref{sec:bias_and_identifiability}. 
 
\section{Systematic biases} \label{sec:bias_and_identifiability}
\begin{figure*}
	\includegraphics[width=\textwidth, height =\textwidth]{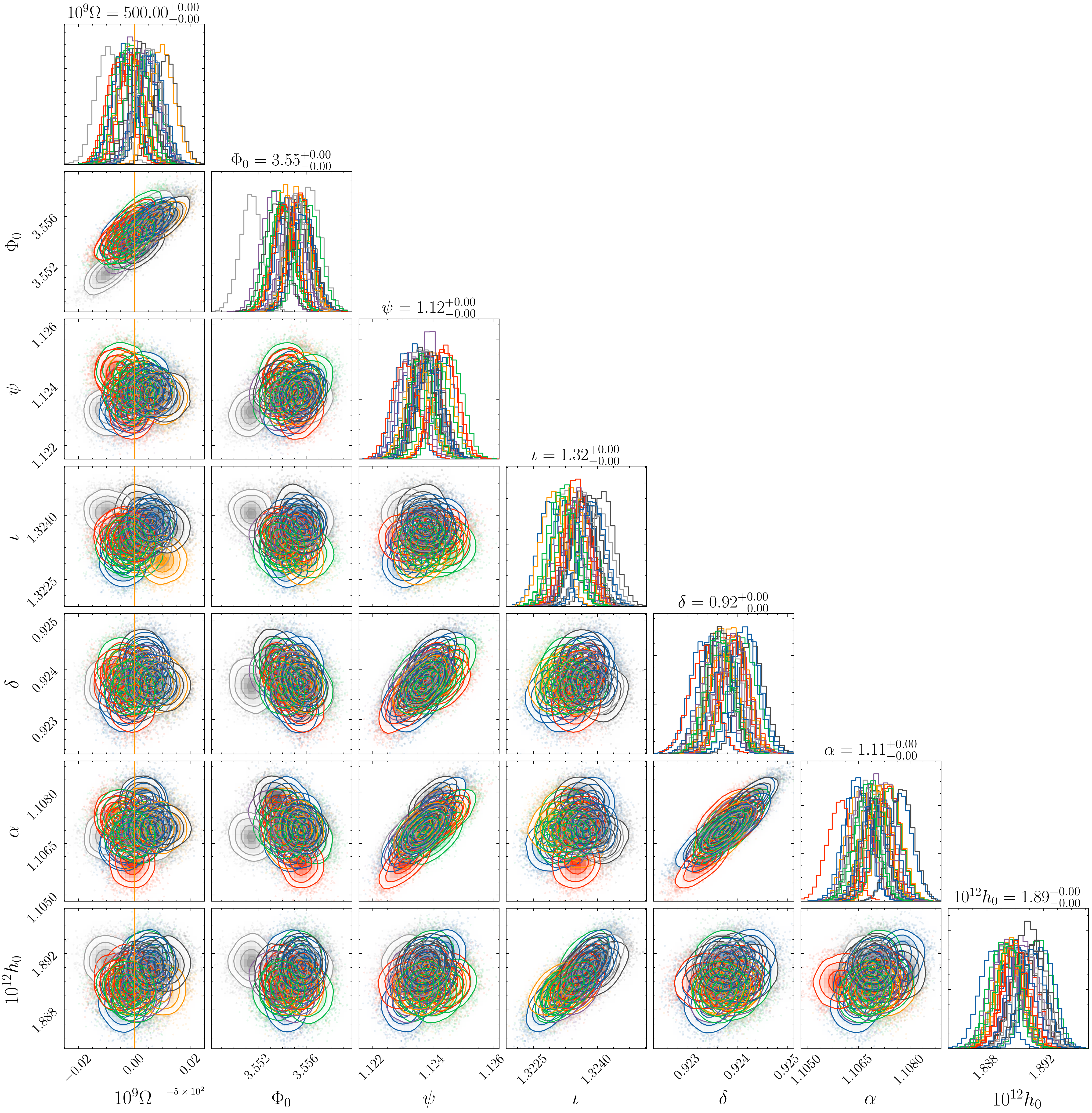}
	\caption{Same as Figure \ref{fig:corner_plot_2} but for a high-SNR system with $h_0 = 10^{-12}$. For six out of seven parameters the true, injected value does not fall within the 90\% credible interval and falls outside the plotted domain. The estimated posteriors are biased away from the injected value due to dropping the pulsar terms. Note that the plotted domain is narrower than in Figure \ref{fig:corner_plot_2}.} 
	\label{fig:bias_for_large_h}
\end{figure*}

\begin{figure*}
	\begin{subfigure}{.48\linewidth}
		\includegraphics[width=\textwidth]{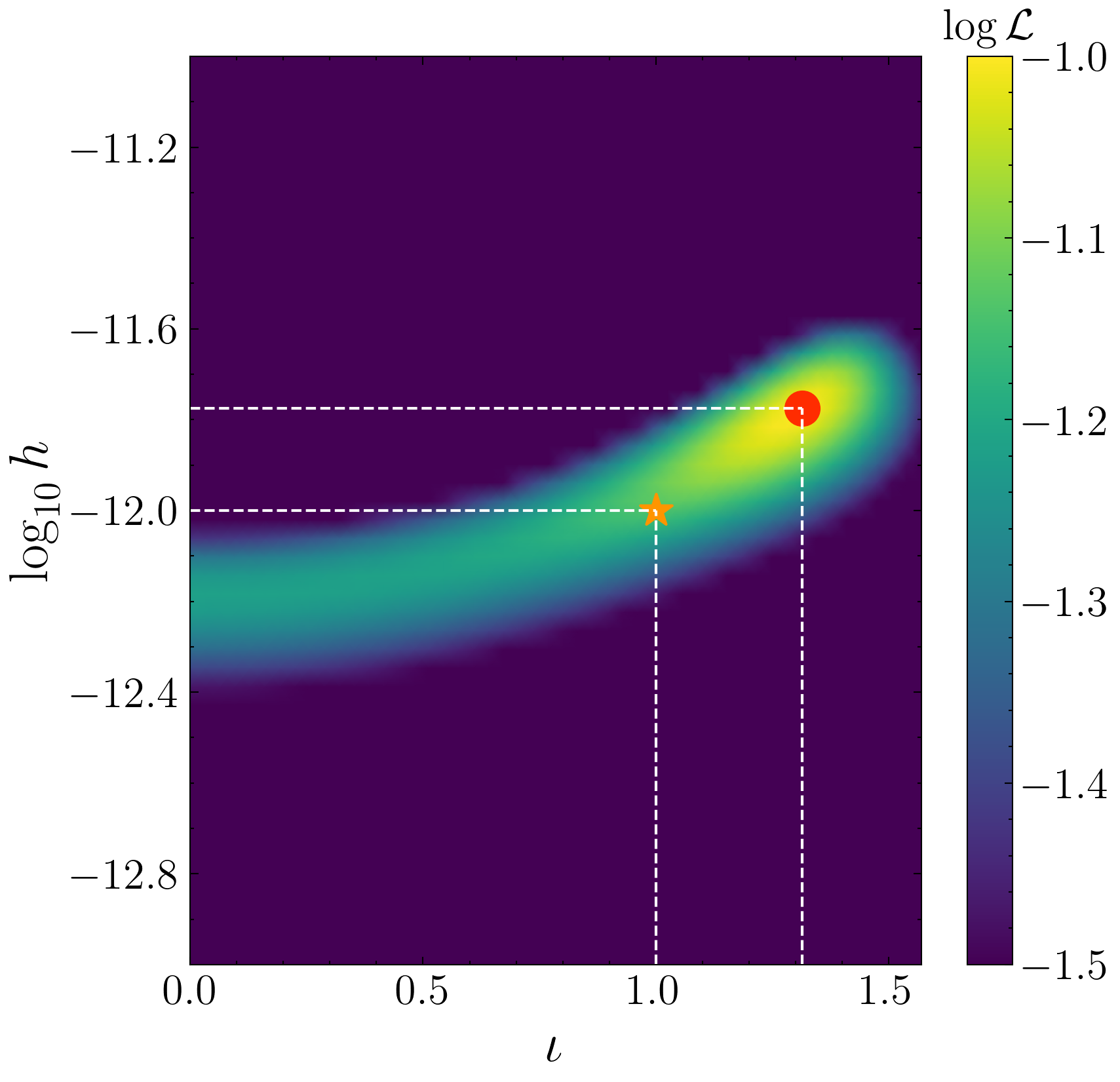}
		\caption{}
		\label{fig:h_iota_biasa}
	\end{subfigure}
	\begin{subfigure}{.48\linewidth}
		\includegraphics[width=\textwidth]{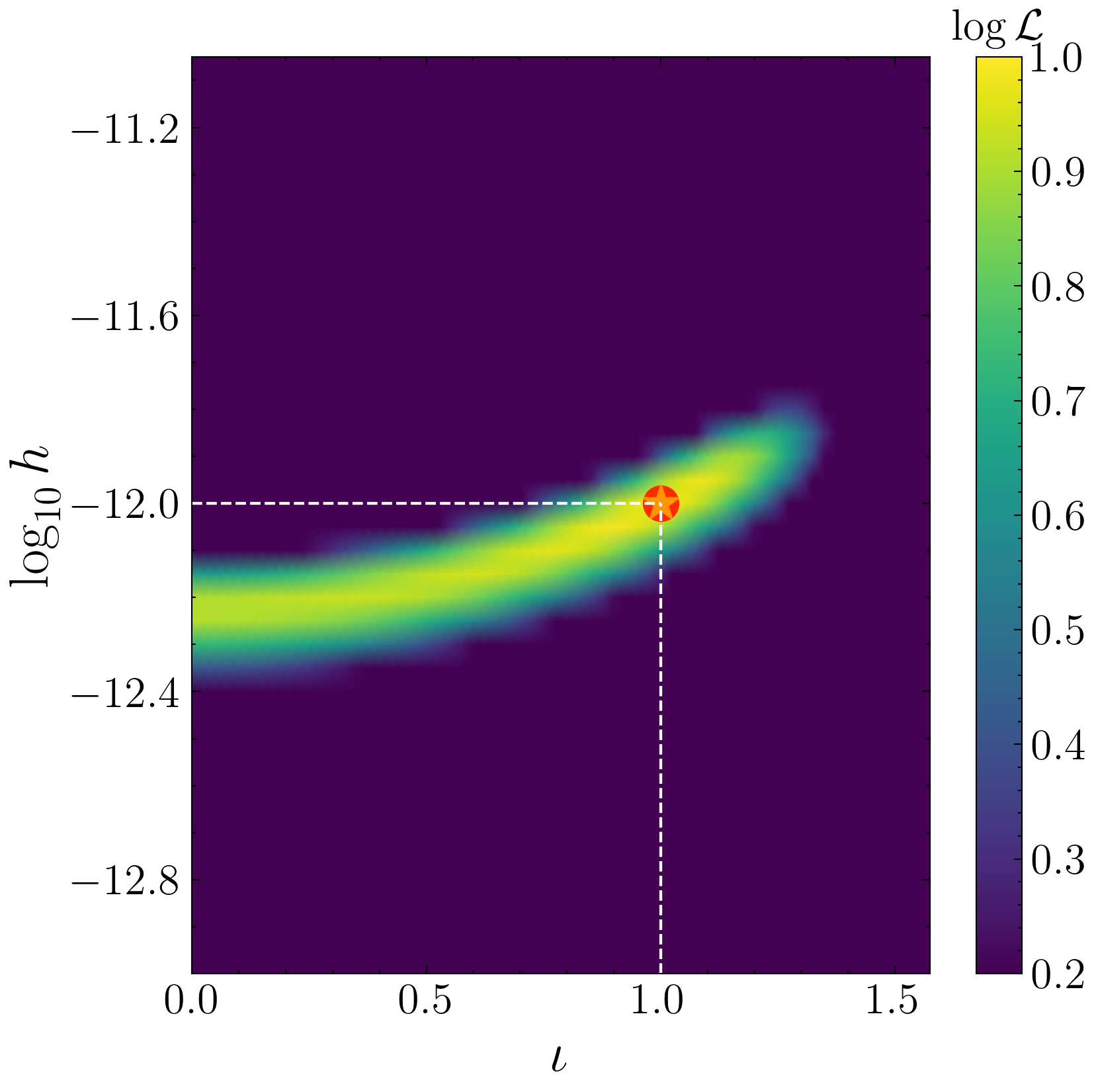}
		\caption{}
		\label{fig:h_iota_biasb}
	\end{subfigure}
	\caption{Log-likelihood $\log \mathcal{L}$ (Equation \ref{eq:likelihood}) contours in the $\iota$-$h_0$ plane calculated based on the Kalman filter using (a) exclusively the Earth terms in Equation \eqref{eq:measuremen_earth} and (b) the Earth terms and the pulsar terms in Equation \eqref{eq:measurement}. The Kalman filter runs on a single realisation of the data. The red point marks the maximum of $\log \mathcal{L}$. The orange star marks the injected parameters, cf. Table \ref{tab:parameters_and_priors}. The dashed white lines label the $(\iota, h_0)$ coordinates of the red point and the orange star. In (a) there is a bias; the red point and the orange star occupy different locations. In (b) there is no bias; the red point and the orange star coincide.}
	\label{fig:h_iota_bias}
\end{figure*}

\begin{figure*}
	\begin{subfigure}{.48\linewidth}
		\includegraphics[width=\textwidth]{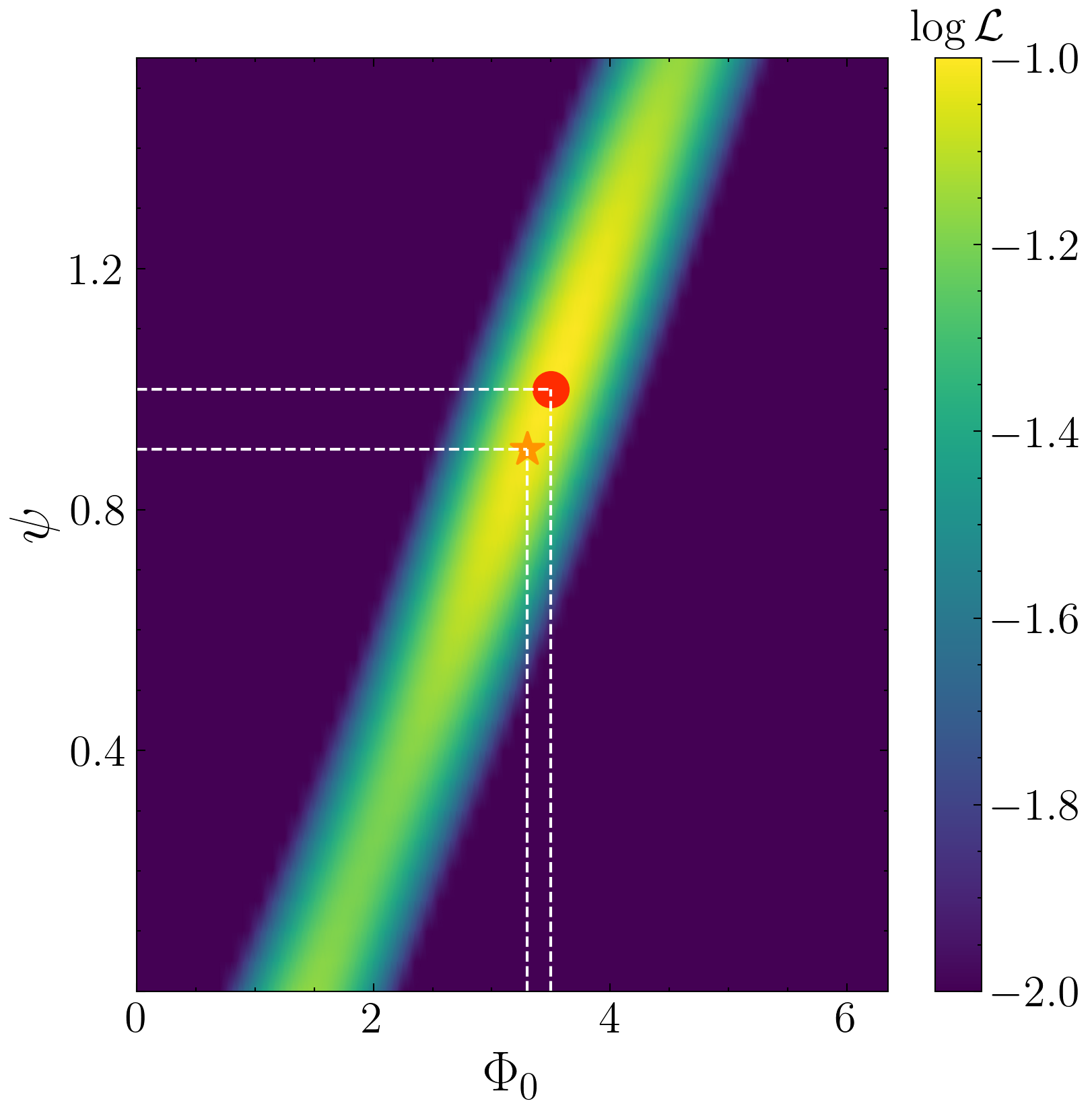}
		\caption{}
		\label{fig:phi_psi_biasa}
	\end{subfigure}
	\begin{subfigure}{.48\linewidth}
		\includegraphics[width=\textwidth]{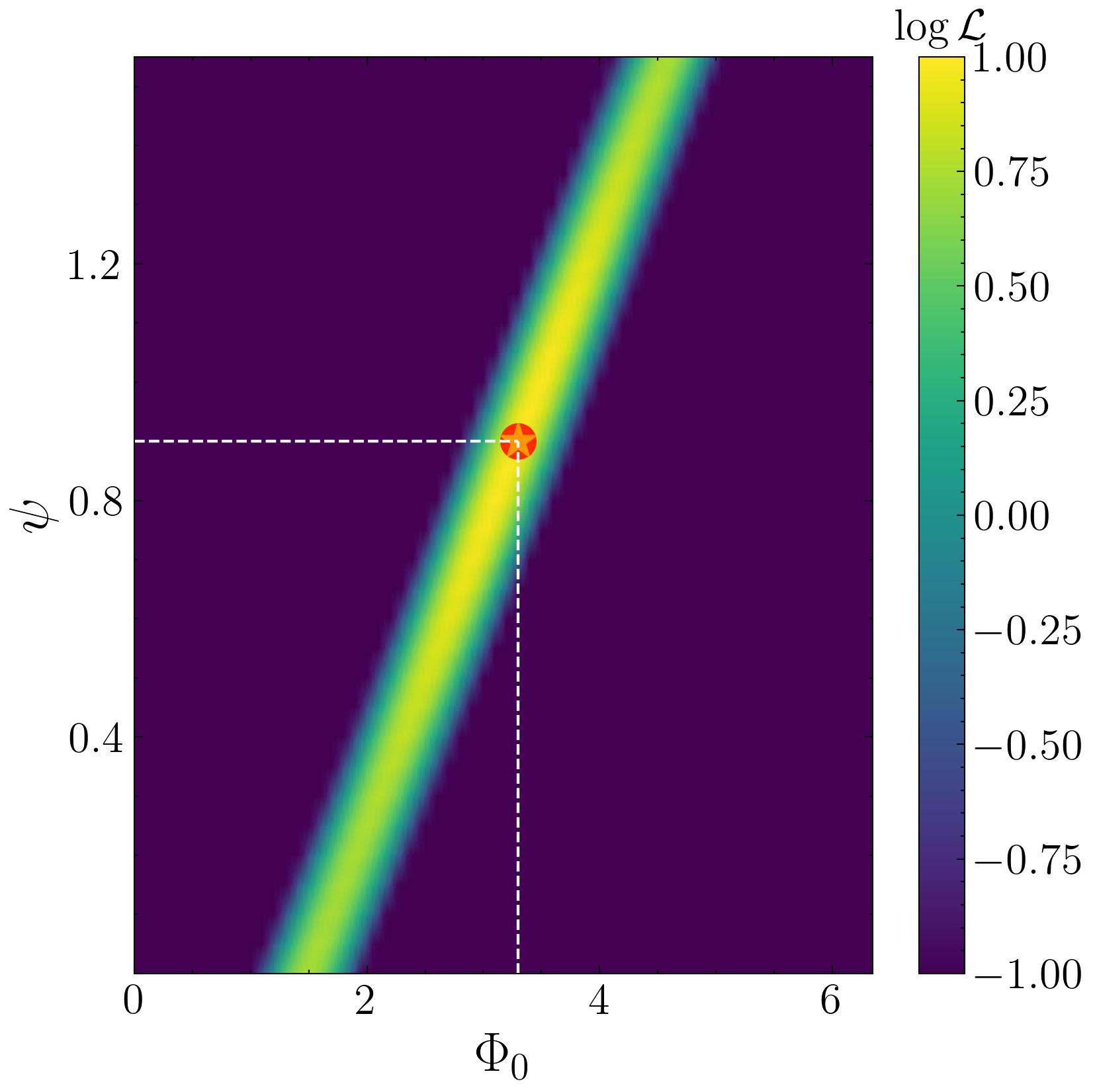}
		\caption{}
		\label{fig:phi_psi_iota_biasb}
	\end{subfigure}
	\caption{Same as Figure \ref{fig:h_iota_bias}, but for  $\log \mathcal{L}$ contours in the $\Phi_0$-$\psi$ plane. Biases are observed in (a), when the Kalman filter runs using exclusively the Earth terms in Equation \ref{eq:measuremen_earth}. Including the pulsar term in (b) remedies the bias and the likelihood maximum coincides with the injected static parameter values.}
	\label{fig:phi_psi_bias}
\end{figure*}

 The tests in Sections \ref{sec:rep_example} and \ref{sec:parameter_space} present preliminary evidence that the inferred value of ${\boldsymbol{\theta}}_{\rm gw}$ (e.g. the medians of the one-dimensional marginalised posteriors) is biased away from the true injected value. The biases result from dropping the pulsar terms from the Kalman filter measurement equation as described in Section \ref{sec:parameter_estim} and are well known in the literature \citep{Zhupulsarterms,Chen2022}.  In this section we further investigate these biases. In order to elucidate the matter without confusion from the measurement noise, we switch to the high-SNR regime and set $h_0 = 10^{-12}$ (c.f. Figure \ref{fig:bayes}) for the representative system in Table \ref{tab:parameters_and_priors}. \newline

Figure \ref{fig:bias_for_large_h} displays a corner plot for $\boldsymbol{\theta}_{\rm gw}$, analogous to Figure \ref{fig:corner_plot_2}, but for high SNR $(h_0 = 10^{-12})$. Except for $\Omega$, the injected values lie outside the 90\% credible interval, and indeed fall outside the plotted domain. The deviation is most severe for $\iota$ with a bias of $\approx 0.3$ rad but is also present to a lesser extent in the static parameters other than $\Omega$. In this section we demonstrate how the bias stems from dropping the pulsar terms described in Section \ref{sec:parameter_estim}. \newline

Figure \ref{fig:h_iota_biasa} displays the $\iota$-$h_0$ contours of the log-likelihood returned by the Kalman filter, i.e. $\log \mathcal{L}(\iota, h_0)$, viz. Equation \eqref{eq:likelihood}. The Kalman filter includes the Earth term in the measurement equation, i.e. using Equation \eqref{eq:measuremen_earth}. The function $\log \mathcal{L}(\iota, h_0)$ is evaluated across the prior domain and all other parameters are held constant at the true injected value of the representative system in Table \ref{tab:parameters_and_priors}. The function $\log \mathcal{L}(\iota, h_0)$ is calculated for a single realisation of the data and the values are normalised with respect to $\max  |\log \mathcal{L}(\iota, h_0) | $. The red point marks the location of $\max \left[ \log \mathcal{L}(\iota, h_0) \right]$. The orange star marks the location of $ \log \mathcal{L}(\iota=1.0, h_0=5\times 10^{-15})$, i.e the location of the injected parameters (cf. Table \ref{tab:parameters_and_priors}). The dashed white lines illustrate the $(\iota, h_0)$ coordinates of the  red point and the orange star. Figure \ref{fig:h_iota_biasb} is identical to Figure \ref{fig:h_iota_biasa}, except that  the Kalman filter includes the Earth and pulsar terms in the measurement equation, i.e. using Equation \eqref{eq:measurement}. \newline

The key observation from Figures \ref{fig:h_iota_biasa} and \ref{fig:h_iota_biasb} is that the location of the maximum ($\iota \approx 1.3 \, {\rm rad}$, $\log_{10} h_0 \approx -11.8$) does not exactly coincide with the true, injected value ($\iota = 1.0 \, {\rm rad}$, $\log_{10} h_0 = -12$ ), when only the Earth terms are included. That is, the red point and the orange star occupy different locations in Figure \ref{fig:h_iota_bias}. This is the cause of the bias in the one-dimensional marginalised posteriors in Figure \ref{fig:bias_for_large_h}. In contrast, the likelihood maximum coincides with the true, injected value, when the pulsar terms are included, i.e. the red point and the orange star are overlaid in Figure \ref{fig:h_iota_biasb}. Similar results where the likelihood maximum is offset from the injected values are obtained for the other parameters, with the exception of $\Omega$. An additional example for likelihood contours in the $\Phi_0$-$\psi$ plane is presented in Figure \ref{fig:phi_psi_bias}. The conclusions are analogous to those drawn in the $\iota$-$h_0$ discussion. \newline

Figures \ref{fig:h_iota_biasa} and \ref{fig:h_iota_biasb} emphasise that the bias observed in Figures \ref{fig:corner_plot_1}, \ref{fig:corner_plot_2} and \ref{fig:bias_for_large_h} is not a numerical convergence problem, whereby the nested sampler gets stuck in a local optimum. It is instead a structural problem rooted in the fact that the model used to generate the synthetic data is different from the model used for parameter inference. Dropping the pulsar terms biases several of the inferred parameters (e.g.\ those that affect the GW amplitude), not just the sky position \citep{Zhupulsarterms,Chen2022}. The bias is not particular to our method; it is shared by all likelihood-based methods that do not include the pulsar terms in the inference model, as is well known in the literature. Quantifying the bias for practical, astronomical PTA continuous wave searches requires a thorough exploration of the SMBHB parameter space, which falls outside the scope of this introductory paper. The bias also depends on the PTA configuration. If the bias is low enough the uncertainty in the one-dimensional marginalised posterior dominates the bias for quiet continuous wave sources with low SNR. This is what we observe for the synthetic data; for example, in Figure \ref{fig:corner_plot_2}, the marginalised posterior broadens, as the SNR decreases, until it overwhelms the shift due to the bias. Conversely the bias dominates if it is high enough, or if the source is sufficiently loud Generalizing the Kalman tracker and nested sampler to correct for the bias will be a key goal of a forthcoming paper.

\section{Computational cost}\label{sec:computation_costs}
Bayesian techniques like nested sampling or Markov chain Monte Carlo are computationally intensive, requiring a large number of likelihood evaluations. In order to readily use these methods, it is important that the central processing unit (CPU) time for a single likelihood evaluation is fast \citep[although some methods for Bayesian inference with expensive likelihoods do exist, e.g.][]{2014InvPr..30a5004B,2023arXiv231208085D,2023JCAP...10..021E}. In traditional PTA analyses, the likelihood function is evaluated on all data simultaneously, with corresponding memory demands of large-matrix multiplication and inversion \citep[see e.g. Section 7 of ][]{2021arXiv210513270T}. In contrast, state-space algorithms are iterative and read only the data at the given time step, with correspondingly smaller memory demands. If state-space algorithms for PTA analysis are to be used in conjunction with Bayesian techniques, it is important that they run quickly. Ideally, they should run at least as fast as traditional PTA analyses, so that they can run in tandem as a cross-check. Moreover, the algorithms must scale favourably with increasing data volume, i.e. the number of pulsars in the array, or the number of time samples. In this section we benchmark the software implementation of the state-space PTA analysis scheme in this paper, as a rough practical guide. We also discuss briefly the scaling of the computational cost as a function of $N$, the number of pulsars in the PTA, and $N_t$, the number of TOAs. More extensive benchmarking is postponed, until the analysis scheme is generalized to ingest TOAs instead of $f_{\rm m}^{(n)}(t)$, and a public, production version of the software is written. \newline

In this paper, as a first pass, we implement a naive version of the Kalman filter, without optimisation. As a benchmark, a single likelihood evaluation implemented in Python for the synthetic data presented in Section \ref{sec:testing} takes $T_{\mathcal{L}} \sim 9 \, {\rm ms}$ of CPU time on a 2.6 GHz  Intel Core i7 processor. A translation into more performant languages such as C++ \citep{andrist2020c++} or Julia \citep{2012arXiv1209.5145B}, the use of Python pre-compilation libraries such as Numba \citep{2015llvm.confE...1L} or JAX \citep{jax2018github}, or additional optimisation \citep{gorelick2014high}, would reduce $T_{\mathcal{L}}$.


It is important to understand how $T_{\mathcal{L}}$ scales with both $N_t$ and $N$. Regarding the former, the theoretical time complexity of the Kalman filter (i.e. its asymptotic behaviour), is $\mathcal{O}(N_t)$, because the Kalman filter is an iterative algorithm. Regarding the latter, the rate-limiting step is set by matrix multiplication, e.g.\ in Equations \eqref{eq:KF_predict} and \eqref{eq:KF_inv} in the Kalman filter's predict and update steps respectively; see Appendix \ref{sec_kalman_general}. The dimension of the Kalman filter matrices, e.g. Equations \eqref{eq:kalman1}--\eqref{eq:kalman_final}, is set by $N$. Matrix multiplication without optimization scales as $\mathcal{O}(N^3)$ \citep{Daum2021}. Modern routines for matrix multiplication reduce the complexity to $\sim \mathcal{O}(N^{2.3})$ \citep{trefethen1997numerical}. Advanced techniques to further reduce the complexity include CKMS recursion \citep{kailath2000linear} or low-rank perturbation methods \citep{doi:10.1080/10618600.2012.760461}. We refer the reader to \cite{8861457} for further information about complexity reduction for the Kalman filter. The memory complexity of the Kalman filter  is $\mathcal{O}(N^{2})$, independent of $N_t$, because the algorithm is iterative. \newline 

\section{Conclusion}\label{sec:discussion}

In this paper we demonstrate a new method for the detection and parameter estimation of GWs from individual, monochromatic SMBHBs in PTA data. The new method is complementary to traditional approaches. We track the evolution of the intrinsic pulsar timing noise explicitly via a state-space method, rather than fitting for the ensemble-averaged power spectral density of the noise. That is, we disentangle statistically the specific time-ordered realisation of the timing noise from the GW-induced modulations, and thereby infer the GW source parameters conditional on the specific observed realisation of the noisy data. We implement a Kalman filter in order to track the intrinsic rotational state evolution of the pulsar and combine it with a Bayesian sampling framework to estimate the posterior distributions of each static parameter, as well as the associated Bayesian evidence (marginalised likelihood) of the model with and without a GW. The favourable time-asymptotic behaviour of the adaptive gain improves the agility of the Kalman tracker compared to alternatives, such as least-squares estimators, and the recursive implementation of the Kalman tracker accelerates the computation. \newline 
 
 We test the new method on synthetic data and find that it detects injected signals successfully and estimates their static parameters accurately with relatively low computational cost. We initially focus on a single, astrophysically representative, SMBHB GW source observed synthetically by the 12.5-year NANOGrav pulsars with $T_{\rm obs} = 10 \, {\rm years}$. The minimum detectable strain is estimated to be $\min(h_0) \approx 2 \times 10^{-15}$ for $\iota=1.0$ rad. We then repeat the parameter estimation exercise for 1000 noise realisations to compute the natural dispersion in the recovered values and hence quantify the statistical accuracy of the method in a real astrophysical application, when the true parameter values are unknown. Consistent posteriors are obtained for most realisations. The median WD is limited to $\lesssim 4\%$ of the width of the prior domain (i.e.\ negligible railing). The median WD divided by the injected value, an approximate measure of the natural dispersion, ranges from  0.2\% for $\Omega$ to 25\% for $h_0$. \newline

Exploration of a broader SMBHB parameter domain at fixed $\iota$ and $h_0$ via 200 randomly sampled parameter vectors reveals a bias in the estimates of each element of the static parameters  $\boldsymbol{\theta}_{\rm gw}$, with the exception of $\Omega$. The bias is examined for a specific SMBHB GW source in the limit of high-SNR. It is greatest for $\iota$, amounting to $\approx 0.3$ rad.  Smaller biases of $\lesssim 0.1$ rad are also observed in $\Phi_0, \psi, \delta$ and $\alpha$. The bias is shown to result from dropping the pulsar terms from the measurement equation in the Kalman filter, consistent with the work of previous authors \citep{Zhupulsarterms,Chen2022}. The computational cost of the method is evaluated; a single likelihood evaluation is found to take $\sim9$ ms which compares favourably with traditional PTA analyses. The runtime of the full PTA analysis is found to be $\sim 1.5 \times 10^{2}$ min for the representative SMBHB GW source. \newline

We emphasize that the Kalman tracker and nested sampler in this paper do not supplant traditional PTA analysis approaches; they complement traditional approaches and are most powerful when used in tandem. Relatedly, it is misleading to ask whether the Kalman tracker and nested sampler are more or less sensitive than traditional approaches for two reasons. First, one must still generalize the Kalman tracker and nested sampler to ingest TOAs directly, as happens traditionally, instead of ingesting a frequency time series, as in this paper. Generalizing the algorithm is a subtle task and is postponed to a forthcoming paper. Second, the Kalman tracker and nested sampler are conditional on a noise model that is related but different to the noise model in traditional approaches. The analysis in this paper assumes a specific, time-ordered realization of a mean-reverting Ornstein-Uhlenbeck process satisfying Equations \eqref{eq:spinevol}--\eqref{eq:xieqn}, whereas traditional analyses assume a stationary Gaussian process described by an ensemble-averaged, power-law power spectral density, whose amplitude and exponent are adjustable. Hence the sensitivities cannot be compared directly. Clarifying the similarities and differences between various approaches promises to be a fruitful avenue of future work. It is also a subject of attention in audio-band GW data analysis involving hidden Markov models applied to data from terrestrial long-baseline interferometers \citep{PhysRevD.102.023006,PhysRevD.105.022002,Abbott_2022SCO,2022PhRvD.106f2002A}. \newline

 The approach in this paper can be extended in at least five ways, enumerated as (i)--(v) below. \newline 
 
(i) A natural first extension is to retain the pulsar terms in Equation \eqref{eq:measurement}. Alternatively, if the method continues to be used with solely the Earth term, in line with standard practice in some published PTA analyses, it would be desirable to evaluate systematically the incurred biases in the model parameters across an astrophysically representative parameter domain, supplementing the results obtained by other authors \citep{Zhupulsarterms,Chen2022,KimpsonP2}.\newline

(ii) In this paper, we consider one specific configuration of the synthetic PTA, namely the same pulsars that make up the 12.5-year NANOGrav (Section \ref{sec:synt_pta}). It is interesting to compare the performance of the method using different pulsar configurations, e.g. those in the PPTA and EPTA. Adding pulsars to PTAs increases the computational cost so it may be advantageous to select a subset of pulsars by exploiting formal optimization techniques from electrical engineering \citep{2023MNRAS.518.1802S}, although the Kalman filter and nested sampler are already cheaper computationally than some other methods. \newline

(iii) In practice different PTA pulsars are observed with different cadences at different times. Extending the Kalman filter to non-uniform time sampling is straightforward \citep{zarchan2000fundamentals}. \newline

(iv) The assumption of a monochromatic source is well-justified astrophysically in various regimes (see Section \ref{sec:plane_gw}) and is an appropriate starting point for this introductory paper. Nevertheless, SMBHBs are not strictly monochromatic. It is interesting to extend the state-space framework such that $f_{\rm gw}$ evolves in time. For $\Delta f_{\rm gw} < 1 / T_{\rm obs}$ (see Equation \eqref{eq:f_evolution}) evolution adds noise incoherently to the pulsar terms, whilst for $\Delta f_{\rm gw} > 1 / T_{\rm obs}$ the pulsar terms induce phase shifts that affect the overall phase coherence \citep{Sesana2010,Perrodin2018}. Careful consideration of the evolution of $f_{\rm gw}$ will be needed when including the pulsar terms in the inference model, as in point (i). \newline

(v) Finally, we assume in this paper that there is only one GW source. However, it may be possible to resolve multiple continuous GW sources concurrently \citep{PhysRevD.85.044034}. The Kalman filter extends naturally to multiple sources; one can modify Equation \eqref{eq:measurement} easily to accommodate a linear superposition of GWs. Taking the logic further, the stochastic background itself is arguably an incoherent sum of many individual GW sources. As long as a way can be found to summarize economically the many static parameters associated with the background sources, it should be possible for a Kalman filter and nested sampler to operate together to detect the stochastic background by generalizing the model selection procedure in Sections \ref{sec:model_selection} and \ref{sec:detection}. Summarizing the parameter set economically, while respecting the mathematical structure of the Kalman filter, is a subtle challenge, which we postpone to a forthcoming paper \citep{KimpsonP3}. If successful, it will complement the traditional approach of cross-correlating pulsar residuals to uncover the Hellings-Downs curve \citep{Hellings,2023ApJ...951L...8A}. \newline

\section*{Acknowledgements}
This research was supported by the Australian Research Council Centre of Excellence for Gravitational Wave Discovery (OzGrav), grant number CE170100004. The numerical calculations were performed on the OzSTAR supercomputer facility at Swinburne University of Technology. The OzSTAR program receives funding in part from the Astronomy National Collaborative Research Infrastructure Strategy (NCRIS) allocation provided by the Australian Government.

\section*{Data Availability}
No new data were generated or analysed in support of this research.

\bibliographystyle{mnras}
\bibliography{example} 

\appendix

\section{Kalman filter} \label{sec:kalman}
In this appendix, we describe the Kalman filter algorithm used in this paper. General recursion relations for the discrete-time Kalman filter are written down for an arbitrary linear dynamical system in Section \ref{sec_kalman_general}. The mapping onto the specific continuous-time state-space model in Section \ref{sec:model} is written down in Section \ref{sec_kalman_specific}. Separately, in Appendix \ref{sec:techniques} we compare the Kalman filter and Ornstein-Uhlenbeck model with traditional PTA data analysis techniques. 

\subsection{Recursion equations}\label{sec_kalman_general}
The linear Kalman filter operates on temporally discrete, noisy measurements $\boldsymbol{Y}_k$, which are related to a set of unobservable discrete system states $\boldsymbol{X}_k$, via a linear transformation
\begin{equation}
	\boldsymbol{Y}_k = \boldsymbol{H}_k \boldsymbol{X}_k + \boldsymbol{v}_k \ ,\label{eq:kalman1}
\end{equation}
where $\boldsymbol{H}_k$ is the measurement matrix or observation model, $\boldsymbol{v}_k$ is a zero-mean Gaussian measurement noise, $ \boldsymbol{v}_k \sim\mathcal{N}(0,\boldsymbol{R}_k)$ with covariance $\boldsymbol{R}_k$, and the subscript $k$ labels the time-step. The Kalman filter evolves the underlying states according to
\begin{equation}
	\boldsymbol{X}_k = \boldsymbol{F}_k \boldsymbol{X}_{k-1} + \boldsymbol{G}_k \boldsymbol{u}_k + \boldsymbol{w}_k \ , \label{eq:kalman2}
\end{equation}
where $\boldsymbol{F}_k$ is the system dynamics matrix, $\boldsymbol{G}_k$ is the control matrix. $\boldsymbol{u}_k$ is the control vector, and $\boldsymbol{w}_k$ is a zero-mean Gaussian process noise, $\boldsymbol{w}_k \sim \mathcal{N}(0,\boldsymbol{Q}_k)$ with covariance $\boldsymbol{Q}_k$ \newline 

The Kalman filter is a recursive estimator with two distinct stages: a ``predict" stage and an ``update" stage. The predict stage predicts $\hat{\boldsymbol{X}}_{k|k-1}$, the estimate of the state at discrete step $k$, given the state estimates from step $k-1$. Specifically, the predict step proceeds as
\begin{align}
	\hat{\boldsymbol{X}}_{k|k-1} &=  \boldsymbol{F}_k \hat{\boldsymbol{X}}_{k-1|k-1} + \boldsymbol{G}_k \boldsymbol{u}_k \ , \\
		\hat{\boldsymbol{P}}_{k|k-1} &=  \boldsymbol{F}_k \hat{\boldsymbol{P}}_{k-1|k-1} \boldsymbol{F}_k^\intercal + \boldsymbol{Q}_k  \ , \label{eq:KF_predict}
\end{align}
where $\hat{\boldsymbol{P}}_{k|k-1}$ is the covariance of the prediction. Note that the predict stage is independent of the measurements. The measurement $\boldsymbol{Y}_k$ is included to update the prediction during the update stage as follows:
\begin{align}
	\boldsymbol{\epsilon}_{k} &= \boldsymbol{Y}_k - \boldsymbol{H}_k \hat{\boldsymbol{X}}_{k|k-1} \label{eq:innovation} \ , \\
	\boldsymbol{S}_k &= \boldsymbol{H}_k \hat{\boldsymbol{P}}_{k|k-1} \boldsymbol{H}_k^\intercal + \boldsymbol{R}_k \ , \label{eq:KF_inv}\\
	\boldsymbol{K}_k &= \hat{\boldsymbol{P}}_{k|k-1} \boldsymbol{H}_k^\intercal \boldsymbol{S}_k^{-1} \ ,\label{eq:kalman gain} \\
		\hat{\boldsymbol{X}}_{k|k} &=\hat{\boldsymbol{X}}_{k|k-1} +\boldsymbol{K}_k  \boldsymbol{\epsilon}_{k}  \ , \label{eq:kalmangainupdate} \\
			\hat{\boldsymbol{P}}_{k|k} &= \left( \boldsymbol{I} - \boldsymbol{K}_k \boldsymbol{H}_k \right) 	\hat{\boldsymbol{P}}_{k|k-1} \ . \label{eq:kalman_final}
\end{align}
Equation \eqref{eq:kalman gain} defines the Kalman gain $\boldsymbol{K}_k$ which is defined so as to minimise the mean squared error in the state estimate, i.e.  $\boldsymbol{K}_k = \text{argmin} \left \{ \boldsymbol{E}[ (\boldsymbol{X}_k - \hat{\boldsymbol{X}}_k)^2 ] \right \}$. For full reviews of the Kalman filter, including its derivation, we refer the reader to \cite{Gelb:1974} and \cite{zarchan2000fundamentals}. \newline 

To apply the Kalman filter in practice means specifying the eight component matrices that make up the ``machinery'' of the filter: $\boldsymbol{X}_k$, $\boldsymbol{Y}_k$, $\boldsymbol{F}_k$, $\boldsymbol{G}_k$, $\boldsymbol{u}_k$, $\boldsymbol{H}_k$, $\boldsymbol{Q}_k$ and $\boldsymbol{R}_k$. In Section \ref{sec_kalman_specific} we describe how the machinery is defined for the state-space model in Section \ref{sec:model}.

\subsection{State-space representation of a PTA analysis}\label{sec_kalman_specific}
We apply the Kalman recursion relations in Section \ref{sec_kalman_general} to the state-space model of a PTA with $N$ pulsars described in Section \ref{sec:model} as follows. \newline

We identify $\boldsymbol{X}(t)$ with a vector of length $N$ composed of the intrinsic pulsar frequency states, i.e. 
\begin{equation}
	\boldsymbol{X}(t) = \left(f_{\rm p}^{(1)}(t), f_{\rm p}^{(2)}(t), ..., f_{\rm p}^{(N)}(t)\right) \ .
\end{equation}
Analogously,  we package the measured pulsar frequencies as
\begin{equation}
	\boldsymbol{Y}(t) = \left(f_{\rm m}^{(1)}(t), f_{\rm m}^{(2)}(t), ..., f_{\rm m}^{(N)}(t) \right) \ .
\end{equation}
The states evolve according to the continuous stochastic differential equation (cf. Equation \eqref{eq:frequency_evolution})
\begin{equation}
	d \boldsymbol{X} = \boldsymbol{A} \boldsymbol{X} dt + \boldsymbol{C}(t) dt + \boldsymbol{\Sigma} d \boldsymbol{B}(t) \ , \label{eq:kalmn2}
\end{equation}
where $\boldsymbol{A}$ is a diagonal $N \times N$ matrix,
\begin{equation}
	\boldsymbol{A} = \text{diag} \left(-\gamma^{(1)}, -\gamma^{(2)}, ..., -\gamma^{(N)}\right) \ ,
\end{equation}
and $\boldsymbol{C}(t)$ is a time-dependent vector with $n$-th component
\begin{equation}
	C^{(n)} =\gamma^{(n)} \left[ f^{(n)} _{\rm em} (t_1) + \dot{f}^{(n)} _{\rm em}(t_1) \, t \right] +  \dot{f}_{\rm em}(t_1)^{(n)} \ .
\end{equation}
The $N \times N$ square matrix $\boldsymbol{\Sigma}$  governs the magnitude of the increments of Brownian motion (Wiener process) $d\boldsymbol{B}(t)$, with
\begin{equation}
	\boldsymbol{\Sigma} = \text{diag} \left(\sigma^{(1)}, \sigma^{(2)}, ..., \sigma^{(N)}\right) \ .
\end{equation}

In the idealized model above, each pulsar's rotational state evolves phenomenologically according to a mean-reverting Ornstein-Uhlenbeck process, described by a Langevin equation, Equation \eqref{eq:kalmn2}, whose general solution is given by \citep{gardiner2009stochastic}
\begin{equation}
	\boldsymbol{X}(t) = e^{\boldsymbol{A} t} \boldsymbol{X}(0) + \int_0^t e^{\boldsymbol{A}(t-t')} \boldsymbol{C}(t') dt' + \int_0^t e^{\boldsymbol{A}(t-t')} \boldsymbol{\Sigma} d\boldsymbol{B}(t') \ . \label{eq:gardenier}
\end{equation} 
From Equation \eqref{eq:gardenier} we construct the discrete, recursive solution for $\boldsymbol{X}(t_k) = \boldsymbol{X}_k$ in the form of Equation \eqref{eq:kalman2}, with
\begin{align}
	\boldsymbol{F}_k &= e^{\boldsymbol{A} \Delta t } \  \\
	&= \text{diag}\left(e^{- \gamma^{(1)} \Delta t},e^{- \gamma^{(2)} \Delta t},...,e^{- \gamma^{(N)} \Delta t} \right) \ ,
\end{align}
\begin{align}
	\boldsymbol{G}_k \boldsymbol{u}_k &= \int_{t_k}^{t_{k+1}}  e^{\boldsymbol{A}\left( t_{k+1} - t' \right)}  \boldsymbol{C}(t') dt' \ , \\
	&= \left(G^{(1)}_k, G^{(2)}_k,...,G^{(N)}_k \right) ,
\end{align}
\begin{equation}
	\boldsymbol{w}_k = \int_{t_k}^{t_{k+1}} e^{\boldsymbol{A}\left( t_{k+1} - t' \right)} \boldsymbol{\Sigma} d \boldsymbol{B}(t') \ ,  \label{eq:appendix_noise}
\end{equation}
\begin{align}
	G_k^{(n)} =&    f^{(n)}_{\rm em}(t_1) + \dot{f}^{(n)}_{\rm em}(t_1)  \left(\Delta t + t_k \right) \nonumber \\ 
	&- e^{-\gamma \Delta t} \left[  f^{(n)}_{\rm em}(t_1) +\dot{f}^{(n)}_{\rm em}(t_1)  t_k \right] \ ,
\end{align}
and $\Delta t = t_{k+1} - t_k$. From Equation \eqref{eq:appendix_noise} the process noise covariance matrix is
\begin{align}
	\boldsymbol{Q}_k \boldsymbol{\delta}_{kj} &= \langle \boldsymbol{\eta}_k \boldsymbol{\eta}_j^\intercal \rangle \\
	&= \text{diag} \left(Q^{(1)}, Q^{(2)},...,Q^{(N)}\right) \ ,
\end{align}
with 
\begin{equation}
Q^{(n)} = \frac{[\sigma^{(n)}]^2}{2 \gamma^{(n)}} \left[ 1 - e^{-2 \gamma^{(n)} \Delta t}\right] \ . \label{eq:a_process_noise}
\end{equation}
The Einstein summation convention is suspended temporarily in the left-hand side of Equation \eqref{eq:a_process_noise}. The two remaining unspecified component matrices of the Kalman filter are the measurement matrix $\boldsymbol{H}_k$ and the measurement covariance matrix $\boldsymbol{R}_k$. These are defined straightforwardly from Equations \eqref{eq:measurement}--\eqref{eq:z_trigonometric}. Specifically, 
$\boldsymbol{H}_k$ is a diagonal matrix where the $n$-th component of the diagonal is given by $g^{(n)}(t_k)$ from Equation \eqref{eq:measurement}. The measurement covariance satisfies $\boldsymbol{R}_k = E \left[ \boldsymbol{v} \boldsymbol{v}^\intercal \right] = \sigma^2_{\rm m}$ for all $k$.

\section{Comparison with traditional PTA analyses}\label{sec:techniques}
In this appendix we compare the state-space formulation of PTA data analysis described in this paper with traditional formulations. In Appendix \ref{appendix:brownian}  we focus on one key difference: the asymptotic behaviour of the adaptive gain in the Kalman filter compared to traditional, least-squares estimators. The adaptive gain controls the fraction of new information that is incorporated into the updated state estimate at every time step and hence controls how nimbly the tracking scheme responds to new data. We illustrate the effect of the adaptive gain on a simple Wiener process as a pedagogical example and then explain how it applies analogously to the PTA problem. In Appendix \ref{appendix:OU} we compare the Ornstein-Uhlenbeck description of the process noise intrinsic to the pulsar, specified by Equations \eqref{eq:frequency_evolution}--\eqref{eq:xieqn}, with the traditional approach in terms of modelling red noise as a Gaussian process with a (broken) power-law power spectral density. We show that the two descriptions are equivalent in certain limits.

\subsection{Adaptive gain}\label{appendix:brownian}
Many traditional PTA analyses fit timing data to a phase model by least-squares estimation. The state-space scheme in this paper achieves the same goal (for a pulse frequency model rather than a phase model, strictly speaking) using a Kalman filter. It is natural to ask how the two approaches differ, if at all. \newline 

To understand the difference from first principles, consider as a simplified pedagogical example the problem of estimating an unobservable state $X$ using a measurement $Y$, where $X$ is generated by a Wiener process, viz.
\begin{align}
	\dot{X} &= w \, , \label{eq:ped1} \\
	Y &= X + v \, ,\label{eq:ped2}
\end{align}
with $w\sim \mathcal{N} \left(0,Q\right)$ and  $v \sim \mathcal{N}\left(0,R\right)$, and where an overdot denotes a derivative with respect to time. We now compare the least-squares estimator and the Kalman filter for the simplified system in Equations \eqref{eq:ped1} and \eqref{eq:ped2} in two cases: (i) $Q=0$ (i.e. the state is constant with no process noise) and (ii) $Q \neq 0$. In order to compare directly the Kalman filter (which is recursive) and the least-squares estimator, we reformulate the batch least-squares regression as a recursive least-squares filter. Batch and recursive least squares are equivalent; both methods minimise the sum of the squared errors and produce identical estimates of the state, $\hat{X}$. \newline 

First, consider $Q=0$. In recursive least-squares the unbiased minimum variate estimate of the state at time-step $k$ is
\begin{equation}
	\hat{X}_{k, \rm RLS} = \frac{1}{k} \sum_{i=1}^{k} Y_k \, .
\end{equation}
When a new data point is ingested at timestep $k+1$, the estimate of the state is updated according to
\begin{align}
		\hat{X}_{k+1, \rm RLS} &= \frac{1}{k+1} \sum_{i=1}^{k+1} Y_k  \\
		&= \hat{X}_{k, \rm RLS} + \frac{1}{k+1} \left(Y_{k+1} -\hat{X}_{k, \rm RLS}\right) \label{eq:RLS_estiamte} \, .
\end{align}
The prefactor $\left(k+1\right)^{-1}$ in Equation \eqref{eq:RLS_estiamte} is the recursive least-squares gain (cf. Equation \eqref{eq:kalman gain}) and the quantity $Y_{k+1} -\hat{X}_{k, \rm RLS}$ is the innovation (cf. Equation \eqref{eq:innovation}). As the number of data points increases the gain tends to zero; in recursive least-squares the additional datum $Y_{k+1}$ becomes less and less important, as $k$ increases and more and more observations are obtained.  As the gain tends to zero, the new observations are asymptotically down-weighted. \newline 

Now apply Kalman filtering to $Q=0$ instead. The state estimate is updated in response to new data according to 
\begin{equation}
	\hat{X}_{k+1, \rm KF} = 	\hat{X}_{k, \rm KF} + \frac{P_0}{R + k P_0} \left(Y_{k+1} -\hat{X}_{k, \rm KF}\right) \label{eq:KF_estimate} \, ,
\end{equation}
where $P_0$ is the initial covariance in $\hat{X}$. In Equation \eqref{eq:KF_estimate}, the quantity $P_0 / (R + k P_0)$ is the Kalman gain. As with recursive least-squares, the gain tends to zero as $k$ increases. Hence, in the special case $Q=0$, the recursive least-squares and Kalman filtering behave equivalently in the long term. \newline 

Now consider $Q \neq 0$. The recursive least-squares estimate of the state proceeds as before via Equation \eqref{eq:RLS_estiamte}. What about the Kalman filter? For the system described by Equations \eqref{eq:ped1} and \eqref{eq:ped2}, the continuous-time Riccati equation for the propagation of the error covariance $P(t)$ is \citep{lewis2017optimal}
\begin{equation}
\dot{P} = Q- \frac{P(t)^2}{R} \, . \label{eq:ricatti}
\end{equation}
The continuous-time Kalman gain is
\begin{equation}
K(t) = \frac{P(t)}{R} \, . \label{eq:kt_cont}
\end{equation}
The solution to Equation \eqref{eq:ricatti} is
\begin{equation}
P(t) = \sqrt{RQ} \left[ \frac{P_0 \cosh \left(t\sqrt{Q/R} \right) +  \sqrt{QR} \sinh \left(t \sqrt{Q/R} \right)}{P_0 \sinh \left(t \sqrt{Q/R} \right) +  \sqrt{QR} \cosh \left(t \sqrt{Q/R} \right)}\right] \, . \label{eq:pt_cont}
\end{equation}
From Equation \eqref{eq:kt_cont} and Equation \eqref{eq:pt_cont} we find $K(t) \rightarrow \sqrt{Q/R} > 0$ as $t$ increases (equivalent to increasing $k$ in the discrete case). That is, the gain tends to a positive definite value, unlike for recursive least squares. Hence the Kalman filter is more responsive asymptotically to additional data than a least-squares estimator. Informally speaking, it is a more nimble tracker, which explains its preferred status in many electrical and mechanical engineering applications \citep{Gelb:1974,zarchan2000fundamentals,byrne2005signal,sarka2013bayesian}. \newline

PTA data analysis is far more complicated than the pedagogical example described by Equations \eqref{eq:ped1} and \eqref{eq:ped2}. Fundamentally, though, the timing noise tracking step can be understood in the same way, as a state estimation problem. In this paper, state estimation is performed with a Kalman filter. In traditional analyses, it is performed through least-squares estimation using timing software such as \textsc{TEMPO2} \citep{tempo2} or \textsc{PINT} \citep{2021ApJ...911...45L}. Specifically, traditional least-squares analyses fit a model of the TOA residuals vector $\delta \boldsymbol{t}$ of the form 
\begin{equation}
	\delta \boldsymbol{t} = \boldsymbol{M} \boldsymbol{\epsilon} + \boldsymbol{\mathcal{F}} \boldsymbol{a} + \boldsymbol{n}
\end{equation}
where $\boldsymbol{M} \boldsymbol{\epsilon}$ represents the deterministic deviation from the least-squares fit to a Taylor-series phase model, $\boldsymbol{\mathcal{F}} \boldsymbol{a}$ represents the stochastic red noise modelled as a sum of sine and cosine Fourier modes with amplitudes drawn from a PSD (usually of power-law form), and $\boldsymbol{n}$ represents the stochastic white noise component, with covariance matrix $\boldsymbol{N}$ \citep{2021arXiv210513270T,2023arXiv230616223J}. Minimizing the squared error is equivalent to maximising the associated likelihood
\begin{equation}
	p(\boldsymbol{r} | \boldsymbol{\epsilon}, \boldsymbol{a}) = \frac{1}{\sqrt{\det \left(2 \pi \boldsymbol{N}\right)}} \exp \left(-\frac{1}{2} \boldsymbol{r}^{\rm T} \boldsymbol{N} \boldsymbol{r}\right) \, ,
\end{equation}
with $\boldsymbol{r} = \delta \boldsymbol{t} -  \boldsymbol{M} \boldsymbol{\epsilon} -\boldsymbol{\mathcal{F}} \boldsymbol{a}$. The above formulation is analogous to the $Q\neq 0$ case analysed above, and the conclusion is the same: the least-squares gain tends to zero asymptotically, whereas the Kalman gain does not. This represents a difference between traditional PTA analyses and the approach in this paper. We refer the reader to \cite{2021arXiv210513270T} for additional details on traditional PTA data analysis methods.

\subsection{Red noise power spectral density}\label{appendix:OU}
In Section \ref{sec:model} the intrinsic achromatic spin wandering of the pulsar, $f_{\rm p}^{(n)}(t)$, is modelled as an Ornstein-Uhlenbeck process governed by Equations \eqref{eq:spinevol}--\eqref{eq:xieqn}. The Ornstein-Uhlenbeck model captures the main qualitative features of a typical PTA pulsar, namely a deterministic secular spin down perturbed by stochastic, small-amplitude, mean-reverting fluctuations. \newline 

In traditional PTA analyses, the red-spectrum timing noise fluctuations are described as a zero-mean Gaussian random process and modelled via a finite decomposition into a Fourier basis (see Appendix \ref{appendix:brownian}). The Fourier coefficients are determined by the power spectral density of the TOA residuals, which is assumed to take the standard form \citep[e.g.][]{2021arXiv210513270T,Goncharov2021},
\begin{equation}
	\rho(f) = \frac{A_{\rm a}^2}{12 \pi^2} \frac{1}{T_{\rm obs}} \left(\frac{f}{1 \text{yr}^{-1}}\right)^{-\gamma} \text{yr}^2 \, , \label{eq:PSD}
\end{equation}
where $f$ denotes the Fourier frequency. The power spectral density is defined by two hyper-parameters, the amplitude $A_{\rm a}$ and the exponent $\gamma$. Traditional PTA analyses seek to estimate the values of the hyper-parameters, rather than the Fourier coefficients themselves, which are drawn randomly by treating Equation \eqref{eq:PSD} as a probability density function. \newline 

Equations \eqref{eq:spinevol}--\eqref{eq:xieqn} are equivalent to specifying a particular form of power spectral density, which is related to but different from Equation \eqref{eq:PSD}. Specifically, Equations \eqref{eq:spinevol}--\eqref{eq:xieqn} lead to a broken power law with $\rho(f) \propto f^{-4}$ for $f \gtrsim \gamma^{(n)}$ and $\rho(f) \propto f^{-2}$ for $f\lesssim \gamma^{(n)}$ approximately. The scalings are obtained by Fourier transforming Equations \eqref{eq:spinevol}--\eqref{eq:xieqn}, applying the Wiener-Khintchine theorem, and converting frequency residuals to phase residuals \citep{Myers2021MNRAS.502.3113M,Meyers2021,2023MNRAS.520.2813A,2024MNRAS.530.4648O} \footnote{The Wiener-Khintchine theorem assumes that the timing noise statistics are stationary, which may not be true in all pulsars. Nonstationarity is a subtle topic, which is deferred to future work.}. The specific form of Equations \eqref{eq:spinevol}--\eqref{eq:xieqn} is one particular choice of stochastic model for pulsar timing noise, whose PSD $\rho(f)$ is broadly consistent with observations for several millisecond pulsars, especially over typical inter-TOA intervals $T_{\rm cad} = t_{k+1} - t_k \sim 1 \, {\rm week}$. Alternative models exist, which include higher derivatives of $f_{\rm p}^{(n)}(t$ \citep{Vargas} or multiple internal stellar components \citep{Myers2021MNRAS.502.3113M,Meyers2021,2023MNRAS.520.2813A,2024MNRAS.530.4648O}. The state-space framework presented in this paper extends straightforwardly to other stochastic models of pulsar spin wandering. Analogously, in traditional PTA analyses, whilst Equation \eqref{eq:PSD} is a default description, multiple alternative parameterised models also exist \citep{Sesana10,2013PhRvD..87j4021L,2015PhRvD..91h4055S,2017MNRAS.468..404C,2017MNRAS.470.1738C,2017PhRvL.118r1102T,2019ApJ...880..116A,2019MNRAS.488..401C}. We refer the interested reader to \cite{2021arXiv210513270T} for additional details on alternative power spectral density models. 


\section{Synthetic PTA specifications}\label{appendix_fiducial}
The synthetic PTA deployed for testing in this paper is constructed to mimic the $N=47$ pulsars from the 12.5-year NANOGrav data release, whose sky positions are plotted in Figure \ref{fig:pulsar_distrib}. The construction recipe is described in full in Section \ref{sec:synt_pta}. In this appendix, we record a complete list of the pulsar parameters in Table \ref{tab:pulsar_values}, to assist the interested reader in implementing the analysis scheme and reproducing the test results in Sections \ref{sec:rep_example}--\ref{sec:bias_and_identifiability}.

\begin{table*}
\begin{tabular}{lccccc}
	\toprule
	JNAME &    $f_{\rm em} (t_1)$ (Hz) &  $\dot{f}^{(n)}_{\rm em}(t_1)\times 10^{16}$ (s$^{-2}$) &  $\alpha$ (deg) &  $\delta$ (deg) &  $d$ (kpc) \\
	\midrule
	J0023+0923 & 327.8 &      -12.3 &   5.8 &    9.4 &   1.8 \\
	J0030+0451 & 205.5 &       -4.3 &   7.6 &    4.9 &   0.3 \\
	J0340+4130 & 303.1 &       -6.5 &  55.1 &   41.5 &   1.6 \\
	J0613-0200 & 326.6 &      -10.2 &  93.4 &   -2.0 &   0.9 \\
	J0636+5128 & 348.6 &       -4.2 &  99.0 &   51.5 &   0.7 \\
	J0645+5158 & 112.9 &       -0.6 & 101.5 &   52.0 &   1.2 \\
	J0740+6620 & 346.5 &      -14.6 & 115.2 &   66.3 &   1.1 \\
	J0931-1902 & 215.6 &       -1.7 & 142.8 &  -19.0 &   3.7 \\
	J1012+5307 & 190.3 &       -6.2 & 153.1 &   53.1 &   0.7 \\
	J1024-0719 & 193.7 &       -7.0 & 156.2 &   -7.3 &   1.2 \\
	J1125+7819 & 238.0 &       -3.9 & 171.5 &   78.3 &   0.9 \\
	J1453+1902 & 172.6 &       -3.5 & 223.4 &   19.0 &   1.3 \\
	J1455-3330 & 125.2 &       -3.8 & 223.9 &  -33.5 &   0.7 \\
	J1600-3053 & 277.9 &       -7.3 & 240.2 &  -30.9 &   1.9 \\
	J1614-2230 & 317.4 &       -9.7 & 243.7 &  -22.5 &   0.7 \\
	J1640+2224 & 316.1 &       -2.8 & 250.1 &   22.4 &   1.5 \\
	J1643-1224 & 216.4 &       -8.6 & 250.9 &  -12.4 &   0.7 \\
	J1713+0747 & 218.8 &       -4.1 & 258.5 &    7.8 &   1.3 \\
	J1738+0333 & 170.9 &       -7.0 & 264.7 &    3.6 &   1.5 \\
	J1741+1351 & 266.9 &      -21.5 & 265.4 &   13.9 &   1.7 \\
	J1744-1134 & 245.4 &       -5.4 & 266.1 &  -11.6 &   0.4 \\
	J1747-4036 & 607.7 &      -48.5 & 267.0 &  -40.6 &   7.1 \\
	J1832-0836 & 367.8 &      -11.2 & 278.1 &   -8.6 &   2.1 \\
	J1853+1303 & 244.4 &       -5.2 & 283.5 &   13.1 &   2.1 \\
	J1857+0943 & 186.5 &       -6.2 & 284.4 &    9.7 &   1.2 \\
	J1903+0327 & 465.1 &      -40.7 & 285.8 &    3.5 &   7.0 \\
	J1909-3744 & 339.3 &      -16.1 & 287.4 &  -37.7 &   1.1 \\
	J1910+1256 & 200.7 &       -3.9 & 287.5 &   12.9 &   1.5 \\
	J1911+1347 & 216.2 &       -7.9 & 288.0 &   13.8 &   1.4 \\
	J1918-0642 & 130.8 &       -4.4 & 289.7 &   -6.7 &   1.1 \\
	J1923+2515 & 264.0 &       -6.7 & 290.8 &   25.3 &   1.2 \\
	J1939+2134 & 641.9 &     -433.1 & 294.9 &   21.6 &   3.5 \\
	J1944+0907 & 192.9 &       -6.4 & 296.0 &    9.1 &   1.2 \\
	J1946+3417 & 315.4 &       -3.1 & 296.6 &   34.3 &   6.9 \\
	J1955+2908 & 163.0 &       -7.9 & 298.9 &   29.1 &   6.3 \\
	J2010-1323 & 191.5 &       -1.8 & 302.7 &  -13.4 &   2.4 \\
	J2017+0603 & 345.3 &       -9.5 & 304.3 &    6.1 &   1.4 \\
	J2033+1734 & 168.1 &       -3.1 & 308.4 &   17.6 &   1.7 \\
	J2043+1711 & 420.2 &       -9.3 & 310.8 &   17.2 &   1.4 \\
	J2145-0750 &  62.3 &       -1.2 & 326.5 &   -7.8 &   0.7 \\
	J2214+3000 & 320.6 &      -15.1 & 333.7 &   30.0 &   0.6 \\
	J2229+2643 & 335.8 &       -1.7 & 337.5 &   26.7 &   1.8 \\
	J2234+0611 & 279.6 &       -9.4 & 338.6 &    6.2 &   1.0 \\
	J2234+0944 & 275.7 &      -15.3 & 338.7 &    9.7 &   1.6 \\
	J2302+4442 & 192.6 &       -5.1 & 345.7 &   44.7 &   0.9 \\
	J2317+1439 & 290.3 &       -2.0 & 349.3 &   14.7 &   1.7 \\
	J2322+2057 & 208.0 &       -4.2 & 350.6 &   21.0 &   1.0 \\
	\bottomrule
\end{tabular}
\caption{List of the fiducial pulsar parameters for the 47 pulsars used to construct the synthetic PTA in Section \ref{sec:synt_pta}. The right ascension and declination of an individual pulsar in J2000 coordinates are labelled by $\alpha$ and $\delta$ respectively. Parameter values are obtained from the ATNF pulsar catalogue \citep{Manchester2005} using the \texttt{psrqpy} package \citep{psrqpy}.}
\label{tab:pulsar_values}
\end{table*}

\section{Dispersion of ${\theta}_{\rm gw}$ estimates}\label{sec:wasserstein}
Every random realization of the noise processes $\xi^{(n)}(t)$ and $\varepsilon^{(n)}(t)$ leads to a different ${\boldsymbol{\theta}}_{\rm gw}$ posterior, when the synthetic data are analysed according to the procedure in Section \ref{sec:methodsummary}. The distance between two posteriors (i.e.\ how similar they are) can be measured by many valid metrics, including those related to the Kolmogorov-Smirnoff test \citep{corder2014nonparametric}. In this paper, we use the WD \citep{Wasserstein,Villani2009} which is popular in machine learning \citep[e.g.][]{2017arXiv170107875A} and other domains. In Section \ref{sec:appendix_overview_WD} we define the WD and summarize its main properties. In Section \ref{sec:appendix_WD_results} we present for reproducibility the WD calculated between every pair of posteriors for each static parameter in $\boldsymbol{\theta}_{\rm gw}$ across the $10^3$ noise realisations of Section  \ref{sec:multiple_noise}. The results are summarised via the median values reported in Table \ref{tab:Wasserstein}.

\subsection{Overview of the WD}\label{sec:appendix_overview_WD}
The WD is a metric that defines a distance between two probability distributions $\mu(x)$ and $\nu(x)$. It has an intuitive interpretation as the lowest total cost with which one can move probability mass from $\mu$ to $\nu$, with respect to a cost function $c(x,y)$. For this reason it is sometimes known as the ``Earth mover's distance''. The $p$-th order Wasserstein distance between two distributions is
\begin{eqnarray}
	W_p(\mu,\nu)= \left[ \inf_{\gamma \in \Gamma(\mu, \nu)}  \int c(x,y)^p d \lambda (x,y)\right]^{1/p} \label{eq:wasserstein} \ ,
\end{eqnarray}
for $p \in [1,\infty)$, where $\lambda(x,y)$ is the transport plan, and $\Gamma(\mu, \nu)$ is the set of all joint probability distributions for $(x,y)$ that have marginals $\mu$ and $\nu$, i.e. $\Gamma(\mu, \nu)$ is the set of couplings of $\mu$ and $\nu$. \newline 

The cost function can be freely chosen to suit the nature of the problem. Often, as in this paper, it is taken to be the absolute value function, $c(x,y) = |x-y|$. In general, $W_p(\mu,\nu)$ with $c(x,y)=|x-y|$ can be computed from $n$ samples by the Hungarian algorithm \citep{Kuhn} in polynomial time $\mathcal{O}(n^3)$ \citep{Villani2009}. However, for $\mu$ and $\nu$ defined on $\mathbb{R}^d$ it is well-known \citep{Dudley} that $E[W_p(\mu,\nu)]$ converges slowly $\propto n^{-1/d}$. In this paper, we calculate the WD exclusively between the one-dimensional marginalized posteriors, setting $d=1$ and obtaining
\begin{eqnarray}
	W_p(\mu,\nu)= \left\{\int_0^1 dz \left[ F_{\mu}^{-1} (z) - F_{\nu}^{-1} (z) \right] \right\}^{1/p} \ , \label{eq:wasserstein}
\end{eqnarray}
where $F_{\mu}^{-1}(z)$ is the inverse cumulative distribution function of $\mu$. \newline

The WD holds certain advantages over valid alternatives like the Kullback–Leibler divergence or the  Kolmogorov-Smirnoff test \citep{gelman2013bayesian,corder2014nonparametric}. It is intuitive, being the minimum cost required to transform one distribution into another. It satisfies the Monge-Kantorovich duality \citep{villani2003topics,Villani2009},
\begin{eqnarray}
	| \boldsymbol{E}(X_{\mu} )-\boldsymbol{E}(Y_{\nu} ) | \leq W_1(\mu, \nu) \ , \label{eq:WDdefn}
\end{eqnarray}
where $X_{\mu}$ and $Y_{\nu}$ are random variates drawn from the distributions $\mu$ and $\nu$ respectively, and $\boldsymbol{E}$ denotes the expected value. It also satisfies desirable properties of a measure of distance, such as symmetry and the triangle inequality. For $p=1$, the WD inherits the units of the underlying distributions. Finally, the WD is versatile; any two distributions can be compared, irrespective of whether they are continuous, discrete, or singular. \newline

\subsection{WD for $\theta_{\rm gw}$} \label{sec:appendix_WD_results}

\begin{figure*}
	\setkeys{Gin}{width=\linewidth}   
	
	\begin{subfigure}[b]{0.22\textwidth}
		\includegraphics[width=\textwidth]{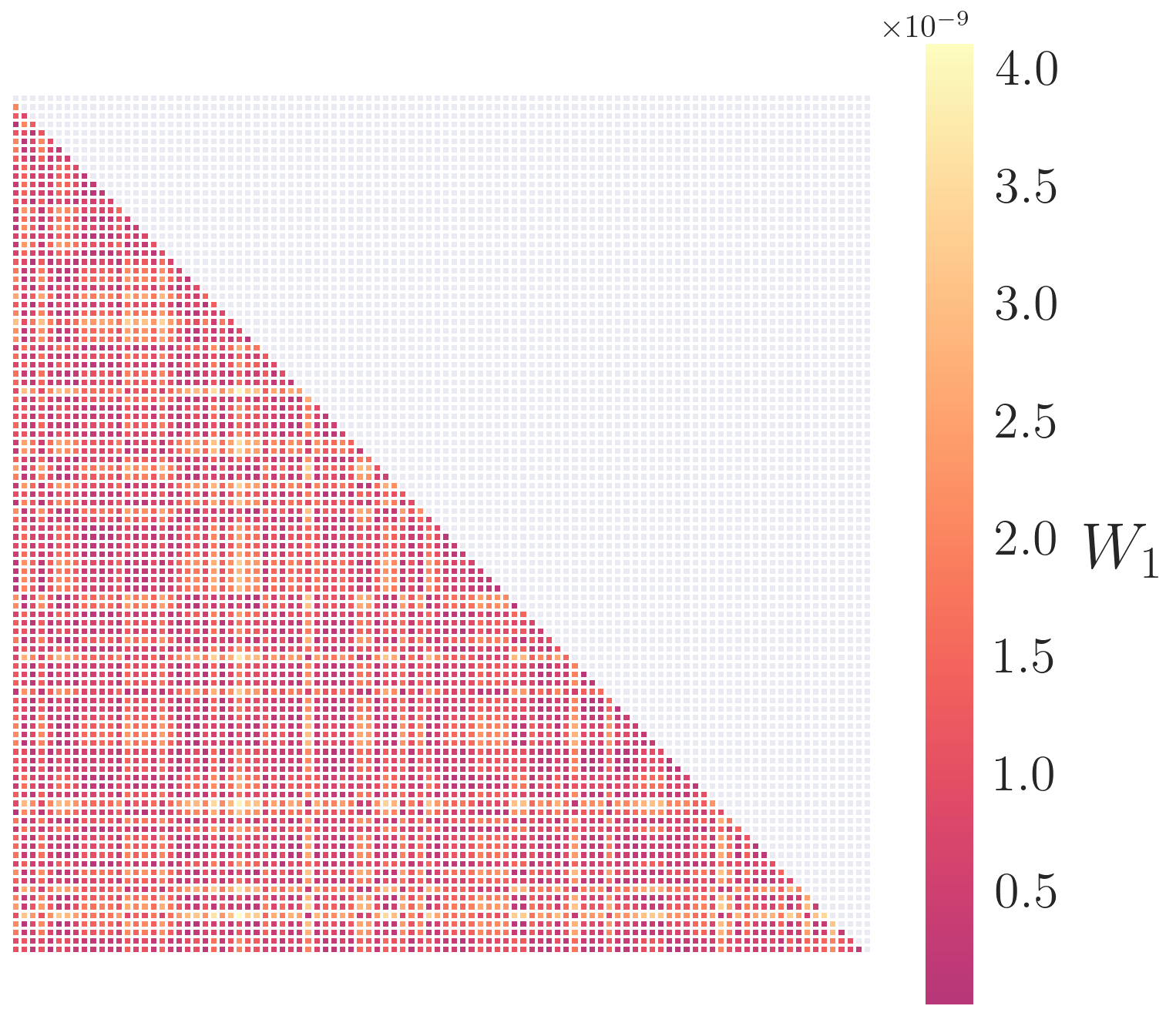}
		\caption{$\Omega$}
	\end{subfigure}
	\hfill
	\begin{subfigure}[b]{0.22\textwidth}
		\includegraphics[width=\textwidth]{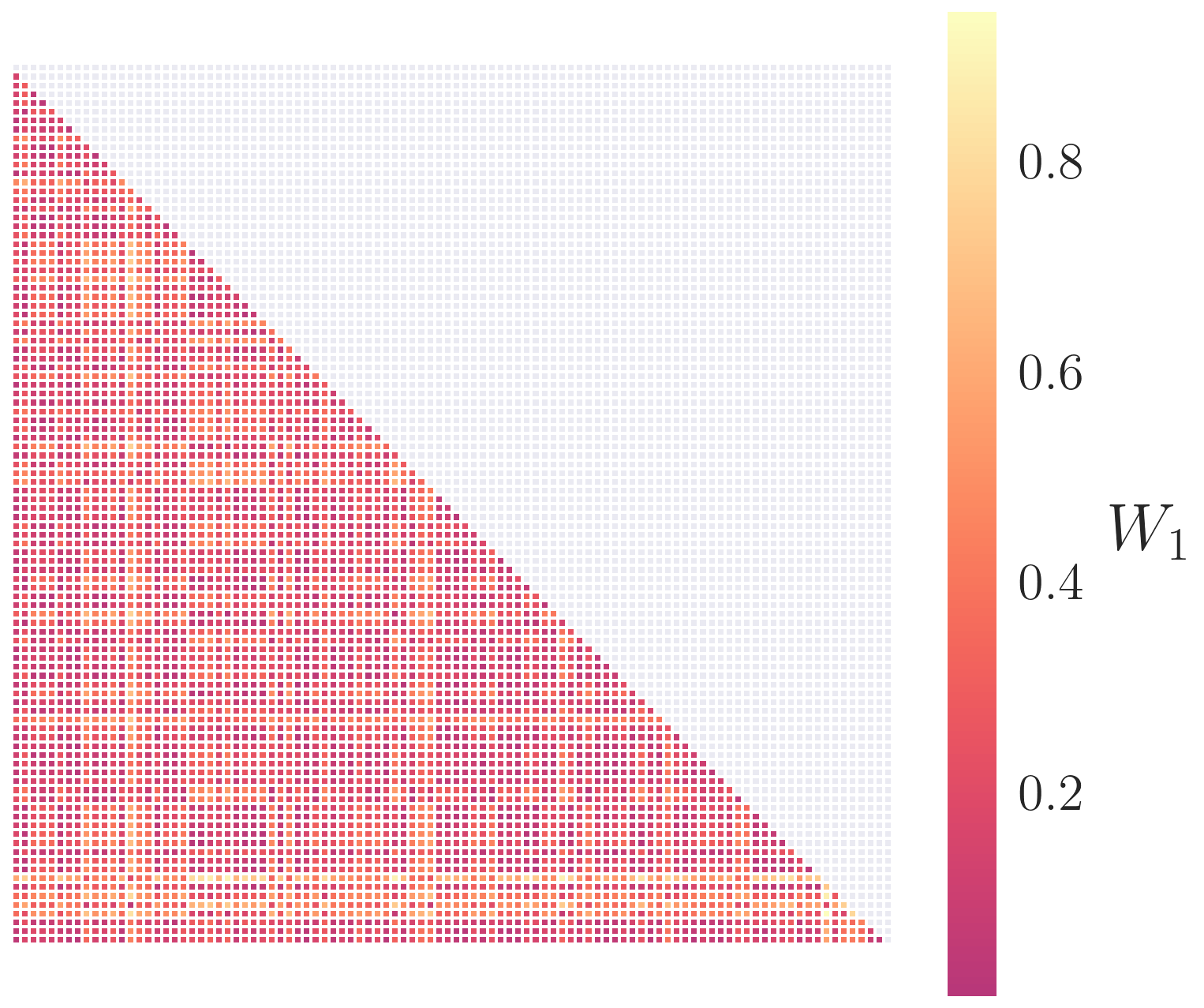}
		\caption{$\Phi_0$}
	\end{subfigure}
	\hfill	
	\begin{subfigure}[b]{0.22\textwidth}
		\includegraphics[width=\textwidth]{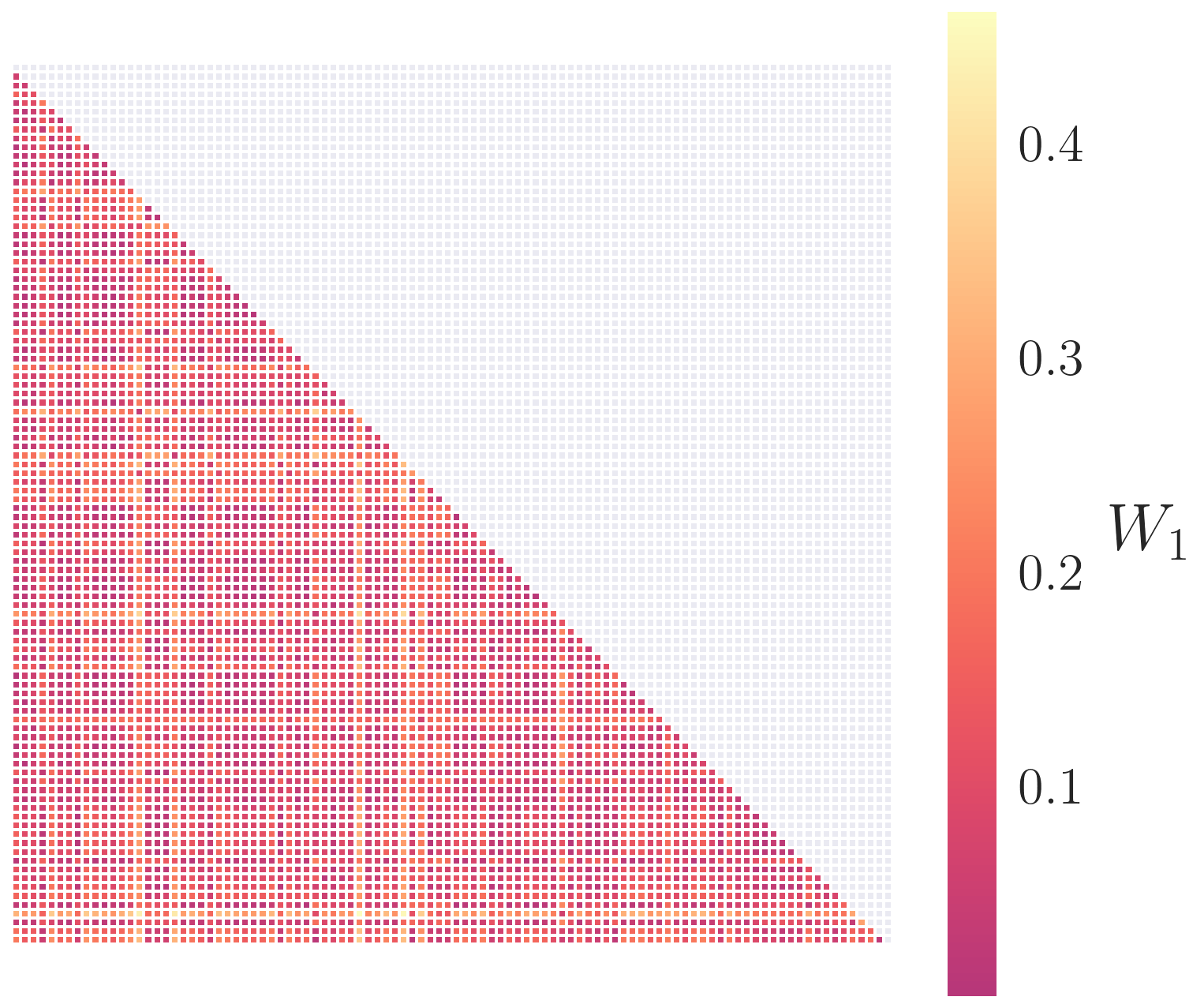}
		\caption{$\psi$} \label{figWD_psi}
	\end{subfigure}
	\hfill
	\begin{subfigure}[b]{0.22\textwidth}
		\includegraphics[width=\textwidth]{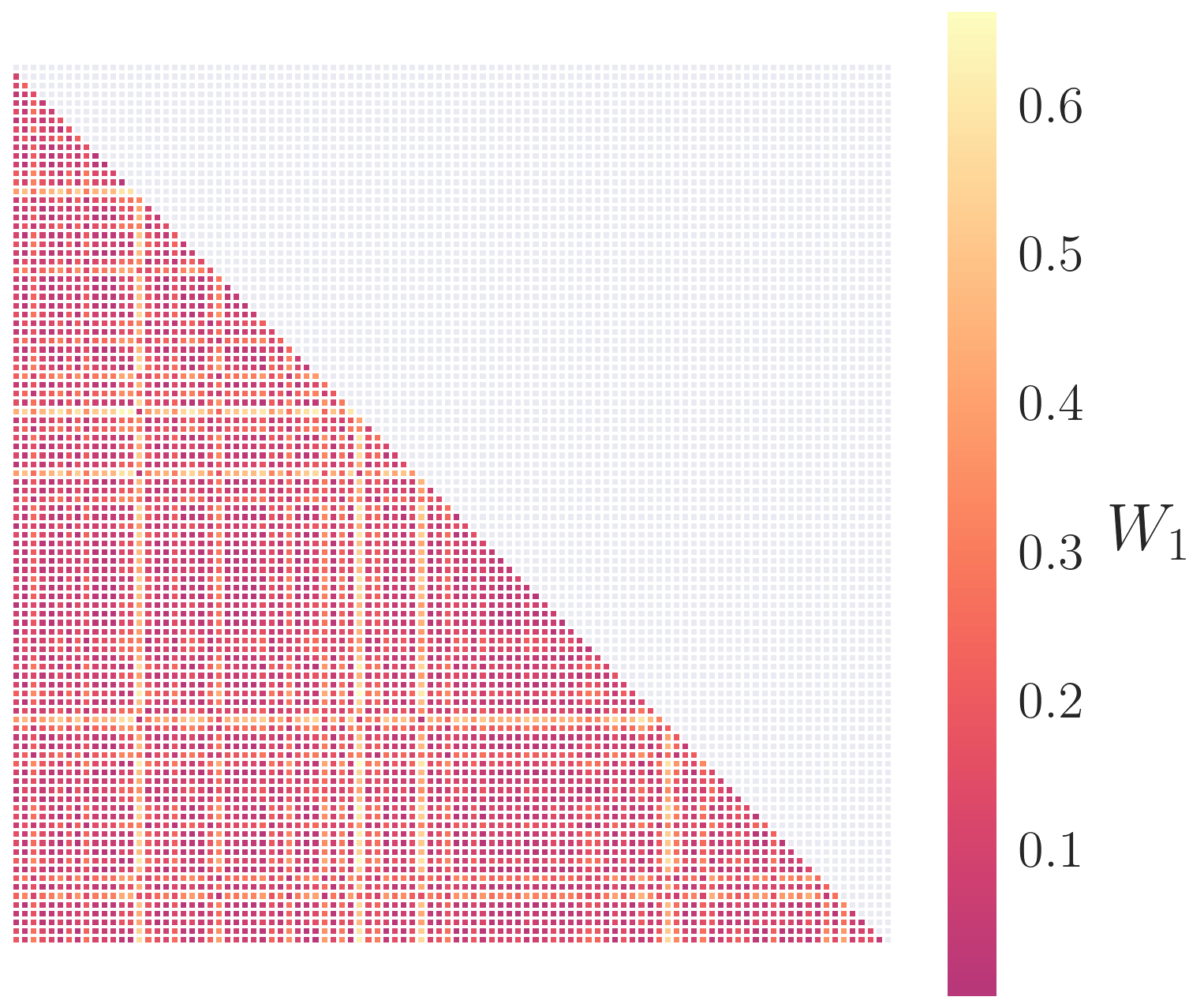}
		\caption{$\iota$}
		\label{subfig:iota}
	\end{subfigure}
	\medskip
	\begin{subfigure}[b]{0.3\textwidth}
		\includegraphics[width=\textwidth]{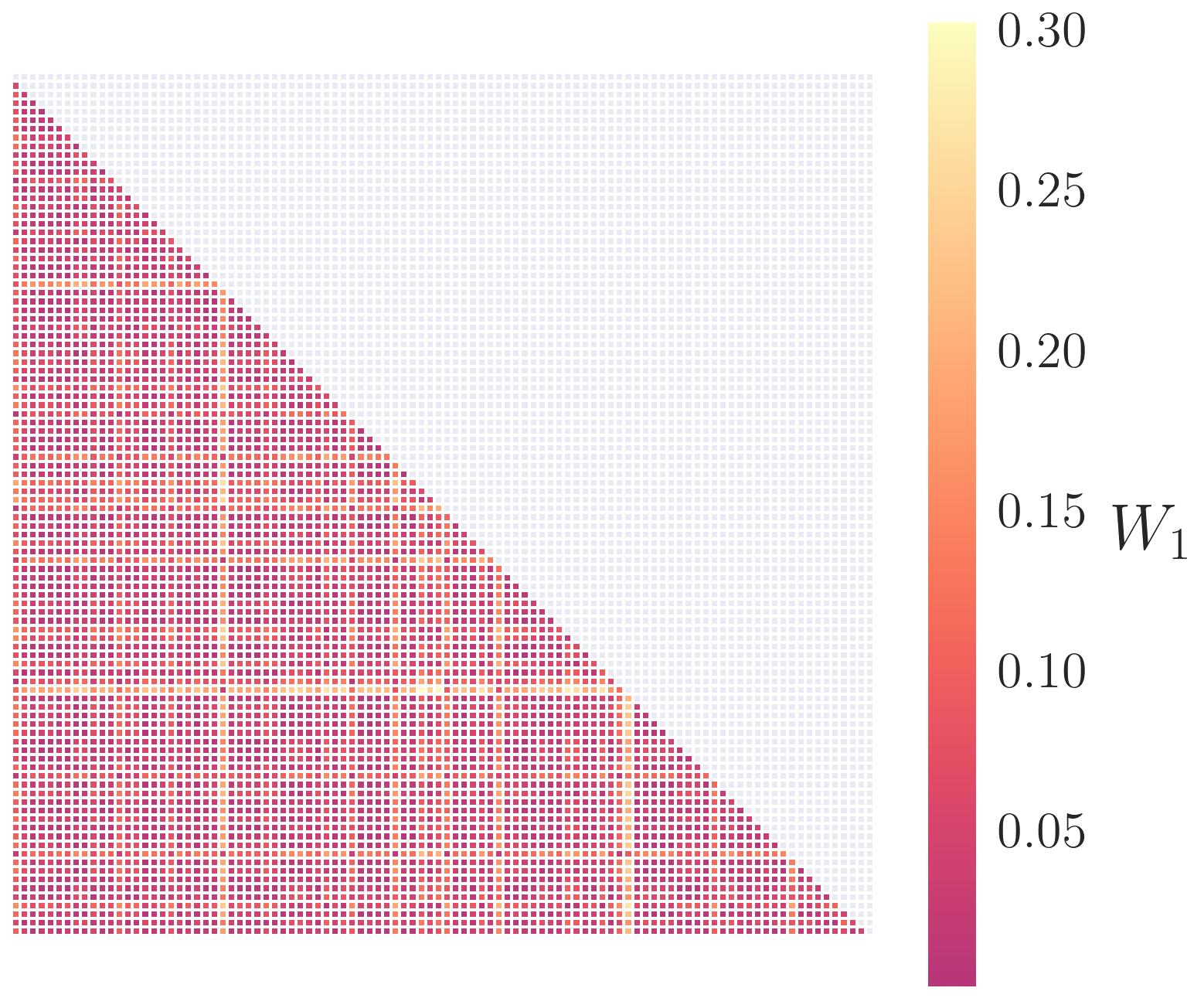}
		\caption{$\delta$}
		\label{subfig:delta}
	\end{subfigure}
	\hfill	
	\begin{subfigure}[b]{0.3\textwidth}
		\includegraphics[width=\textwidth]{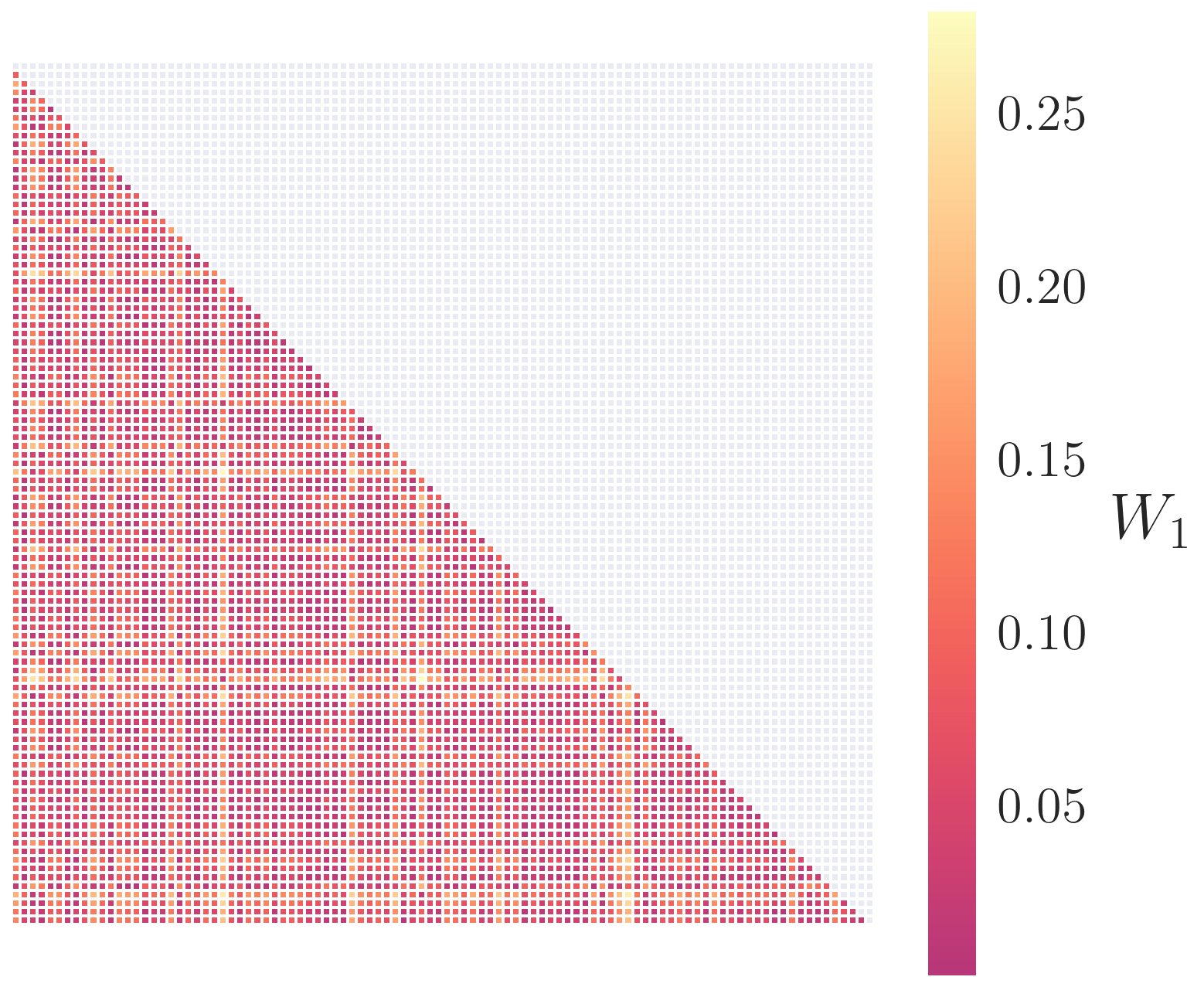}
		\caption{$\alpha$}
		\label{subfig:alpha}
	\end{subfigure}
	\hfill	
	\begin{subfigure}[b]{0.3\textwidth}
		\includegraphics[width=\textwidth]{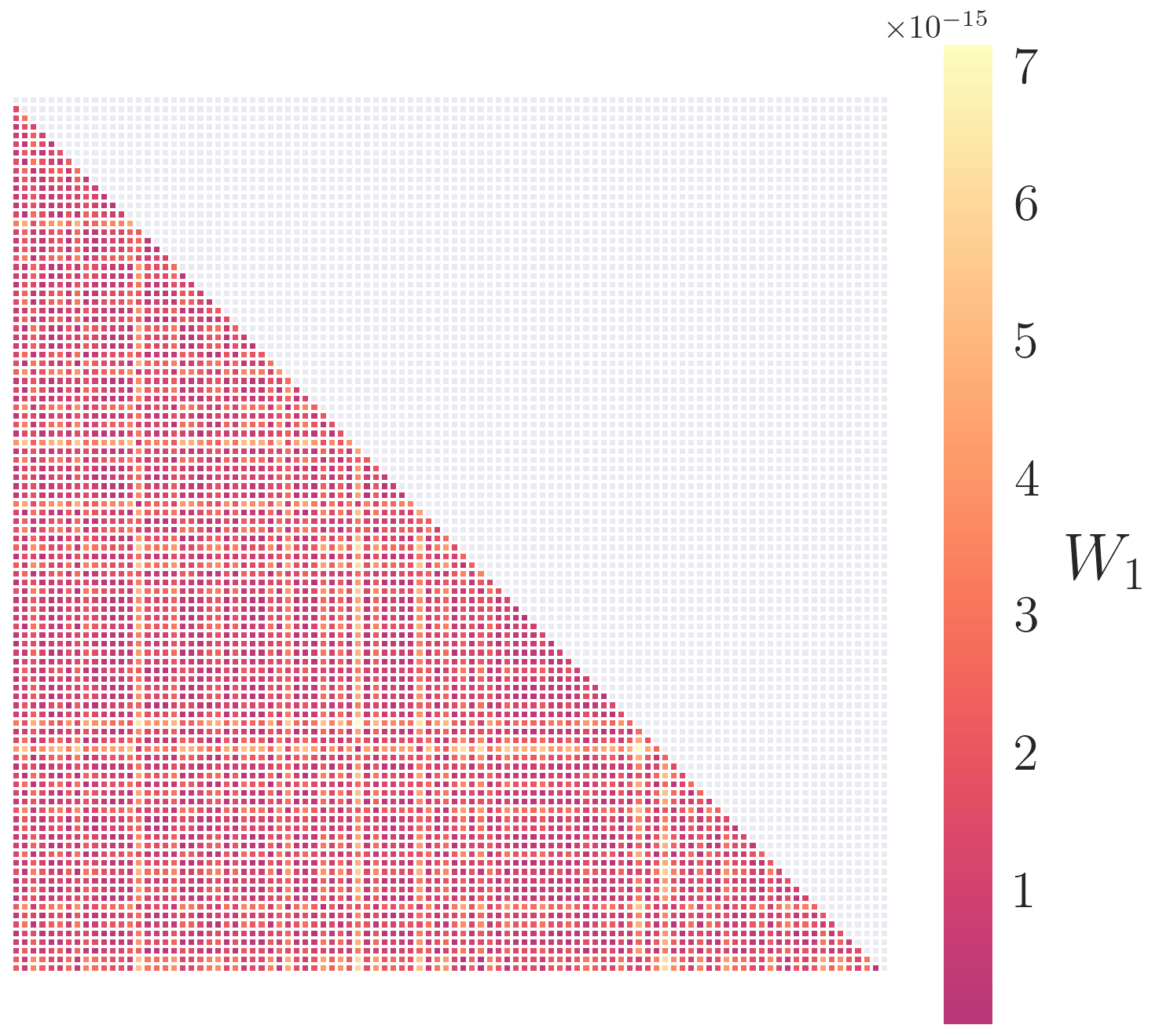}
		\caption{$h$}
		\label{subfig:h}
	\end{subfigure}
	\caption{First moment of the WD, $W_1$ (colour scale), calculated between each pair of one-dimensional posteriors for the representative system in Table \ref{tab:parameters_and_priors} across $10^3$ realisations as discussed in Section \ref{sec:appendix_WD_results}, for each component of $\boldsymbol{\theta}_{\rm gw}$. $W_1$ provides an upper bound on the difference in expected values between any two probability distributions. $W_1$ is generally low across all parameters and posteriors. $\boldsymbol{W}_1(\theta)$ is the vector for variable $\theta$ of $100 \choose 2$  $W_1$ values that are plotted in each panel. The Pearson correlation coefficient between $\boldsymbol{W}_1(\iota)$ and $\boldsymbol{W}_1(h)$ (i.e. the  $\iota$-$h$ panels) is $0.75$, and between the $\boldsymbol{W}_1(\delta)$ and $\boldsymbol{W}_1(\alpha)$ (i.e. the $\delta$-$\alpha$ panels) is $0.60$.} \label{fig:pairwise_wasserstein}
\end{figure*}
Table \ref{tab:Wasserstein} summarizes the median value of the first moment of the WD, $W_1$, between the $5\times 10^5$ pairs of one-dimensional posteriors across the $10^3$ realisations analysed in Section \ref{sec:multiple_noise}, for each of the seven parameters in ${\boldsymbol{\theta}}_{\rm gw}$. To assist with reproducibility, we present in Figure \ref{fig:pairwise_wasserstein} the WD values for a subset (numbering $10^2$) of the realisations. We plot a subset of realisations to avoid overcrowding the figure. The subset corresponds to the 100 realisations plotted in Figure \ref{fig:corner_plot_2}. Figure \ref{fig:pairwise_wasserstein} contains seven subplots, one for each element of ${\boldsymbol{\theta}}_{\rm gw}$. Each subplot is a lower triangular heat-map, where each point denotes the $W_1$ value between a pair of one-dimensional posteriors for that element. Lower values of $W_1$ are magenta; higher values of $W_1$ are yellow. Note that the heat-map colour scale is not the same in every subplot. Instead the colour scale is set by the minimum and maximum $W_1$ values for each parameter. Normally it would be preferable to set the colour scale by the domain of the prior, but doing so renders the heat-map uniform, because the WD is generally much smaller than the prior domain (cf. Table \ref{tab:Wasserstein}). \newline

Figure \ref{fig:pairwise_wasserstein} is consistent with the summary results presented in Table \ref{tab:Wasserstein}, and agrees with the conclusions drawn in Section \ref{sec:multiple_noise}, namely that the nested sampling scheme repeatedly converges to similar posteriors, for different realisations of the data, although dispersion remains. The $W_1$ value is small compared to the width of the prior, cf. Figure \ref{fig:corner_plot_2} and $W_{1 \rm, median, prior}$ in Table \ref{tab:Wasserstein}. The $W_1$ value is significant compared to the injection value (up to 140 \% for $h_0$, typically $\lessapprox$ 10 \% for other parameters; see Table \ref{tab:Wasserstein}), cf. $W_{1 \rm, median, inj}$ in Table \ref{tab:Wasserstein}. Qualitatively, Figure \ref{fig:pairwise_wasserstein} exhibits some correlated structure between panels. Specifically Figure \ref{subfig:iota} and Figure \ref{subfig:h} ($\iota$ and $h$ respectively) appear correlated, as do Figure \ref{subfig:delta} and Figure \ref{subfig:alpha} ($\delta$ and $\alpha$ respectively). Quantitively, we define $\boldsymbol{W}_1(\theta)$ as the vector of length $100 \choose 2$ $W_1$ values for variable $\theta$ that are plotted in each panel of Figure \ref{fig:pairwise_wasserstein}. The Pearson correlation coefficient between $\boldsymbol{W}_1(\iota)$ and $\boldsymbol{W}_1(h)$ is $0.75$, and between the $\boldsymbol{W}_1(\delta)$ and $\boldsymbol{W}_1(\alpha)$ is $0.60$.

\bsp	
\label{lastpage}
\end{document}